\documentclass[a4paper,11pt]{article}

\usepackage{amsmath,amsthm,amsbsy,amssymb,amsfonts,graphicx,mathrsfs,bm,cite,color}
\usepackage{hyperref}
\usepackage[utf8]{inputenc}
\usepackage[all]{xy}
\usepackage{longtable}

\makeatletter
\catcode`\@=11
\@addtoreset{equation}{section}

\makeatother

\makeatletter

\@addtoreset{table}{section}
\makeatother

\renewcommand{\thefootnote}{\arabic{footnote}}

\newcommand{\Exp}[1]{\operatorname{e}^{#1}}
\newcommand{\abs}[1]{\lvert {#1} \rvert}
\newcommand{\rmd}{{\mathrm{d}}}
\newcommand{\nn}{\nonumber}

\newcommand{\cA}{\mathcal A}\newcommand{\cB}{\mathcal B}

\newcommand{\cE}{\mathcal E}
\newcommand{\cH}{\mathcal H}

\newcommand{\cM}{\mathcal M}\newcommand{\cN}{\mathcal N}

\newcommand{\cT}{\mathcal T}

\newcommand{\sfh}{\mathsf{h}}

\newcommand{\sfk}{\mathsf{k}}
\newcommand{\sfm}{\mathsf{m}}
\newcommand{\sfn}{\mathsf{n}}
\newcommand{\sfp}{\mathsf{p}}
\newcommand{\sfq}{\mathsf{q}}
\newcommand{\sfr}{\mathsf{r}}
\newcommand{\sfs}{\mathsf{s}}
\newcommand{\sft}{\mathsf{t}}
\newcommand{\sfx}{\mathsf{x}}
\newcommand{\sfy}{\mathsf{y}}

\newcommand{\sfD}{\mathsf{D}}
\newcommand{\sfE}{\mathsf{E}}
\newcommand{\sfF}{\mathsf{F}}
\newcommand{\sfH}{\mathsf{H}}
\newcommand{\sfI}{\mathsf{I}}
\newcommand{\sfJ}{\mathsf{J}}
\newcommand{\sfK}{\mathsf{K}}
\newcommand{\sfL}{\mathsf{L}}
\newcommand{\sfM}{\mathsf{M}}
\newcommand{\sfN}{\mathsf{N}}
\newcommand{\sfR}{\mathsf{R}}
\newcommand{\sfT}{\mathsf{T}}
\newcommand{\sfY}{\mathsf{Y}}

\makeatletter
\newcommand*{\rmT}{{\mathpalette\@transpose{}}}
\newcommand*{\@transpose}[2]{\raisebox{\depth}{$\m@th#1\intercal$}}
\makeatother

\newcommand{\SL}{\text{SL}}
\newcommand{\OO}{\text{O}}

\newcommand{\bdelta}{{\boldsymbol\delta}}

\newcommand{\tr}{\text{tr}}

\newcommand{\MA}{\hat{A}}

\newcommand{\MG}{\hat{G}}
\newcommand{\MN}{\hat{N}}
\newcommand{\Mg}{\hat{g}}

\renewcommand{\AA}{\mathscr{A}}
\newcommand{\AB}{\mathscr{B}}
\newcommand{\AC}{\mathscr{C}}

\newcommand{\AG}{\mathscr{G}}

\newcommand{\APhi}{\Phi}
\newcommand{\Ag}{g}

\newcommand{\BA}{\mathsf{A}}
\newcommand{\BB}{\mathsf{B}}
\newcommand{\BC}{\mathsf{C}}

\newcommand{\BG}{\mathsf{G}}

\newcommand{\BN}{\mathsf{N}}
\newcommand{\BPhi}{\varphi}
\newcommand{\Bg}{\mathsf{g}}
\newcommand{\Bm}{\mathsf{m}}

\newcommand{\By}{\mathsf{y}}
\newcommand{\Ay}{y}
\newcommand{\Az}{z}

\newcommand{\GT}{\mathbb{T}}

\allowdisplaybreaks[3]

\begin{document}

\begin{titlepage}
\renewcommand{\thefootnote}{\fnsymbol{footnote}}

%\vspace*{-1cm} 
%\begin{flushright}
%
%\end{flushright}

\vspace*{1.0cm}

\begin{center}
\Large\textbf{Exotic branes and mixed-symmetry potentials II: \linebreak Duality rules and exceptional $p$-form gauge fields}
\end{center}

\vspace{1.0cm}

\centerline{
{\large Jos\'e J.~Fern\'andez-Melgarejo$^{a}$}%
\footnote{E-mail address: \texttt{jj.fernandezmelgarejo@um.es}},
{\large Yuho Sakatani$^{b}$}%
\footnote{E-mail address: \texttt{yuho@koto.kpu-m.ac.jp}}, 
{\large Shozo Uehara$^{b}$}%
\footnote{E-mail address: \texttt{uehara@koto.kpu-m.ac.jp}}
}

\vspace{0.2cm}

\begin{center}
${}^a${\it Departamento de F\'isica, Universidad de Murcia,}\\
{\it Campus de Espinardo, 30100 Murcia, Spain}

\vspace*{1mm}

${}^b${\it Department of Physics, Kyoto Prefectural University of Medicine,}\\
{\it Kyoto 606-0823, Japan}
\end{center}

\vspace*{1mm}

\begin{abstract}
In $U$-duality-manifest formulations, supergravity fields are packaged into covariant objects such as the generalized metric and $p$-form fields $\cA_p^{I_p}$. While a parameterization of the generalized metric in terms of supergravity fields is known for $U$-duality groups $E_n$ with $n\leq 8$, a parameterization of $\cA_p^{I_p}$ has not been fully determined. In this paper, we propose a systematic method to determine the parameterization of $\cA_p^{I_p}$, which necessarily involves mixed-symmetry potentials. We also show how to systematically obtain the $T$- and $S$-duality transformation rules of the mixed-symmetry potentials entering the multiplet. As the simplest non-trivial application, we find the parameterization and the duality rules associated with the dual graviton. Additionally, we show that the 1-form field $\cA_1^{I_1}$ can be regarded as the generalized graviphoton in the exceptional spacetime.
\end{abstract}
\thispagestyle{empty}
\end{titlepage}

\tableofcontents

\setcounter{footnote}{0}

\newpage

\section{Introduction}

In the previous paper \cite{1907.07177}, we conducted a detailed survey of mixed-symmetry potentials in 11D and type II supergravities. 
By considering their reduction to $d$ dimensions, they yield various $p$-form fields $\cA_p^{I_p}$, which transform covariantly under $E_n$ $U$-duality transformation ($n=11-d$). 
In the $U$-duality-covariant formulation of supergravity known as exceptional field theory (EFT) \cite{1308.1673,1312.0614,1312.4542,1406.3348} (see \cite{hep-th:0307098,hep-th:0312247,hep-th:0406150,0712.1795,0901.1581,1008.1763,1110.3930,1111.0459,1208.5884} for earlier fundamental works), and the $U$-duality-manifest approaches to brane actions \cite{1009.4657,1712.07115,1712.10316,1802.00442}, the $p$-form fields $\cA_p^{I_p}$ play an important role in providing $U$-duality-covariant descriptions. 
However, to make contact with the standard descriptions in supergravity and brane actions, explicit parameterizations of $\cA_p^{I_p}$ are needed. 
In this paper, we propose a systematic method to determine the parameterization of $\cA_p^{I_p}$ by utilizing the equivalence between M-theory (or type IIA theory) and type IIB theory. 
In our method, in addition to the parameterization of the $p$-form fields, the duality transformation rules of various potentials can also be obtained. 
As the first non-trivial example, we obtain the $T$- and $S$-duality rules for the dual graviton, Eqs.~\eqref{eq:Busc-B6A-B6B}--\eqref{eq:Busc-A71A-A71B} and \eqref{eq:dual-graviton-S-dual}, respectively.

In EFT, the fundamental fields are the generalized metric $\cM_{IJ}$ and $p$-form fields $\cA_p^{I_p}$, as well as certain auxiliary fields. 
For $E_n$ EFT with $n\leq 8$\,, the parameterization of the generalized metric has been determined in \cite{1111.0459,1303.2035} by means of the bosonic fields in 11D supergravity. 
The parameterization in terms of type IIB supergravity has been determined in \cite{1405.7894,1612.08738} for $E_n$ EFT with $n\leq 7$\,. 
They are nothing more than the two different parameterizations of the same object $\cM_{IJ}$, and as was concretely realized in \cite{1701.07819}, we can relate the two parameterizations through some redefinitions of fields. 
As was shown in \cite{1701.07819}, by rewriting the M-theory fields in terms of type IIA fields, these field redefinitions are precisely the $T$-duality transformation rule. 
However, the analysis of \cite{1701.07819} is limited to the $E_n$ EFT with $n\leq 7$\,, where the generalized metric contains only the standard $p$-form potentials. 
In this paper, we extend their analysis to the case of $E_8$ EFT, and find the $T$-duality and $S$-duality rules for the dual graviton.
This gives a non-trivial check of our duality rules for the dual graviton mentioned in the first paragraph. 

If we look at the explicit parameterization of the 1-form field $\cA_1^{I}$\,, its first component $\cA_1^{i}$ is the graviphoton. 
In 11D, the graviphoton is defined as $\MA_\mu^i \equiv \bm{\Mg}_{\mu\nu}\, \Mg^{\nu i}$\,, by using the 11D inverse metric $\Mg^{\hat{M}\hat{N}}$ and the metric $\bm{\Mg}_{\mu\nu}$ in the external spacetime. 
In this paper, we propose that the 1-form field $\cA_1^{I}$ can be regarded as a generalized graviphoton in the exceptional spacetime
\begin{align}
 \cA_\mu^I = \bm{m}_{\mu\nu}\, \cM^{\nu I} \,,\qquad 
 \bm{m} \equiv (\cM^{\mu\nu})^{-1} \,,
\end{align}
where $\cM^{\hat{I}\hat{J}}$ is the inverse generalized metric in $E_{11}$ EFT (see \cite{1907.02080} for a recent work) and $\hat{I}$ is the index for the $l_1$-representation \cite{hep-th:0307098} of $E_{11}$ that contains $\{\mu,\,I\}$ as a subset. 
We also find that the parameterizations of the higher $p$-form fields $\cA_p^{I_p}$ ($p\geq 2$) can be easily obtained from that of the 1-form $\cA_1^{I}$ through a simple antisymmetrization of indices. 

\section{Parameterization of the 1-form $\cA_1^{I}$}
\label{sec:1-form}

In this section, we explain our method to determine the parameterizations of the 1-form $\cA_1^{I}$\,. 
The index $I$ transforms in a fundamental representation of the $E_{n}$ algebra with a Dynkin label $[1,0,\dotsc,0]$, known as the vector representation or the particle multiplet. 
Our approach relies on the existence of two equivalent descriptions of EFT by deleting different nodes of the $E_n$ Dynkin diagram, M-theory and type IIB theory (see \cite{1907.07177} and references therein for details):
\begin{align}
\scalebox{0.7}{
 \xygraph{
    *\cir<6pt>{} ([]!{+(0,-.4)} {\alpha_1}) - [r]
    *\cir<6pt>{} ([]!{+(0,-.4)} {\alpha_2}) - [r]
    \cdots ([]!{+(0,0.8)} {\text{M-theory}}) ([]!{+(0,-.4)} {}) - [r]
    *\cir<6pt>{} ([]!{+(0,-.4)} {\alpha_{n-4}}) - [r]
    *\cir<6pt>{} ([]!{+(0,-.4)} {\alpha_{n-3}})
(
        - [u] *{\scalebox{2}{$\times$}}*\cir<6pt>{} ([]!{+(.5,0)} {\alpha_{n}}),
        - [r] *\cir<6pt>{} ([]!{+(0,-.4)} {\alpha_{n-2}})
        - [r] *\cir<6pt>{} ([]!{+(0,-.4)} {\alpha_{n-1}})
)}\quad ,
 \qquad\qquad 
 \xygraph{
    *\cir<6pt>{} ([]!{+(0,-.4)} {\alpha_1}) - [r]
    *\cir<6pt>{} ([]!{+(0,-.4)} {\alpha_2}) - [r]
    \cdots ([]!{+(0,0.8)} {\text{Type IIB theory}}) ([]!{+(0,-.4)} {}) - [r]
    *\cir<6pt>{} ([]!{+(0,-.4)} {\alpha_{n-4}}) - [r]
    *\cir<6pt>{} ([]!{+(0,-.4)} {\alpha_{n-3}})
(
        - [u] *\cir<6pt>{} ([]!{+(.5,0)} {\alpha_{n}}),
        - [r] *{\scalebox{2}{$\times$}}*\cir<6pt>{} ([]!{+(0,-.4)} {\alpha_{n-2}})
        - [r] *\cir<6pt>{} ([]!{+(0,-.4)} {\alpha_{n-1}})
)}\quad.
}
\label{eq:Dynkin-decomposition}
\end{align}
As is explained in the accompanying paper \cite{1907.07177}, in terms of M-theory, the 1-form field $\cA_1^{I}$ is decomposed into $\SL(n)$ tensors as follows:
\begin{align}
 \bigl(\cA_\mu^{I}\bigr)= \bigl(\cA_\mu^i,\,\tfrac{\cA_{\mu; i_1i_2}}{\sqrt{2!}},\,\tfrac{\cA_{\mu; i_1\cdots i_5}}{\sqrt{5!}},\,\tfrac{\cA_{\mu; i_1\cdots i_7,i}}{\sqrt{7!}},\dotsc\bigr)\,,
\label{eq:M-decomp}
\end{align}
where $i,j=d,\dotsc,9,\Az$ are indices of the fundamental representation of $\SL(n)$. 
On the other hand, in terms of type IIB theory, the 1-form field is decomposed into $\SL(n-1)\times\SL(2)$ tensors as follows:
\begin{align}
 \bigl(\bm{\cA}_\mu^{\sfI}\bigr) &= \bigl(\bm{\cA}_\mu^\sfm,\, \bm{\cA}_{\mu; \sfm}^\alpha ,\,\tfrac{\bm{\cA}_{\mu; \sfm_1\sfm_2\sfm_3}}{\sqrt{3!}},\,\tfrac{\bm{\cA}_{\mu; \sfm_1\cdots \sfm_5}^\alpha}{\sqrt{5!}},\,\tfrac{\bm{\cA}_{\mu; \sfm_1\cdots \sfm_6,\sfm}}{\sqrt{6!}}, \dotsc\bigr) \,,
\label{eq:IIB-decomp}
\end{align}
where $\alpha,\beta =1,2$ are the $\SL(2)$ $S$-duality indices and $\sfm,\sfn=d,\dotsc,9$ are indices of the fundamental representation of $\SL(n-1)$. 
In order to stress the difference between the two parameterizations, we have denoted the 1-form in type IIB parameterization by $\bm{\cA}_\mu^{\sfI}$\,. 
Although we know the tensor structures of each component, it is not obvious how to determine the explicit parameterization in terms of the standard supergravity fields, which is the main subject of this paper.

As demonstrated in \cite{1701.07819}, the two decompositions \eqref{eq:M-decomp} and \eqref{eq:IIB-decomp} can be related by using the equivalence between M-theory on $T^2$ with coordinates $(x^\alpha)=(x^\Ay,\,x^\Az)$ and type IIB theory on $S^1$ with a coordinate $\sfx^\By$:
\begin{align}
\begin{gathered}
 \xymatrix{
 \text{M-theory/}T^2 \ar[d]_{\text{compactification on $x^\Az$}} \ar@{<->}[drrr]^{\text{our map}} & & & \\
 \text{ Type IIA theory/$S^1$ } \ar@{<->}[rrr]_{\text{$T$-duality along $x^\Ay$/$\sfx^\By$}} & & & \text{ Type IIB theory/$S^1$ } .\\
}
\end{gathered}
\label{eq:our-map}
\end{align}
Here, $x^\Az$ is a coordinate along the M-theory circle, and the coordinate $x^\Ay$ in M-theory (or type IIA theory) is mapped to the coordinate $\sfx^\By$ in type IIB theory under the $T$-duality. 
By using the map, we can rewrite various quantities in M-theory in terms of type IIB supergravity.

\subsection{Supergravity fields}

In order to discuss the parameterization, we will briefly explain the supergravity fields considered in this paper. 
We basically follow the convention of \cite{1701.07819}. 

\paragraph{11D supergravity:}
In 11D supergravity, we consider the following bosonic fields,
\begin{align}
 \{\Mg_{\hat{M}\hat{N}},\,\MA_{\hat 3},\,\MA_{\hat 6},\,\MA_{\hat 8,\hat 1}\}\qquad \bigl(\hat{M},\hat{N}=0,\dotsc,9,\Az\bigr)\,. 
\end{align}
The standard potentials $\MA_3$ and $\MA_6$ couple to the M2-brane and M5-brane, respectively, while the dual graviton $\MA_{8,1}$ couples to the Kaluza--Klein monopole $6^1$ (sometimes called MKK) \cite{hep-th:9802199}. 
When we consider a compactification to $d$ dimensions, the 11D metric $\Mg_{\hat{M}\hat{N}}$ is decomposed as
\begin{align}
 (\Mg_{\hat{M}\hat{N}}) = \begin{pmatrix}
 \bm{\Mg}_{\mu\nu} +\MA_\mu^k\,\MG_{kl}\,\MA_\nu^l & -\MA_\mu^k\,\MG_{kj} \\
 -\MG_{ik}\,\MA_\nu^k & \MG_{ij}
 \end{pmatrix} \qquad (\mu,\nu=0,\dotsc,d-1),
\end{align}
where we have defined the graviphoton as $\MA_\mu^i \equiv -\Mg_{\mu k}\,\MG^{ki} = \bm{\Mg}_{\mu\nu}\, \Mg^{\nu i}$\,. 

\paragraph{Type IIA supergravity:}

When we consider type IIA supergravity, we use the following standard 11D--10D map:
\begin{align}
\begin{split}
 &(\Mg_{\hat{M}\hat{N}})\equiv \begin{pmatrix}
 \Mg_{MN} & \Mg_{M\Az} \\ \Mg_{\Az N} & \Mg_{\Az\Az}
 \end{pmatrix}
 =\begin{pmatrix}
 \Exp{-\frac{2}{3}\,\APhi}\,\Ag_{MN}+\Exp{\frac{4}{3}\,\APhi}\,\AC_M\, \AC_N & \Exp{\frac{4}{3}\,\APhi}\, \AC_M \\ \Exp{\frac{4}{3}\,\APhi}\, \AC_N & \Exp{\frac{4}{3}\,\APhi}
 \end{pmatrix} \,,
\\
 &\MA_{\hat{3}} = \AC_3 + \AB_2\wedge \rmd x^\Az \,,\qquad 
  \MA_{\hat{6}} = \AB_6 + \bigl(\AC_5 - \tfrac{1}{2!}\, \AC_3\wedge \AB_2\bigr)\wedge \rmd x^\Az \,,
\end{split}
\label{eq:11D-10D}
\end{align}
where we have added the hat to the subscript, like $\MA_{\hat{p}}$\,, to stress that it is a $p$-form in 11D. 
In our convention, the dual graviton $\MA_{\hat{8},\hat{1}}=\{\MA_{\hat{8}, 1},\,\MA_{\hat{8}, \Az}\}$ follows the 11D--10D map,
\begin{align}
 \MA_{\hat{8}, 1} = \AA_{8, 1} + \AA_{7, 1}\wedge \rmd x^{\Az} \,, \qquad
 \MA_{\hat{8}, \Az} = \AA_{8} + \bigl(\AC_7 - \tfrac{1}{3!}\,\AC_3\wedge\AB_2\wedge\AB_2\bigr)\wedge\rmd x^{\Az} \,,
\end{align}
where $\MA_{\hat{8}, \Az}$ corresponds to $\hat{\tilde{N}}$ studied in \cite{hep-th:9802199}. 
The metric and the graviphoton are defined as
\begin{align}
 (\Ag_{MN}) = \begin{pmatrix}
 \bm{\Ag}_{\mu\nu} +\AA_\mu^p\,\AG_{pq}\,\AA_\nu^q & -\AA_\mu^p\,\AG_{pn} \\
 -\AG_{mp}\,\AA_\nu^p & \AG_{mn}
 \end{pmatrix},\qquad 
 \AA_\mu^m \equiv \bm{\Ag}_{\mu\nu}\,\Ag^{\nu m} \,. 
\end{align}
Then we find the 11D--10D map for the graviphoton:
\begin{align}
 \MA_\mu^m = \AA_\mu^m \,, \qquad 
 \MA_\mu^\Az = -\bigl(\AC_\mu + \AA_\mu^p\,\AC_p\bigr)\,. 
\label{eq:11D-10D-photon}
\end{align}

\paragraph{Type IIB supergravity:}
In type IIB theory, in addition to the standard Einstein-frame metric $\Bg_{MN}$, we consider the following $\SL(2)$ $S$-duality-covariant tensors:
\begin{align}
 &(\Bm_{\alpha\beta}) \equiv \Exp{\BPhi} \begin{pmatrix} \Exp{-2\,\BPhi} +(\BC_0)^2 & \BC_0 \\ \BC_0 & 1 \end{pmatrix},\qquad 
 (\BA^\alpha_2) \equiv \begin{pmatrix} \BB_2 \\\ -\BC_2\end{pmatrix} ,
\\
 &\BA_4 \equiv \BC_4 - \frac{1}{2}\,\BC_2\wedge \BB_2\,,\qquad
 (\BA^\alpha_6) \equiv \begin{pmatrix} \BC_6 - \BC_4\wedge \BB_2 +\frac{1}{3}\, \BB_2\wedge \BC_2 \wedge \BB_2 \\ -\bigl(\BB_6 - \BC_4\wedge \BC_2 + \frac{1}{6}\,\BB_2\wedge \BC_2 \wedge \BC_2\bigr) \end{pmatrix} .
\end{align}
We also consider the dual graviton $\BA_{7,1}$\,, whose behavior under duality transformations is to be determined. 
Upon compactification to $d$ dimensions, the graviphoton is introduced as
\begin{align}
 (\Bg_{MN}) = \begin{pmatrix}
 \bm{\Bg}_{\mu\nu} + \BA_\mu^\sfp\, \BG_{\sfp\sfq}\, \BA_\nu^\sfq & - \BA_\mu^\sfp\, \BG_{\sfp\sfn} \\
 -\BG_{\sfm\sfp}\, \BA_\nu^\sfp & \BG_{\sfm\sfn}
 \end{pmatrix} ,\qquad \BA_\mu^\sfm \equiv \bm{\Bg}_{\mu\nu}\,\Bg^{\nu \sfm}\,. 
\end{align}

\subsection{Strategy: Linear map}

Here, let us explain the detailed procedure, how to determine the parameterization of the 1-form in both the M-theory and type IIB languages:
\begin{align}
 (\cA_\mu^I) = {\footnotesize\begin{pmatrix}
 \cA_\mu^i \\
 \frac{\cA_{\mu; i_1i_2}}{\sqrt{2!}} \\
 \frac{\cA_{\mu; i_1\cdots i_5}}{\sqrt{5!}} \\
 \frac{\cA_{\mu; i_1\cdots i_7,i}}{\sqrt{7!}} \\
 \vdots
\end{pmatrix}},\qquad\quad
 (\bm{\cA}_\mu^\sfI) = {\footnotesize\begin{pmatrix}
 \bm{\cA}_\mu^\sfm \\
 \bm{\cA}^\alpha_{\mu; \sfm} \\
 \frac{\bm{\cA}_{\mu; \sfm_1\sfm_2\sfm_3}}{\sqrt{3!}} \\
 \frac{\bm{\cA}^\alpha_{\mu; \sfm_1\cdots \sfm_5}}{\sqrt{5!}} \\
 \tfrac{\bm{\cA}_{\mu; \sfm_1\cdots \sfm_6,\sfm}}{\sqrt{6!}} \\
 \vdots
 \end{pmatrix}},
\label{eq:1-form-MB}
\end{align}
where ellipses stand for the rest of the components that complete the $U$-duality multiplet that potentially involve further mixed-symmetry potentials. 

To determine the parameterization, we make the following modest assumptions:
\begin{itemize}
\item The M-theory fields $\cA_{1; p,q,r,\dotsc}$ and the type IIB fields $\bm{\cA}^{\alpha_1\cdots\alpha_s}_{1; p,q,r,\dotsc}$ are respectively parameterized by the following fields:
\begin{alignat}{2}
 &\text{M-theory: }&\quad &\{\MA_\mu^i ,\, \MA_{\hat{3}},\, \MA_{\hat{6}},\,\MA_{\hat{8},\hat{1}},\,\dotsc \}\,, 
\\
 &\text{Type IIB theory: }&\quad &\{\BA_\mu^\sfm ,\, \BA^\alpha_2,\, \BA_4,\, \BA^\alpha_6,\,\BA_{7,1},\,\dotsc \}\,.
\end{alignat}

\item The top form is normalized with weight one:
\begin{align}
\begin{split}
 \cA_{\mu ; p,q,r,\dotsc} &= \MA_{\mu p,q,r,\dotsc} + \text{(sum of products of potentials)} \,,
\\
 \bm{\cA}^{\alpha_1\cdots\alpha_s}_{\mu ; p,q,r,\dotsc} &= \BA^{\alpha_1\cdots\alpha_s}_{\mu p,q,r,\dotsc} + \text{(sum of products of potentials)} \,.
\end{split}
\end{align}
\end{itemize}
According to these, the first components of the 1-forms should be, respectively,
\begin{align}
 \cA_\mu^i = \MA_\mu^i\quad (\text{M-theory})\,,\qquad \bm{\cA}_\mu^\sfm = \BA_\mu^\sfm\quad (\text{type IIB})\,. 
\label{eq:graviphoton-M-B}
\end{align}
In the following, we explain the procedure to determine the components with higher level, which is based on \cite{1701.07819}. 
In order to utilize the map \eqref{eq:our-map}, we decompose the physical coordinates on the $n$-torus in M-theory as $(x^i)=(x^a,\,x^\alpha)$ ($a,b=1,\dotsc,n-2$) and those on the $(n-1)$-torus in type IIB theory as $(\sfx^\sfm)=(\sfx^a,\,\sfx^\By)$\,. 
Under the decomposition, the 1-form fields \eqref{eq:1-form-MB} are decomposed into $\SL(n-2)\times \SL(2)$ tensors as follows:
\begin{align}
 {\footnotesize
 (\cA_\mu^I) = \begin{pmatrix}
 \cA_\mu^a \\ \cA_\mu^\alpha \\ \hline
 \frac{\cA_{\mu; a_1a_2}}{\sqrt{2!}} \\ \cA_{\mu; a \alpha} \\
 \cA_{\mu; \Ay\Az} \\ \hline \frac{\cA_{\mu; a_1\cdots a_5}}{\sqrt{5!}} \\
 \frac{\cA_{\mu; a_1\cdots a_4\alpha}}{\sqrt{4!}} \\
 \frac{\cA_{\mu; a_1a_2a_3 \Ay\Az}}{\sqrt{3!}} \\ \hline
 \frac{\cA_{\mu; a_1\cdots a_5 \Ay\Az,a}}{\sqrt{5!}} \\
 \frac{\cA_{\mu; a_1\cdots a_5 \Ay\Az,\alpha}}{\sqrt{5!}} \\
 \vdots
 \end{pmatrix},\qquad\quad 
 (\cA_\mu^\sfI) = \begin{pmatrix}
 \bm{\cA}_\mu^a \\
 \bm{\cA}_\mu^{\By} \\ \hline
 \bm{\cA}^\alpha_{\mu; a} \\ \bm{\cA}^\alpha_{\mu; \By} \\ \hline
 \frac{\bm{\cA}_{\mu; a_1a_2a_3}}{\sqrt{3!}} \\ \frac{\bm{\cA}_{\mu; a_1a_2 \By}}{\sqrt{2!}} \\ \hline
 \frac{\bm{\cA}^\alpha_{\mu; a_1\cdots a_5}}{\sqrt{5!}} \\
 \frac{\bm{\cA}^\alpha_{\mu; a_1\cdots a_4\By}}{\sqrt{4!}} \\ \hline
 \frac{\bm{\cA}_{\mu; a_1\cdots a_5\By,a}}{\sqrt{5!}} \\
 \frac{\bm{\cA}_{\mu; a_1\cdots a_5\By,\By}}{\sqrt{5!}} \\
 \vdots
 \end{pmatrix} , 
}
\label{eq:SL(n-2)xSL(2)}
\end{align}
where toroidal directions (either compactified or $T$-dualized) are shown explicitly. 
In terms of the Dynkin diagram given in \eqref{eq:Dynkin-decomposition}, in M-theory we have first performed the level decomposition associated with the node $\alpha_n$. 
Secondly, we have done the level decomposition associated with $\alpha_{n-2}$\,. 
On the other hand, in type IIB theory the order is reversed. 
In the end, we obtain the same decomposition. 
Indeed, the set of $\SL(n-2)\times \SL(2)$ tensors appearing in \eqref{eq:SL(n-2)xSL(2)} has the same structure. 
Then, we make the following identifications \cite{1701.07819}:
\begin{align}
{\footnotesize
 \begin{pmatrix}
 \cA_\mu^a \\ \cA_\mu^\alpha \\ \hline
 \frac{\cA_{\mu a_1a_2}}{\sqrt{2!}} \\
 \cA_{\mu; a \alpha} \\
 \cA_{\mu; \Ay\Az} \\ \hline \frac{\cA_{\mu; a_1\cdots a_5}}{\sqrt{5!}} \\
 \frac{\cA_{\mu; a_1\cdots a_4\alpha}}{\sqrt{4!}} \\
 \frac{\cA_{\mu; a_1a_2a_3 \Ay\Az}}{\sqrt{3!}} \\ \hline
 \frac{\cA_{\mu; a_1\cdots a_5 \Ay\Az,a}}{\sqrt{5!}} \\
 \frac{\cA_{\mu; a_1\cdots a_5 \Ay\Az,\alpha}}{\sqrt{5!}} \\
 \vdots
 \end{pmatrix}_{\text{M}}
 =\
\begin{pmatrix}
 \bm{\cA}_\mu^a \\
 \bm{\cA}_{\mu; \By}^\alpha \\ \hline
 \frac{\bm{\cA}_{\mu; a_1a_2\By}}{\sqrt{2!}} \\
 \bm{\cA}_{\mu; a}^\beta \,\epsilon_{\beta\alpha} \\
 \bm{\cA}_\mu^\By \\ \hline
 \frac{\bm{\cA}_{\mu; a_1\cdots a_5\By,\By}}{\sqrt{5!}} \\
 \frac{\bm{\cA}^\beta_{\mu; a_1\cdots a_4} \epsilon_{\beta\alpha}}{\sqrt{4!}} \\
 \frac{\bm{\cA}_{\mu; a_1a_2a_3}}{\sqrt{3!}} \\ \hline
 \frac{\bm{\cA}_{\mu; a_1\cdots a_5\By,a}}{\sqrt{5!}} \\
 \frac{\bm{\cA}_{\mu; a_1\cdots a_5}^\beta\,\epsilon_{\beta\alpha}}{\sqrt{5!}} \\
 \vdots
\end{pmatrix}_{\text{IIB}}
\ \text{or}\quad
 \begin{pmatrix}
 \bm{\cA}_\mu^a \\
 \bm{\cA}_\mu^{\By} \\ \hline
 \bm{\cA}^\alpha_{\mu; a} \\ \bm{\cA}^\alpha_{\mu; \By} \\ \hline
 \frac{\bm{\cA}_{\mu; a_1a_2a_3}}{\sqrt{3!}} \\ \frac{\bm{\cA}_{\mu; a_1a_2 \By}}{\sqrt{2!}} \\ \hline
 \frac{\bm{\cA}^\alpha_{\mu; a_1\cdots a_5}}{\sqrt{5!}} \\
 \frac{\bm{\cA}^\alpha_{\mu; a_1\cdots a_4\By}}{\sqrt{4!}} \\ \hline
 \frac{\bm{\cA}_{\mu; a_1\cdots a_5\By,a}}{\sqrt{5!}} \\
 \frac{\bm{\cA}_{\mu; a_1\cdots a_5\By,\By}}{\sqrt{5!}} \\
 \vdots
 \end{pmatrix}_{\text{IIB}}
 = \begin{pmatrix}
 \cA_\mu^a \\
 \cA_{\mu; \Ay\Az} \\ \hline
 \epsilon^{\alpha\beta}\,\cA_{\mu; a\beta} \\
 \cA_\mu^\alpha \\ \hline
 \frac{\cA_{\mu; a_1a_2a_3\Ay\Az}}{\sqrt{3!}} \\
 \frac{\cA_{\mu; a_1a_2}}{\sqrt{2!}} \\ \hline
 \frac{\epsilon^{\alpha\beta}\,\cA_{\mu; a_1\cdots a_5 \Ay\Az,\beta}}{\sqrt{5!}} \\
 \frac{\epsilon^{\alpha\beta}\,\cA_{\mu; a_1\cdots a_4\beta}}{\sqrt{4!}} \\ \hline
 \frac{\cA_{\mu; a_1\cdots a_5 \Ay\Az,a}}{\sqrt{5!}} \\
 \frac{\cA_{\mu; a_1\cdots a_5}}{\sqrt{5!}} \\
 \vdots
\end{pmatrix}_{\text{M}} \,,
}
\label{eq:M-B-map}
\end{align}
where we have defined
\begin{align}
 \epsilon \equiv (\epsilon^{\alpha\beta})\equiv(\epsilon_{\alpha\beta})\equiv \begin{pmatrix} 0 & 1 \\ -1 & 0 \end{pmatrix}.
\end{align}
We will refer to the set of linear relations established in \eqref{eq:M-B-map} as the \emph{linear map}. 
Actually, by using a constant matrix $S^I{}_\sfJ$, it can be rewritten as
\begin{align}
 \cA_\mu^I = S^I{}_\sfJ\, \bm{\cA}_\mu^\sfJ \,,\qquad \bm{\cA}_\mu^\sfI = (S^{-1})^\sfI{}_J\,\cA_\mu^J \,. 
\label{eq:linear-map-A}
\end{align}
We note that this identification was originally proposed in \cite{hep-th:0402140} in the context of $E_{11}$\,. 

Now, for simplicity, we assume the standard $T$-duality rule for the NS--NS fields\footnote{This assumption is not necessary in the approach discussed in Section \ref{sec:generalized-metric}.}
\begin{align}
\begin{split}
 \Ag_{AB} &\overset{\text{A--B}}{=} \Bg_{AB} - \frac{\Bg_{A \By}\,\Bg_{B \By}-\BB_{A \By}\,\BB_{B \By}}{\Bg_{\By\By}}\,,\qquad 
 \Ag_{A \Ay}\overset{\text{A--B}}{=}-\frac{\BB_{A \By}}{\Bg_{\By\By}}\,,\qquad 
 \Ag_{\Ay\Ay}\overset{\text{A--B}}{=}\frac{1}{\Bg_{\By\By}}\,,
\\
 \AB_{AB} &\overset{\text{A--B}}{=} \BB_{AB} - \frac{\BB_{A\By}\,\Bg_{B\By}-\Bg_{A\By}\,\BB_{B\By}}{\Bg_{\By\By}}\,,\qquad 
 \AB_{A\Ay} \overset{\text{A--B}}{=} -\frac{\Bg_{A\By}}{\Bg_{\By\By}} \,, 
\end{split}
\label{eq:Buscher-NS}
\end{align}
where $A,B=\{\mu,\,a\}$ (i.e.~nine directions except the $T$-dual direction $x^\Ay$ or $\sfx^\By$). 
From these, we obtain the $T$-duality rule for the graviphoton:
\begin{align}
 \AA_\mu^a \overset{\text{A--B}}{=} \BA_\mu^a\,,\qquad 
 \AA_\mu^\Ay \overset{\text{A--B}}{=} \BB_{\mu \By} + \BA_\mu^\sfp\,\BB_{\sfp\By} \,. 
\label{eq:Buscher-graviphoton}
\end{align}
By using the 11D--10D relation \eqref{eq:11D-10D-photon}, the first rule gives
\begin{align}
 \cA_\mu^a = \MA_\mu^a \overset{\text{M--A}}{=} \AA_\mu^a \overset{\text{A--B}}{=} \BA_\mu^a = \bm{\cA}_\mu^a \,,
\end{align}
which is nothing but the first row of \eqref{eq:M-B-map}. 

\subsection{Detailed procedures}

We will continue this process by considering the index structure. 
The second components of the 1-form in the M-theory and the type IIB parameterization are generically expanded as
\begin{align}
 \cA_{\mu; i_1i_2} = \MA_{\mu i_1i_2} + c_1\, \MA_\mu^k\,\MA_{ki_1i_2} \,,\qquad
 \bm{\cA}_{\mu; \sfm}^\alpha = \BA_{\mu \sfm}^\alpha + c_2\, \BA_\mu^\sfp\,\BA^\alpha_{\sfp\sfm} \,,
\end{align}
where $c_1$ and $c_2$ are parameters to be determined. 
From $\cA_\mu^\alpha \overset{\text{M--B}}{=} \bm{\cA}_{\mu; \By}^\alpha$ in \eqref{eq:M-B-map}, we have
\begin{align}
 \MA_\mu^\alpha = \cA_\mu^\alpha \overset{\text{M--B}}{=} \bm{\cA}_{\mu; \By}^\alpha = \BA_{\mu\By}^\alpha + c_2\, \BA_\mu^\sfp\,\BA^\alpha_{\sfp\By}\,. 
\label{eq:graviphoton-rule2}
\end{align}
On the other hand, the second rule of \eqref{eq:Buscher-graviphoton} and the 11D--10D relation \eqref{eq:11D-10D-photon} gives
\begin{align}
 \MA_\mu^\Ay \overset{\text{M--B}}{=} \AA_\mu^\Ay \overset{\text{A--B}}{=} \BB_{\mu \By} + \BA_\mu^\sfp\,\BB_{\sfp\By} \,,
\end{align}
and by comparing this with the $\alpha=\Ay$ component of \eqref{eq:graviphoton-rule2}\,, we find $c_2=1$\,. 

Similarly, the map $\cA_{\mu; \Ay\Az}\overset{\text{M--B}}{=}\bm{\cA}_\mu^\By$ in \eqref{eq:M-B-map} gives
\begin{align}
 \cA_{\mu; \Ay\Az} = \MA_{\mu \Ay\Az} + c_1\, \MA_\mu^a\,\MA_{a\Ay\Az} \overset{\text{M--B}}{=} \bm{\cA}_\mu^\By \,. 
\label{eq:c1-identification}
\end{align}
On the other hand, the second line of \eqref{eq:Buscher-NS} and the 11D--10D relation \eqref{eq:11D-10D} give
\begin{align}
 \MA_{AB\Az} &\overset{\text{M--B}}{=} \BB_{AB} - \frac{\BB_{A\By}\,\Bg_{B\By}-\Bg_{A\By}\,\BB_{B\By}}{\Bg_{\By\By}}\,,\qquad 
 \MA_{A\Ay\Az} \overset{\text{M--B}}{=} -\frac{\Bg_{A\By}}{\Bg_{\By\By}} \,.
\label{eq:A3-rule}
\end{align}
By substituting the second relation into the left-hand side of \eqref{eq:c1-identification} and using $\Bg_{\mu\By}=-(\BA_\mu^a\,\Bg_{a\By} + \BA_\mu^\By\,\Bg_{\By\By})$\,, we obtain
\begin{align}
 \frac{\BA_\mu^a\,\Bg_{a\By} + \BA_\mu^\By\,\Bg_{\By\By} - c_1\,\BA_\mu^a\,\Bg_{a\By}}{\Bg_{\By\By}} \overset{\text{B--M}}{=} \MA_{\mu \Ay\Az} + c_1\, \MA_\mu^a\,\MA_{a\Ay\Az} \overset{\text{M--B}}{=} \bm{\cA}_\mu^\By =\BA_\mu^\By \,,
\end{align}
which shows $c_1=1$\,. 
Thus, the parameterizations of $\cA_{\mu; i_1i_2}$ and $\bm{\cA}_{\mu; \sfm}^\alpha$ are determined as
\begin{align}
 \cA_{\mu; ij} = \MA_{\mu ij} + \MA_\mu^k\,\MA_{kij} \,,\qquad
 \bm{\cA}_{\mu; \sfm}^\alpha = \BA_{\mu\sfm}^\alpha + \BA_\mu^\sfp\,\BA^\alpha_{\sfp\sfm} \,.
\end{align}

In order to determine the parameterization of further components of the 1-forms, the $T$-duality rules \eqref{eq:Buscher-NS} are not enough and we need additional $T$-duality rules. 
To find the $T$-duality rules, we assume that
\begin{itemize}
\item the $T$-duality rules have the 9D covariance (in the nine directions $x^A$ orthogonal to the $T$-duality direction $\sfx^\By$)\,; 

\item the metric appears in the $T$-duality rule only through the combination $\frac{\Bg_{A\By}}{\Bg_{\By\By}}$ and the graviphoton does not appear explicitly. 
\end{itemize}
By using these assumptions, we obtain the set of standard $T$-duality rules. 

For example, from the $\alpha=\Az$ component of \eqref{eq:graviphoton-rule2}, we find
\begin{align}
 \MA_\mu^\Az \overset{\text{M--B}}{=} -\BC_{\mu\By} - \BA_\mu^\sfp\,\BC_{\sfp\By}\,.
\end{align}
In terms of the type IIA field, this is equivalent to
\begin{align}
 \AC_\mu + \AA_\mu^a\,\AC_a + \AA_\mu^\Ay\,\AC_\Ay \overset{\text{A--B}}{=} \BC_{\mu\By} + \BA_\mu^a\,\BC_{a\By}\,,
\end{align}
and by using the identity $\Ag_{\mu \Ay}=-(\AA_\mu^a\,\Ag_{ay} + \AA_\mu^\Ay\,\Ag_{\Ay\Ay})$\,, we obtain
\begin{align}
 \Bigl(\AC_\mu - \frac{\Ag_{\mu \Ay}}{\Ag_{\Ay\Ay}}\,\AC_\Ay\Bigr) + \AA_\mu^a\,\Bigl(\AC_a - \frac{\Ag_{a\Ay}}{\Ag_{\Ay\Ay}}\,\AC_\Ay\Bigr) \overset{\text{A--B}}{=} \BC_{\mu\By} + \BA_\mu^a\,\BC_{a\By}\,. 
\end{align}
From the assumption that the $T$-duality rule does not contain the graviphoton explicitly, this implies the standard $T$-duality rule:
\begin{align}
 \BC_{A\By} \overset{\text{B-A}}{=} \AC_A - \frac{\AC_\Ay\,\Ag_{A\Ay}}{\Ag_{\Ay\Ay}} \,,
\end{align}
or conversely,
\begin{align}
 \AC_A \overset{\text{A--B}}{=} \BC_{A\By} - \BC_0\,\BB_{A\By} \,, 
\end{align}
where we have employed the standard rule $\AC_\Ay \overset{\text{A--B}}{=} \BC_0$\,. 

Similarly, if we consider the linear map $\cA_{\mu; a\alpha} \overset{\text{M--B}}{=} \bm{\cA}_{\mu; a}^\beta\,\epsilon_{\beta\alpha}$ in \eqref{eq:M-B-map}, we find
\begin{align}
 \MA_{\mu a\alpha} + \MA_\mu^k\,\MA_{k a\alpha} \overset{\text{M--B}}{=} \bigl(\BA_{\mu a}^\beta + \BA_\mu^\sfp\,\BA^\beta_{\sfp a}\bigr)\,\epsilon_{\beta\alpha} \,.
\label{eq:3-form-2-form}
\end{align}
In particular, for $\alpha=\Ay$\,, we obtain a map between the type IIA/IIB fields,
\begin{align}
 \AC_{\mu a\Ay} + \AA_\mu^b\,\AC_{ba\Ay} -\bigl(\AC_\mu + \AA_\mu^m\,\AC_m\bigr)\,\AB_{a\Ay} \overset{\text{A--B}}{=} \BC_{\mu a} + \BA_\mu^b\,\BC_{ba} + \BA_\mu^\By\,\BC_{\By a} \,,
\end{align}
and this is equivalent to
\begin{align}
\begin{split}
 &\AC_{\mu a\Ay} - \AC_\mu \,\AB_{a\Ay} + \frac{\AC_\Ay\,\AB_{a\Ay}\, \Ag_{\mu \Ay}}{\Ag_{\Ay\Ay}}
 + \AA_\mu^b\,\Bigl(\AC_{ba\Ay} - \AC_b\,\AB_{a\Ay} + \frac{\AC_\Ay\,\AB_{a\Ay}\, \Ag_{b\Ay}}{\Ag_{\Ay\Ay}} \Bigr)
\\
 &\overset{\text{A--B}}{=}
 \BC_{\mu a} - \frac{\BC_{\By a}\,\Bg_{\mu\By}}{\Bg_{\By\By}}
 + \BA_\mu^b\,\Bigl(\BC_{ba} - \frac{\BC_{\By a}\,\Bg_{b\By}}{\Bg_{\By\By}}\Bigr) \,.
\end{split}
\end{align}
Then, we find the $T$-duality rule
\begin{align}
 \AC_{AB\Ay} \overset{\text{A--B}}{=} \BC_{AB} -2\, \frac{\BC_{[A|\By}\,\Bg_{|B]\By}}{\Bg_{\By\By}} \,.
\end{align}

\paragraph*{Further steps\\}

We can further proceed by considering a general expansion of the $\SL(2)$ singlet $\bm{\cA}_{\mu; \sfm_1\sfm_2\sfm_3}$\,,
\begin{align}
 \bm{\cA}_{\mu; \sfm_1\sfm_2\sfm_3} &= \BA_{\mu \sfm_1\sfm_2\sfm_3} + c_3\, \epsilon_{\alpha\beta}\,\BA_{\mu[\sfm_1}^\alpha\,\BA_{\sfm_2\sfm_3]}^\beta
\nn\\
 &\quad + c_4\, \BA_\mu^\sfp\,\BA_{\sfp \sfm_1\sfm_2\sfm_3} + c_5\, \epsilon_{\alpha\beta}\,\BA_\mu^\sfp\,\BA_{\sfp[\sfm_1}^\alpha\,\BA_{\sfm_2\sfm_3]}^\beta \,.
\end{align}
Similarly, unknown $T$-duality rules can also be expanded by considering possible 9D covariant expressions with parameters. 
Then, the consistency with the linear map \eqref{eq:M-B-map} determines all of the parameters. 
In this manner, by using the linear map \eqref{eq:M-B-map}, we can find both the parameterization of $\cA_{\mu}^{I}$ and $T$-duality rules for the gauge potentials one after another. 

\subsection{Results}
\label{sec:1-form-results}

By continuing the above procedure, we have determined the M-theory parameterization as
\begin{align}
 (\cA_\mu^{I})
 = \begin{pmatrix}
 \cA_\mu^i\\
 \frac{\cA_{\mu; i_1i_2}}{\sqrt{2!}}\\
 \frac{\cA_{\mu; i_1\cdots i_5}}{\sqrt{5!}}\\
 \frac{\cA_{\mu; i_1\cdots i_7,i}}{\sqrt{7!}}\\
 \vdots
 \end{pmatrix}
 = \begin{pmatrix}
 \MA_\mu^i \\
 \frac{1}{\sqrt{2!}}\,\bigl(\MN_{\mu; i_1i_2}+\MA_\mu^k\,\MN_{k; i_1i_2}\bigr) \\
 \frac{1}{\sqrt{5!}}\,\bigl(\MN_{\mu; i_1\cdots i_5}+\MA_\mu^k\, \MN_{k; i_1\cdots i_5} \bigr) \\
 \frac{1}{\sqrt{7!}}\, \bigl(\MN_{\mu; i_1\cdots i_7,i} + \MA_{\mu}^k\,\MN_{k; i_1\cdots i_7,i}\bigr)\\
 \vdots
 \end{pmatrix}.
\label{eq:cA-M}
\end{align}
Remarkably, the two tensors $\MN_{\mu; p,q,r,\dotsc}$ and $\MN_{k; p,q,r,\dotsc}$ in each row can be regarded as particular components of 11D-covariant tensors:
\begin{align}
\begin{split}
 \MN_{\hat{M}_1;\hat{M}_2\hat{M}_3} &=\MA_{\hat{M}_1\hat{M}_2\hat{M}_3} \,,
\\
 \MN_{\hat{M}_1;\hat{M}_2\cdots \hat{M}_6} &= \MA_{\hat{M}_1 \cdots \hat{M}_6} - 5\,\MA_{\hat{M}_1 [\hat{M}_2\hat{M}_3}\, \MA_{\hat{M}_4\hat{M}_5\hat{M}_6]}\,,
\\
 \MN_{\hat{M}_1;\hat{M}_2\cdots \hat{M}_8, \hat{N}}
 &\simeq \MA_{\hat{M}_1\cdots \hat{M}_8, \hat{N}} - 21\,\bigl(\MA_{\hat{M}_1[\hat{M}_2\cdots \hat{M}_6}\,\MA_{\hat{M}_7\hat{M}_8]\hat{N}} -\MA_{\hat{M}_1[\hat{M}_2\cdots \hat{M}_6}\,\MA_{\hat{M}_7\hat{M}_8\hat{N}]} \bigr)
\\
 &\quad +35\,\MA_{\hat{M}_1[\hat{M}_2\hat{M}_3}\,\MA_{\hat{M}_4\hat{M}_5\hat{M}_6}\,\MA_{\hat{M}_7\hat{M}_8]\hat{N}} \,,
\end{split}
\label{eq:N-M}
\end{align}
where the meaning of the equivalence $\simeq$ is explained below. 

As discussed in \cite{1108.5067,1109.2025,1201.5819,1303.0221} (see also \cite{1907.07177}), for any mixed-symmetry potential, not all of the components couple to supersymmetric branes. 
For the dual graviton $\cA_{\mu; i_1\cdots i_7, i}$\,, only the components satisfying
\begin{align}
 i\in \{i_1,\dotsc, i_7\} 
\label{eq:SUSY-rule}
\end{align}
couple to supersymmetric branes. 
The components that do not couple to supersymmetric branes correspond to the $E_{11}$ roots $\alpha$ satisfying $\alpha\cdot\alpha<2$\,, and they are not connected to the standard $p$-form potentials under $T$-duality and $S$-duality. 
Since our procedure to determine the parameterization is based on $T$-duality and $S$-duality, it can only provide the parameterization of the components that couple to supersymmetric branes. 
In this sense, it is more honest to express the last equation of \eqref{eq:N-M} as
\begin{align}
 \MN_{\hat{M}_1;\hat{M}_2\cdots \hat{M}_7x, x}
 &= \MA_{\hat{M}_1\cdots \hat{M}_7x, x} - 21\,\MA_{\hat{M}_1[\hat{M}_2\cdots \hat{M}_6}\,\MA_{\hat{M}_7x]x} +35\,\MA_{\hat{M}_1[\hat{M}_2\hat{M}_3}\,\MA_{\hat{M}_4\hat{M}_5\hat{M}_6}\,\MA_{\hat{M}_7x]x}
\nn\\
 &= \MA_{\hat{M}_1\cdots \hat{M}_7x, x} - 15\,\MA_{\hat{M}_1x[\hat{M}_2\cdots \hat{M}_5}\,\MA_{\hat{M}_6\hat{M}_7]x}
\nn\\
 &\quad -10\,\MA_{\hat{M}_1x[\hat{M}_2}\,\MA_{\hat{M}_3\hat{M}_4\hat{M}_5}\,\MA_{\hat{M}_6\hat{M}_7]x}
 +15\,\MA_{\hat{M}_1[\hat{M}_2\hat{M}_3}\,\MA_{\hat{M}_4\hat{M}_5|x|}\,\MA_{\hat{M}_6\hat{M}_7]x} \,.
\end{align}
In this paper, equalities that hold under the restriction \eqref{eq:SUSY-rule} are denoted by $\simeq$\,. 
The parameterizations of mixed-symmetry potentials that do not satisfy the restriction \eqref{eq:SUSY-rule} are not determined in this paper.

Now we turn to the results in type IIB theory. 
The parameterization takes the form
\begin{align}
 (\bm{\cA}_\mu^{\sfI})
 = \begin{pmatrix}
 \BA_\mu^\sfm \\
 \bigl(\BN^\alpha_{\mu;\sfm} + \BA_\mu^\sfp\, \BN^\alpha_{\sfp;\sfm}\bigr) \\
 \frac{1}{\sqrt{3!}}\, \bigl(\BN_{\mu;\sfm_1\sfm_2\sfm_3} + \BA_\mu^\sfp\,\BN_{\sfp;\sfm_1\sfm_2\sfm_3} \bigr) \\
 \frac{1}{\sqrt{5!}}\,\bigl(\BN^\alpha_{\mu;\sfm_1\cdots \sfm_5} + \BA_\mu^\sfp\,\BN^\alpha_{\sfp;\sfm_1\cdots \sfm_5} \bigr) \\
 \frac{1}{\sqrt{6!}}\,\bigl(\BN_{\mu;\sfm_1\cdots \sfm_6,\sfm} + \BA_{\mu}^\sfp\,\BN_{\sfp;\sfm_1\cdots \sfm_6,\sfm}\bigr) \\
 \vdots
 \end{pmatrix},
\label{eq:cA-B}
\end{align}
where
\begin{align}
\begin{split}
 \BN_{M_1;M_2}^\alpha &=\BA_{M_1M_2}^\alpha\,,
\\
 \BN_{M_1;M_2M_3M_4} &\equiv \BA_{M_1\cdots M_4} - \frac{3}{2}\,\epsilon_{\gamma\delta}\,\BA^\gamma_{M_1[M_2}\,\BA^\delta_{M_3M_4]}
\\
 &=\BC_{M_1\cdots M_4} - 3\,\BC_{M_1[M_2}\,\BB_{M_3M_4]}\,,
\\
 \BN^\alpha_{M_1;M_2\cdots M_6} &\equiv \BA^\alpha_{M_1 \cdots M_6} + 5\,\BA^\alpha_{M_1[M_2}\, \BA_{M_3\cdots M_6]}
 + 5\,\epsilon_{\gamma\delta}\,\BA^\gamma_{M_1[M_2}\,\BA^\delta_{M_3M_4}\,\BA^\alpha_{M_5M_6]}
\\
 &= \begin{pmatrix}
 \BC_{M_1\cdots M_6} -10\,\BC_{M_1 [M_2M_3M_4}\,\BB_{M_5M_6]} + 15\,\BC_{M_1 [M_2}\,\BB_{M_3M_4}\,\BB_{M_5M_6]}
\\
 -\bigl(\BB_{M_1\cdots M_6} -10\,\BC_{M_1 [M_2M_3M_4}\,\BC_{M_5M_6]}\bigr)
\end{pmatrix},
\\
 \BN_{M_1;M_2\cdots M_7,N} &\simeq 
 \BA_{M_1\cdots M_7,N}
 + 6\,\BB_{M_1 [M_2\cdots M_6}\,\BB_{M_7] N}
 - 6\,\BC_{M_1 [M_2}\,\BC_{M_3 \cdots M_7] N}
\\
 &\ 
 -60\,\BC_{M_1 [M_2 M_3 M_4}\,\BC_{M_5 M_6}\, \BB_{M_7]N}
 +10\, \BC_{M_1 [M_2M_3M_4}\, \BC_{M_5M_6M_7]N}
\\
 &\ 
 + \frac{45}{2}\, \bigl(\BB_{M_1 [M_2}\, \BB_{M_3 M_4}\, \BC_{M_5 M_6}\, \BC_{M_7]N}
   - \BC_{M_1 [M_2}\, \BC_{M_3 M_4}\, \BB_{M_5 M_6}\, \BB_{M_7]N}\bigr) \,.
\end{split}
\label{eq:N-B}
\end{align}
The last component $\BN_{1;6,1}$ is relatively long, and the $S$-duality invariance is not clear. 
However, this is because of the definition of the dual graviton $\BA_{7,1}$\,. 
As we will see later (in Eq.~\eqref{eq:L-T-B}), a certain redefinition of $\BA_{7,1}$ makes the expression of $\BN_{1;6,1}$ simpler. 

\subsection{$T$-duality rule}
\label{sec:T-dual}

In addition to the parameterizations, we have obtained the $T$-duality rules as follows:
\begin{align}
\begin{split}
 \Ag_{AB} &\overset{\text{A--B}}{=} \Bg_{AB} - \frac{\Bg_{A\By}\,\Bg_{B\By}-\BB_{A\By}\,\BB_{B\By}}{\Bg_{\By\By}}\,,\qquad 
 \Ag_{A \Ay}\overset{\text{A--B}}{=}-\frac{\BB_{A \By}}{\Bg_{\By\By}}\,, 
\\
 \AA_\mu^a &\overset{\text{A--B}}{=} \BA_\mu^a\,,\qquad 
 \AA_\mu^\Ay \overset{\text{A--B}}{=} \BB_{\mu\By} + \BA_\mu^{\sfp}\,\BB_{\sfp\By} \,,
\\
 \AB_{AB} &\overset{\text{A--B}}{=} \BB_{AB} - \frac{\BB_{A \By}\,\Bg_{B \By}-\Bg_{A\By}\,\BB_{B\By}}{\Bg_{\By\By}}\,,\qquad 
 \AB_{A\Ay} \overset{\text{A--B}}{=} -\frac{\Bg_{A\By}}{\Bg_{\By\By}} \,,
\\
 \AC_{A_1\cdots A_{n-1}\Ay}&\overset{\text{A--B}}{=} \BC_{A_1\cdots A_{n-1}} - (n-1)\,\frac{\BC_{[A_1\cdots A_{n-2}|\By|}\,\Bg_{A_{n-1}]\By}}{\Bg_{\By\By}}\,,
\\
 \AC_{A_1\cdots A_n} &\overset{\text{A--B}}{=} \BC_{A_1\cdots A_n\By} - n\, \BC_{[A_1\cdots A_{n-1}}\, \BB_{A_n]\By} - n\,(n-1)\,\frac{\BC_{[A_1\cdots A_{n-2}|\By|}\, \BB_{A_{n-1}|\By|}\,\Bg_{A_n]\By}}{\Bg_{\By\By}}\,. 
\end{split}
\label{eq:Busc-standard}
\end{align}
For the 6-form potential $B_6$ and the dual graviton $A_{7, 1}$\,, we find
\begin{align}
 \AB_{A_1\cdots A_5\Ay}&\overset{\text{A--B}}{=} \BB_{A_1\cdots A_5 \By} -5\,\BA_{[A_1\cdots A_4}\,\BC_{A_5]\By} -5\,\BA_{[A_1A_2A_3|\By|}\,\BC_{A_4A_5]}
\nn\\
 &\quad\ 
 - \frac{45}{2}\, \BC_{[A_1A_2}\,\BB_{A_3A_4}\,\BC_{A_5]\By}
 - \frac{15}{2}\, \BC_{[A_1A_2}\,\BC_{A_3A_4}\,\BB_{A_5]\By}
\nn\\
 &\quad\ 
 - \frac{10\,\BA_{[A_1\cdots A_3|\By|}\,\BC_{A_4|\By|}\Bg_{A_5]\By}}{\Bg_{\By\By}}
 - \frac{15\,\BC_{[A_1A_2}\, \BB_{A_3|\By|}\,\BC_{A_4|\By|}\,\Bg_{A_5]\By}}{\Bg_{\By\By}} \,,
\label{eq:Busc-B6A-B6B}
\\
 \AB_{A_1\cdots A_6}&\overset{\text{A--B}}{=}
 \BA_{A_1 \cdots A_6 \By, \By}
 -6 \, \BB_{[A_1\cdots A_5|\By|}\, \BB_{A_6] \By}
\nn\\
 &\quad\ 
 +30 \, \BA_{[A_1 A_2 A_3|\By}\, \BC_{|A_4 A_5}\, \BB_{A_6] \By}
 +30 \, \BA_{[A_1 \cdots A_4}\, \BC_{A_5|\By|}\, \BB_{A_6] \By}
\nn\\
 &\quad\ 
 - \frac{315}{2} \, \BB_{[A_1 A_2}\, \BB_{A_3|\By|}\, \BC_{A_4 A_5}\, \BC_{A_6] \By}
 + \frac{60\, \BA_{[A_1 A_2 A_3|\By}\, \BB_{A_4|\By}\, \BC_{A_5|\By}\, \Bg_{|A_6]\By}}{\Bg_{\By\By}} \,,
\\
 \AA_{A_1\cdots A_6\Ay, \Ay}&\overset{\text{A--B}}{=} \BB_{A_1\cdots A_6}
 - 30\, \BB_{[A_1A_2}\,\BC_{A_3A_4}\, \BC_{A_5A_6]}
 - \frac{6\,\BB_{[A_1\cdots A_5|\By}\,\Bg_{|A_6]\By}}{\Bg_{\By\By}}
 + \frac{20\, \BA_{[A_1A_2A_3|\By|}\, \BC_{A_4A_5}\,\Bg_{A_6]\By}}{\Bg_{\By\By}}
\nn\\
 &\quad\ 
 + \frac{30\, \BB_{[A_1|\By|}\, \BC_{A_2A_3}\, \BC_{A_4A_5}\,\Bg_{A_6]\By}}{\Bg_{\By\By}}
 + \frac{150\, \BB_{[A_1A_2}\,\BC_{A_3A_4}\, \BC_{A_5|\By|}\,\Bg_{A_6]\By}}{\Bg_{\By\By}}\,,
\label{eq:Busc-A71A-B6B}
\\
 \AA_{A_1\cdots A_6\Ay, B}&\overset{\text{A--B}}{\simeq} \BA_{A_1\cdots A_6\By, B}
 - 6\,\BC_{[A_1\cdots A_5|\By|}\,\BC_{A_6]B}
 - 6\,\BB_{[A_1\cdots A_5|B|}\,\BB_{A_6]\By}
\nn\\
 &\quad\ 
 + 10\,\BC_{[A_1A_2A_3|B|}\,\BC_{A_4A_5A_6]\By}
 + 20\,\BC_{[A_1A_2A_3|\By|}\,\BB_{A_4|B|}\,\BC_{A_5A_6]}
\nn\\
 &\quad\ 
 + 40\,\BC_{[A_1A_2A_3|\By|}\,\BB_{A_4A_5}\,\BC_{A_6]B}
 + 30\,\BB_{[A_1A_2}\,\BC_{A_3A_4}\, \BC_{A_5A_6]}\,\BB_{B\By} 
\nn\\
 &\quad\ 
 +\frac{45}{2}\,\BB_{[A_1A_2}\,\BC_{A_3A_4}\,\bigl(-\BB_{A_5|\By|}\,\BC_{A_6]B} + \BC_{A_5|\By|}\,\BB_{A_6]B}\bigr)
\nn\\
 &\quad\ 
 -\frac{\BA_{A_1 \cdots A_6 \By, \By}\,\Bg_{B\By}}{\Bg_{\By\By}}
 + \frac{6\, \BC_{[A_1 \cdots A_5 |\By|}\, \BC_{A_6]\By}\,\Bg_{B\By}}{\Bg_{\By\By}}
 -\frac{6\, \BC_{B\By}\, \BC_{[A_1 \cdots A_5 |\By|}\,\Bg_{|A_6]\By}}{\Bg_{\By\By}}
\nn\\
 &\quad\ 
 +\frac{6 \, \BB_{B\By}\, \BB_{[A_1\cdots A_5|\By}\, \Bg_{|A_6]\By}}{\Bg_{\By\By}}
 -\frac{10\, \BC_{[A_1 A_2| B\By}\, \BC_{|A_3 A_4 A_5 |\By|}\,\Bg_{A_6] \By}}{\Bg_{\By\By}}
\nn\\
 &\quad\ 
 +\frac{120\, \BC_{[A_1 A_2 |B\By|}\, \BB_{A_3 A_4}\, \BC_{A_5|\By|}\, \Bg_{A_6]\By}}{\Bg_{\By\By}}
 -\frac{20\, \BB_{B\By}\,\BC_{[A_1 A_2 A_3|\By|}\, \BC_{A_4 A_5}\,\Bg_{A_6]\By}}{\Bg_{\By\By}}
\nn\\
 &\quad\ 
 +\frac{40\, \BC_{[A_1 A_2 A_3|\By|}\, \BB_{A_4|B|}\, \BC_{A_5|\By|}\,\Bg_{A_6] \By}}{\Bg_{\By\By}}
 -\frac{120\, \BB_{B\By}\,\BB_{[A_1 A_2}\, \BC_{A_3 A_4}\, \BC_{A_5|\By}\,\Bg_{|A_6] \By}}{\Bg_{\By\By}}
\nn\\
 &\quad\ 
 +\frac{45}{2}\,\frac{\BB_{[A_1 A_2}\, \BB_{A_3|\By|}\, \BC_{A_4 A_5}\, \BC_{A_6]\By}\, \Bg_{B\By}}{\Bg_{\By\By}} \,.
\label{eq:Busc-A71A-A71B}
\end{align}
The $T$-duality rules \eqref{eq:Busc-B6A-B6B} and \eqref{eq:Busc-A71A-B6B} coincide with the known results \cite{hep-th:9806169} (see Appendix A therein), for which the following identification of supergravity fields is needed:
\begin{align}
 \begin{pmatrix} \mathsf{g}_{\mu\nu} \\ B \\ C^{(1)} \\ C^{(3)} \\ C^{(5)} \\ \tilde{B} \\ N \end{pmatrix}_{\!\!\!{\text{(IIA)}\atop\text{\cite{hep-th:9806169}}}}
 =\begin{pmatrix}
 \Ag_{MN}\\ \AB_2\\ \AC_1\\ \AC_3\\ \AC_5\\ - \AB_6 \\ \AA_{7,1} \end{pmatrix}_{\!\!\!{\text{(IIA)}\atop\text{here}}} ,\quad
 \begin{pmatrix}
 g_{\mu\nu}\\ \cB \\ C^{(0)}\\ C^{(2)}\\ C^{(4)}\\ C^{(6)}\\ \tilde{\cB} 
\end{pmatrix}_{\!\!\!{\text{(IIB)}\atop\text{\cite{hep-th:9806169}}}}
 = \begin{pmatrix}
 \Bg_{MN}\\ \BB_2\\ - \BC_0 \\ - \BC_2\\ - \BA_4\\ -\bigl( \BC_6-\tfrac{1}{4}\, \BB_2\wedge \BB_2\wedge \BC_2\bigr) \\ -\bigl( \BB_6-\tfrac{1}{4}\, \BC_2\wedge \BC_2\wedge \BB_2\bigr) 
\end{pmatrix}_{{\!\!\!{\text{(IIB)}\atop\text{here}}}} .
\end{align}
On the other hand, \eqref{eq:Busc-A71A-A71B} has been obtained in \cite{hep-th:9908094}, where $\BB_2=0$ and $\BC_2=0$ are assumed. 
If we truncate $\BB_2$ and $\BC_2$\,, we have $\BA_4=\BC_4$ and the $T$-duality rule \eqref{eq:Busc-A71A-A71B} reduces to
\begin{align}
 \AA_{A_1\cdots A_6\Ay,B}&\overset{\text{A--B}}{\simeq} \BA_{A_1\cdots A_6\By,B} + 10\,\BA_{[A_1A_2A_3|B|}\,\BA_{A_4A_5A_6]\By}
\nn\\
 &\quad\ 
  -\frac{\BA_{A_1 \cdots A_6 \By, \By}\,\Bg_{B\By}}{\Bg_{\By\By}}
  -\frac{10\, \BA_{[A_1 A_2| B\By}\, \BA_{|A_3 A_4 A_5 |\By|}\,\Bg_{A_6] \By}}{\Bg_{\By\By}} \,.
\end{align}
More explicitly, according to the restriction \eqref{eq:SUSY-rule}, the direction $x\equiv B$ must be contained in $\{A_1,\dotsc, A_6\}$ and by choosing $A_6=x$\,, we have
\begin{align}
 \AA_{A_1\cdots A_5xy, x}&\overset{\text{A--B}}{=} \BA_{A_1\cdots A_5x\By, x} + 5\,\BA_{[A_1A_2A_3|x|}\,\BA_{A_4A_5]x\By} -\frac{\BA_{A_1 \cdots A_5x\By, \By}\,\Bg_{x\By}}{\Bg_{\By\By}}
\nn\\
 &\quad\ 
  +\frac{5\, \BA_{[A_1 A_2| x\By}\, \BA_{|A_3 A_4|x\By|}\,\Bg_{A_5] \By}}{\Bg_{\By\By}}
  -\frac{5\, \BA_{[A_1 A_2| x\By}\, \BA_{|A_3 A_4 A_5]\By}\,\Bg_{x\By}}{3\,\Bg_{\By\By}} \,.
\end{align}
If we denote $k\equiv \partial_x$ and $h\equiv \partial_{\sfy}$\,, and define
\begin{align}
 N^{(7)}_{M_1\cdots M_6}\equiv \AA_{M_1\cdots M_7, x}\,,\quad 
 \sfN^{(7)}_{M_1\cdots M_6}\equiv \BA_{M_1\cdots M_7, x}\,,\quad 
 \cN^{(7)}_{M_1\cdots M_7}\equiv \BA_{M_1\cdots M_7, \By}\,, 
\end{align}
the result of \cite{hep-th:9908094} (see Eq.~(5.13)) is precisely reproduced:
\begin{align}
 (\iota_k N^{(7)})_{A_1\cdots A_5\Ay}&\overset{\text{A--B}}{=} (\iota_k\iota_h \sfN^{(7)})_{A_1\cdots A_5} - 5\,(\iota_k \BA)_{[A_1A_2A_3}\,(\iota_k\iota_h \BA)_{A_4A_5]}
  - \frac{(\iota_k\iota_h \cN^{(7)})_{A_1 \cdots A_5}\,\Bg_{x\By}}{\Bg_{\By\By}}
\nn\\
 & -\frac{5\, (\iota_k\iota_h \BA)_{[A_1 A_2}\, (\iota_k\iota_h \BA)_{A_3 A_4}\,\Bg_{A_5] \By}}{\Bg_{\By\By}}
  + \frac{5\, (\iota_h \BA)_{[A_1 A_2 A_3}\,(\iota_k\iota_h \BA)_{A_4 A_5]}\, \Bg_{x\By}}{3\,\Bg_{\By\By}} \,.
\end{align}
This shows that $\BA_{7,1}$ corresponds to the 7-form $\sfN^{(7)}$ or $\cN^{(7)}$ of \cite{hep-th:9908094} under $\BB_2=0$ and $\BC_2=0$. 

In the above computation, we have shown only $T$-duality transformations from type IIB to type IIA, but we can easily find the inverse map. 
The standard rules \eqref{eq:Busc-standard} have the same form even for the map from type IIA to type IIB:
\begin{align}
 \Bg_{AB} &\overset{\text{B--A}}{=} \Ag_{AB} - \frac{\Ag_{A\Ay}\,\Ag_{B\Ay}-\AB_{A\Ay}\,\AB_{B\Ay}}{\Ag_{\Ay\Ay}}\,,\qquad 
 \Bg_{A \By}\overset{\text{B--A}}{=}-\frac{\AB_{A \Ay}}{\Ag_{\Ay\Ay}}\,,
\nn\\
 \BA_\mu^a &\overset{\text{B--A}}{=} \AA_\mu^a\,,\qquad 
 \BA_\mu^\By \overset{\text{B--A}}{=} \AB_{\mu \Ay} + \AA_\mu^{p}\,\AB_{p\Ay} \,,
\nn\\
 \BB_{AB} &\overset{\text{B--A}}{=} \AB_{AB} - \frac{\AB_{A \Ay}\,\Ag_{B \Ay}-\Ag_{A \Ay}\,\AB_{B \Ay}}{\Ag_{\Ay\Ay}}\,,\qquad 
 \BB_{Ay} \overset{\text{B--A}}{=} -\frac{\Ag_{A\Ay}}{\Ag_{\Ay\Ay}} \,,
\nn\\
 \BC_{A_1\cdots A_{n-1}\By}&\overset{\text{B--A}}{=} \AC_{A_1\cdots A_{n-1}} - (n-1)\,\frac{\AC_{[A_1\cdots A_{n-2}|\Ay|}\,\Ag_{A_{n-1}]\Ay}}{\Ag_{\Ay\Ay}}\,,
\\
 \BC_{A_1\cdots A_n} &\overset{\text{B--A}}{=} \AC_{A_1\cdots A_n\Ay} - n\, \AC_{[A_1\cdots A_{n-1}}\, \AB_{A_n]\Ay} - n\,(n-1)\,\frac{\AC_{[A_1\cdots A_{n-2}|\Ay|}\, \AB_{A_{n-1}|\Ay|}\,\Ag_{A_n]\Ay}}{\Ag_{\Ay\Ay}}\,. 
\nn
\end{align}
Regarding the 6-form potential and the dual graviton, the results are as follows:
\begin{align}
 \BB_{A_1\cdots A_5\By}
 &\overset{\text{B--A}}{=}\AB_{A_1\cdots A_5\Ay}
 +5\,\AC_{[A_1\cdots A_4|\Ay|}\,\AC_{A_5]}
 +5\,\AC_{[A_1A_2A_3}\,\AC_{A_4A_5]\Ay}
\nn\\
 &\quad -\frac{15\,\AC_{[A_1A_2|\Ay|}\,\AC_{A_3A_4|\Ay|}\,\Ag_{A_5]\Ay}}{\Ag_{\Ay\Ay}}
 -\frac{5\,\AC_{\Ay}\,\AC_{[A_1\cdots A_4|\Ay|}\,\Ag_{A_5]\Ay}}{\Ag_{\Ay\Ay}} \,,
\\
 \BB_{A_1\cdots A_6}
 &\overset{\text{B--A}}{=}
 \AA_{A_1\cdots A_6\Ay,\Ay}
 -6\,\AB_{[A_1\cdots A_5\Ay}\,\AB_{A_6]\Ay}
 -30\,\AC_{[A_1\cdots A_4|\Ay|}\,\AC_{A_5}\,\AB_{A_6]\Ay}
\nn\\
 &\quad -10\,\AC_{[A_1A_2A_3}\,\AC_{A_4A_5|\Ay|}\,\AB_{A_6]\Ay}
 +30\,\AC_{[A_1A_2|\Ay|}\,\AC_{A_3A_4|\Ay|}\,\AB_{A_5A_6]}
\nn\\
 &\quad -\frac{30\,\AC_{\Ay}\,\AC_{[A_1\cdots A_4|\Ay|}\,\AB_{A_5|\Ay|}\,\Ag_{A_6]\Ay}}{\Ag_{\Ay\Ay}}
 -\frac{90\,\AC_{[A_1A_2|\Ay|}\,\AC_{A_3A_4|\Ay|}\,\AB_{A_5|\Ay|}\,\Ag_{A_6]\Ay}}{\Ag_{\Ay\Ay}}\,,
\\
 \BA_{A_1\cdots A_6\By,\By}
 &\overset{\text{B--A}}{=} \AB_{A_1\cdots A_6}
 -\frac{6\,\AB_{[A_1\cdots A_5|\Ay|}\,\Ag_{A_6]\Ay}}{\Ag_{\Ay\Ay}}
 +\frac{45}{2}\, \frac{\AC_{[A_1A_2|\Ay|}\,\AB_{A_3A_4}\,\AC_{A_5}\,\Ag_{A_6]\Ay}}{\Ag_{\Ay\Ay}}\,,
\\
 \BA_{A_1\cdots A_6\By, B}
 &\overset{\text{B--A}}{\simeq}
 \AA_{A_1\cdots A_6\Ay, B}
 -6\,\AB_{[A_1\cdots A_5|B|}\,\AB_{A_6]\Ay}
 +6\,\AC_{[A_1\cdots A_5}\,\AC_{A_6] B\Ay}
\nn\\
 &\quad -\frac{15}{2} \,\AC_{[A_1\cdots A_4|\Ay|}\,\AC_{A_5A_6]B}
 +20\,\AC_{[A_1A_2A_3}\,\AC_{A_4A_5|B|}\,\AB_{A_6]\Ay}
\nn\\
 &\quad -20\,\AC_{[A_1A_2A_3}\,\AC_{A_4A_5|\Ay|}\,\AB_{A_6]B}
 -40\,\AC_{[A_1A_2A_3}\,\AC_{A_4|B\Ay|}\,\AB_{A_5A_6]}
\nn\\
 &\quad -\frac{45}{2} \,\AC_{[A_1A_2|\Ay|}\,\AC_{A_3}\,\AB_{A_4|B|}\,\AB_{A_5A_6]}
 -\frac{6\,\AA_{[A_1\cdots A_5|B\Ay, \Ay}\,\Ag_{|A_6]\Ay}}{\Ag_{\Ay\Ay}}
\\
 &\quad 
 +\frac{6\,\AB_{B\Ay}\,\AB_{[A_1\cdots A_5|\Ay|}\,\Ag_{A_6]\Ay}}{\Ag_{\Ay\Ay}}
 +\frac{15\,\AC_{[A_1\cdots A_4|\Ay|}\,\AC_{A_5|B\Ay|}\,\Ag_{A_6]\Ay}}{\Ag_{\Ay\Ay}}
\nn\\
 &\quad 
 +\frac{15}{2}\,\frac{\AC_{[A_1\cdots A_4|\Ay|}\,\AC_{A_5A_6]\Ay}\,\Ag_{B\Ay}}{\Ag_{\Ay\Ay}}
 -\frac{20\,\AC_{[A_1A_2A_3}\,\AC_{A_4A_5|\Ay|}\,\AB_{A_6]\Ay}\,\Ag_{B\Ay}}{\Ag_{\Ay\Ay}}
\nn\\
 &\quad 
 -\frac{195}{8} \,\frac{\AC_{[A_1A_2|\Ay|}\,\AC_{A_3A_4|\Ay|}\,\AB_{A_5A_6]}\,\Ag_{B\Ay}}{\Ag_{\Ay\Ay}}
 -\frac{45}{4} \,\frac{\AC_{[A_1A_2|\Ay|}\,\AC_{A_3A_4|\Ay|}\,\AB_{A_5|B|}\,\Ag_{A_6]\Ay}}{\Ag_{\Ay\Ay}}
\nn\\
 &\quad 
 -\frac{45}{2} \,\frac{\AB_{B\Ay}\,\AC_{[A_1A_2|\Ay|}\,\AC_{A_3}\,\AB_{A_4A_5}\,\Ag_{A_6]\Ay}}{\Ag_{\Ay\Ay}}
 -\frac{45}{2} \,\frac{\AC_{\Ay}\,\AC_{[A_1A_2|\Ay|}\,\AB_{A_3A_4}\,\AB_{A_5|B|}\,\Ag_{A_6]\Ay}}{\Ag_{\Ay\Ay}}\,.
\nn
\end{align}

Now, let us comment more on the restriction rule. 
In the $T$-duality rule \eqref{eq:Busc-A71A-A71B}, we are assuming that $B$ is contained in $\{A_1,\cdots ,A_6\}$\,. 
When the restriction is removed, we expect that the right-hand side of the $T$-duality rule is modified. 
In general, the components that do not satisfy the restriction are in the same orbit as the $(\alpha\neq\beta)$-component of the type IIB potential $\AA^{\alpha\beta}_8$\,, which is electric-magnetic dual to the 0-form potential $\Bm_{\alpha\beta}$\,. 
Therefore, it will be possible that $\AA^{\alpha\beta}_{A_1 \cdots A_6B\By}$ appears on the right-hand side of \eqref{eq:Busc-A71A-A71B}. 

\subsection{$S$-duality rule}

The standard $S$-duality transformation rules are reproduced as follows:
\begin{align}
\begin{split}
 &\Bg'_{MN} =\Bg_{MN}\,,\quad \BA'^\sfm_\mu = \BA_\mu^\sfm\,,\quad 
 \BC'_0 = - \frac{\BC_0}{(\BC_0)^2+\Exp{-2\BPhi}}\,,\quad \Exp{-\BPhi'} = \frac{\Exp{-\BPhi'}}{(\BC_0)^2+\Exp{-2\BPhi}}\,,
\\
 &\BB'_2 =- \BC_2\,,\quad \BC'_2 = \BB_2\,,\quad \BC'_4 = \BC_4 - \BB_2\wedge \BC_2 \,, \quad 
 \BA'_4 = \BA_4\,,
\\
 &\BC'_6 = - \BB_6 + \frac{1}{2}\, \BB_2\wedge \BC_2\wedge \BC_2 \,, \quad 
 \BB'_6 = \BC_6 - \frac{1}{2}\, \BC_2\wedge \BB_2\wedge \BB_2\,,\quad 
 \BA'^1_6 = \BA^2_6\,,\quad \BA'^2_6 = - \BA^1_6\,. 
\end{split}
\end{align}
From the $S$-duality invariance of $\bm{\cA}_{\mu; \sfm_1\cdots \sfm_6, \sfm}$, we also find
\begin{align}
 \BA'_{M_1\cdots M_7,M}&\simeq \BA_{M_1\cdots M_7,M} + 7\,\bigl(\BB_{[M_1\cdots M_6}\,\BB_{M_7]M}-\BB_{[M_1\cdots M_6}\,\BB_{M_7M]}\bigr)
\nn\\
 &\quad + 7\,\bigl(\BC_{[M_1\cdots M_6}\,\BC_{M_7]M} - \BC_{[M_1\cdots M_6}\,\BC_{M_7M]} \bigr)
\nn\\
 &\quad - \frac{105}{2}\,\bigl(\BA_{[M_1\cdots M_4}\,\BB_{M_5M_6}\,\BC_{M_7]M} - \BA_{[M_1\cdots M_4}\,\BC_{M_5M_6}\,\BB_{M_7M]}\bigr)
\nn\\
 &\quad +\frac{945}{4}\,\bigl(\BB_{[M_1M_2}\,\BB_{M_3M_4}\,\BC_{M_5M_6}\,\BC_{M_7]M} -\BB_{[M_1M_2}\,\BB_{M_3M_4}\,\BC_{M_5M_6}\,\BC_{M_7M]} \bigr)\,.
\label{eq:dual-graviton-S-dual}
\end{align}

\section{Another approach based on the generalized metric}
\label{sec:generalized-metric}

In this section, we discuss another derivation of the $T$-/$S$-duality transformation rule for the dual graviton, which is based on the generalized metric. 
We also explain another method to determine the parameterization of the 1-form $\cA_\mu^I$\,. 

In $d$ dimensions, scalar fields are packaged into a $U$-duality-covariant object called the generalized metric, denoted as $\cM_{IJ}$ and $\sfM_{\sfI \sfJ}$ in M-theory and type IIB, respectively. 
The generalized vielbeins, $\cE^I{}_J$ and $\sfE^\sfI{}_\sfJ$ respectively, are defined such that
\begin{align}
 \cM_{IJ} \equiv \delta_{KL}\,\cE^K{}_I \,\cE^L{}_J \,, \qquad
 \sfM_{\sfI\sfJ} \equiv \delta_{\sfK\sfL}\,\sfE^\sfK{}_\sfI \,\sfE^\sfL{}_\sfJ \,.
\label{eq:gen-metric}
\end{align}
According to \cite{1111.0459}, the generalized vielbein can be constructed as follows. 
We first consider the positive-root generators of the $E_n$ algebra, which are summarized as
\begin{align}
 \{E_{\bm{\alpha}}\} = \{K^i{}_j \ (i<j),\, R^{i_1i_2i_3} ,\, R^{i_1\cdots i_6} ,\, R^{i_1\cdots i_8,i} ,\dotsc \bigr\} 
\end{align}
in the M-theory parameterization and as
\begin{align}
 \{\sfE_{\bm{\alpha}}\} = \{\sfK^\sfm{}_\sfn\ (\sfm<\sfn),\, \sfR_{22},\,\sfR_\alpha^{\sfm_1\sfm_2},\,\sfR^{\sfm_1\cdots \sfm_4},\,\sfR^{\sfm_1\cdots \sfm_6}_\alpha,\,\sfR^{\sfm_1\cdots \sfm_7,\sfm},\dotsc\} 
\end{align}
in the type IIB parameterization. 
We also consider the Cartan generators,
\begin{align}
 \{H_k\} = \{K^d{}_d - K^{d+1}{}_{d+1},\, \dotsc,\, K^9{}_9 - K^\Az{}_\Az,\, K^8{}_8 + K^9{}_9 + K^\Az{}_\Az + \tfrac{1}{3}\,D \} 
\end{align}
in the M-theory parameterization ($D\equiv K^i{}_i$) and
\begin{align}
 \{\sfH_\sfk\} = \{\sfK^d{}_d - \sfK^{d+1}{}_{d+1},\, \dotsc,\, \sfK^7{}_7 - \sfK^8{}_8,\, \sfK^8{}_8 + \sfK^9{}_9 -\tfrac{1}{4}\,\sfD -\sfR_{12},\, 2\,\sfR_{12},\, \sfK^8{}_8 - \sfK^9{}_9 \} 
\end{align}
in the type IIB parameterization ($\sfD\equiv K^\sfm{}_\sfm$). 
Then, we prepare the matrix representations of these generators in the vector representation. 
In the M-theory parameterization, the matrix representations have been obtained in \cite{1111.0459} for $n\leq 7$ and in \cite{1303.2035} for $n=8$\,. 
In the type IIB parameterization, they have been determined in \cite{1405.7894,1612.08738} for $n\leq 7$. 
The results for $n=8$ are given in Appendix \ref{app:generators}.
Then, we define the generalized vielbein in the M-theory parameterization as
\begin{align}
\begin{split}
 &\cE \equiv (\cE^I{}_J) \equiv \hat{\cE}\, L\,,\qquad \hat{\cE}\equiv \Exp{h^k\,H_k} \Exp{\sum_{i<j} h_i{}^j\,K^i{}_j}\,,
\\
 &L\equiv (L^I{}_J) \equiv \Exp{\frac{1}{3!}\,\bm{\MA}_{i_1i_2i_3}\,R^{i_1i_2i_3}} \Exp{\frac{1}{6!}\,\bm{\MA}_{i_1\cdots i_6}\,R^{i_1\cdots i_6}} \Exp{\frac{1}{8!}\,\bm{\MA}_{i_1\cdots i_8,i}\,R^{i_1\cdots i_8,i}} \cdots \,, 
\end{split}
\label{eq:L-M}
\end{align}
and the generalized vielbein in the type IIB parameterization as
\begin{align}
\begin{split}
 &\sfE \equiv (\sfE^\sfI{}_\sfJ) \equiv \hat{\sfE}\, \sfL\,,\qquad \hat{\sfE}\equiv \Exp{\sfh^\sfk\,\sfH_\sfk} \Exp{\sum_{\sfm<\sfn} \sfh_\sfm{}^\sfn\,\sfK^\sfm{}_\sfn}\,,
\\
 &\sfL \equiv (\sfL^\sfI{}_\sfJ) \equiv \Exp{\frac{1}{2!}\,\bm{\BA}^\alpha_{\sfm_1\sfm_2}\,\sfR_\alpha^{\sfm_1\sfm_2}} \Exp{\frac{1}{4!}\,\bm{\BA}_{\sfm_1\cdots\sfm_4}\,\sfR^{\sfm_1\cdots\sfm_4}} \Exp{\frac{1}{6!}\,\bm{\BA}^\alpha_{\sfm_1\cdots\sfm_6}\,\sfR_\alpha^{\sfm_1\cdots\sfm_6}}\Exp{\frac{1}{7!}\,\bm{\BA}_{\sfm_1\cdots\sfm_7,\sfm}\,\sfR^{\sfm_1\cdots\sfm_7,\sfm}} \cdots \,.
\end{split}
\label{eq:L-B}
\end{align}
The objects $\bm{\MA}$ and $\bm{\BA}$ are again the M-theory and type IIB fields respectively, expressed in a new basis. 
Namely, $\bm{\MA}$ and $\bm{\BA}$ are respectively related to $\MA$ and $\BA$ by field redefinitions, as we will show in this section.
The ellipses in both parameterizations disappear for $n\leq 8$\,. 

The generalized metrics \eqref{eq:gen-metric} are then expressed as
\begin{align}
 \cM_{IJ} \equiv (\cE^\rmT\,\cE)_{IJ} = (L^\rmT\,\hat{\cM}\,L)_{IJ} \,,\qquad 
 \sfM_{\sfI\sfJ} \equiv (\sfE^\rmT\,\sfE)_{\sfI\sfJ} = (\sfL^\rmT\,\hat{\sfM}\,\sfL)_{\sfI\sfJ} \,,
\end{align}
where we have defined the untwisted metrics as
\begin{align}
 \hat{\cM}_{IJ} \equiv (\hat{\cE}^\rmT\,\hat{\cE})_{IJ}\,,\qquad 
 \hat{\sfM}_{\sfI\sfJ} \equiv (\hat{\sfE}^\rmT\,\hat{\sfE})_{\sfI\sfJ} \,,
\end{align}
which are parameterized by the supergravity fields $h^k$ (vielbein) and $\sfh^k$ (vielbein and $\sfm_{\alpha\beta}$)\,. 
Explicitly, the untwisted metrics take the following form:
\begin{align}
 \hat{\cM} &\equiv \abs{\MG}^{\frac{1}{d-2}}\,{\footnotesize
 \begin{pmatrix}
 \MG_{ij} & 0 & 0 & 0 & \\
 0 & \MG^{i_1i_2, j_1j_2} & 0 & 0 & \cdots \\
 0 & 0 & \MG^{i_1\cdots i_5, j_1\cdots j_5} & 0 & \\
 0 & 0 & 0 & \MG^{i_1\cdots i_7, j_1\cdots j_7}\,\MG^{ij} & \\
   & \vdots & & & \ddots
 \end{pmatrix}},
\\
 \hat{\sfM} &\equiv \abs{\BG}^{\frac{1}{d-2}}{\arraycolsep=0.5mm {\footnotesize \begin{pmatrix}
 \BG_{\sfm\sfn} & 0 & 0 & 0 & 0 & \\
 0 & \Bm_{\alpha\beta} \,\BG^{\sfm\sfn} & 0 & 0 & 0 & \\
 0 & 0 & \BG^{\sfm_1\sfm_2\sfm_3, \sfn_1\sfn_2\sfn_3} & 0 & 0 & \cdots \\
 0 & 0 & 0 & \Bm_{\alpha\beta} \,\BG^{\sfm_1\cdots \sfm_5, \sfn_1\cdots \sfn_5} & 0 & \\
 0 & 0 & 0 & 0 & \BG^{\sfm_1\cdots\sfm_6, \sfn_1\cdots \sfn_6}\,\BG^{\sfm\sfn} & \\
   & & \vdots & & & \ddots
\end{pmatrix}}} ,
\end{align}
where
\begin{align}
\begin{split}
\begin{alignedat}{2}
 \MG^{i_1\cdots i_p, j_1\cdots j_p} &\equiv \delta^{j_1\cdots j_p}_{k_1\cdots k_p}\,\MG^{i_1k_1}\cdots \MG^{i_pk_p}\,,&\qquad 
 \abs{\MG}&\equiv \det(\MG_{ij})\,,
\\
 \BG^{\sfm_1\cdots \sfm_p, \sfn_1\cdots \sfn_p} &\equiv \delta^{\sfn_1\cdots \sfn_p}_{\sfq_1\cdots \sfq_p}\,\BG^{\sfm_1\sfq_1}\cdots \BG^{\sfm_p\sfq_p}\,,&\qquad 
 \abs{\BG}&\equiv \det(\BG_{\sfm\sfn})\,.
\end{alignedat}
\end{split}
\end{align}
On the other hand, the twist matrices $L$ and $\sfL$ contain various gauge potentials, which can be computed by using the matrix representations of the $E_n$ generators given in Appendix \ref{app:generators}. 

As we have introduced the parameterization of the generalized metrics, let us explain the procedure to obtain the duality rules, which has been proposed in \cite{1701.07819} for $n\leq 7$.

\subsection{Linear map between generalized metrics}

Here, we explain how to determine the duality transformation rules from the generalized metric. 
As we have discussed in Section \ref{sec:1-form}, in the M-theory and type IIB parameterizations, we are using different bases, which are related through the linear map \eqref{eq:linear-map-A}. 
Accordingly, the generalized metrics in the two parameterizations are related as
\begin{align}
 \sfM_{\sfI\sfJ} = (S^\rmT)_\sfI{}^K\,\cM_{KL}\,S^L{}_\sfJ \,.
\label{eq:linear-map-metric}
\end{align}
The explicit form of $S^I{}_\sfJ$ has been obtained in \cite{1701.07819} only for $n\leq 7$\,, and in this paper, we extend the result to be applicable to $n\leq 8$\,. 
Under the linear map, the generalized coordinates are transformed as
\begin{align}
 x^I = S^I{}_\sfJ\, \sfx^\sfJ \,,\qquad \sfx^\sfI = (S^{-1})^\sfI{}_J\,x^J \,.
\label{eq:E8-linear-map}
\end{align}
Since the matrix size of $S^I{}_\sfJ$ is very large ($21\times 21$), we show the linear map as follows:
\begin{align*}
{\footnotesize
 \underbrace{\begin{pmatrix}
 x^a \\[-0.2mm]
 x^\alpha \\ \hline
 \frac{y_{a_1a_2}}{\sqrt{2!}} \\[-0.2mm]
 y_{a \alpha} \\[-0.2mm]
 y_{\Ay\Az} \\ \hline
 \frac{y_{a_1\cdots a_5}}{\sqrt{5!}} \\[-0.2mm]
 \frac{y_{a_1\cdots a_4\alpha}}{\sqrt{4!}} \\[-0.2mm]
 \frac{y_{a_1a_2a_3 \Ay\Az}}{\sqrt{3!}} \\ \hline
 \frac{y_{a_1\cdots a_6\alpha,a}}{\sqrt{6!}} \\[-0.2mm]
 \frac{y_{a_1\cdots a_6(\alpha,\beta)}}{\sqrt{6!}} \\[-0.2mm]
 \frac{\epsilon^{\alpha\beta}\,y_{a_1\cdots a_6[\alpha,\beta]}}{\sqrt{2!\,6!}} \\[-0.2mm]
 \frac{y_{a_1\cdots a_5\Ay\Az,a}}{\sqrt{5!}} \\[-0.2mm]
 \frac{y_{a_1\cdots a_5 \Ay\Az,\alpha}}{\sqrt{5!}} \\ \hline
 \frac{y_{a_1\cdots a_6\Ay\Az,b_1b_2b_3}}{\sqrt{6!\,3!}} \\[-0.2mm]
 \frac{y_{a_1\cdots a_6\Ay\Az,b_1b_2\alpha}}{\sqrt{6!\,2!}} \\[-0.2mm]
 \frac{y_{a_1\cdots a_6\Ay\Az,a\Ay\Az}}{\sqrt{6!}} \\ \hline
 \frac{y_{a_1\cdots a_6\Ay\Az,b_1\cdots b_6}}{\sqrt{6!\,6!}} \\[-0.2mm]
 \frac{y_{a_1\cdots a_6\Ay\Az,b_1\cdots b_5\alpha}}{\sqrt{6!\,5!}} \\[-0.2mm]
 \frac{y_{a_1\cdots a_6\Ay\Az,b_1\cdots b_4\Ay\Az}}{\sqrt{6!\,4!}} \\ \hline
 \frac{y_{a_1\cdots a_6\Ay\Az,b_1\cdots b_6\Ay\Az,a}}{\sqrt{6!\,6!}}\\[-0.2mm]
 \frac{y_{a_1\cdots a_6\Ay\Az,b_1\cdots b_6\Ay\Az,\alpha}}{\sqrt{6!\,6!}}
 \end{pmatrix}}_{x^I}
 =
 \underbrace{\begin{pmatrix}
 \sfx^a \\[-0.5mm]
 \By_\By^\alpha \\ \hline
 \frac{\By_{a_1a_2\By}}{\sqrt{2!}} \\[-0.5mm]
 \By_a^\beta\,\epsilon_{\beta\alpha} \\[-0.5mm]
 \sfx^\By \\ \hline
 \frac{\By_{a_1\cdots a_5\By,\By}}{\sqrt{5!}} \\[-0.5mm]
 \frac{\By^\beta_{a_1\cdots a_4}\,\epsilon_{\beta\alpha}}{\sqrt{4!}} \\[-0.5mm]
 \frac{\By_{a_1a_2a_3}}{\sqrt{3!}} \\ \hline
 \frac{\By^\beta_{a_1\cdots a_6\By,a\By}\,\epsilon_{\beta\alpha}}{\sqrt{6!}} \\[-0.5mm]
 \frac{\epsilon_{\alpha\gamma}\,\epsilon_{\beta\delta}\,\By^{\gamma\delta}_{a_1\cdots a_6\By}}{\sqrt{6!}} \\[-0.5mm]
 \sfY_{a_1\cdots a_6} \\[-0.5mm]
 \sfY_{a_1\cdots a_5,a} \\[-0.5mm]
 \frac{\By_{a_1\cdots a_5}^\beta\,\epsilon_{\beta\alpha}}{\sqrt{5!}} \\ \hline
 \frac{\By_{a_1\cdots a_6\By,b_1b_2b_3\By}}{\sqrt{6!\,3!}} \\[-0.5mm]
 \frac{\By^\gamma_{a_1\cdots a_6\By,b_1b_2}\,\epsilon_{\beta\alpha}}{\sqrt{6!\,2!}} \\[-0.5mm]
 \frac{\By_{a_1\cdots a_6,a}}{\sqrt{6!}} \\ \hline
 \frac{\By_{a_1\cdots a_6\By,b_1\cdots b_6\By,\By}}{\sqrt{6!\,6!}} \\[-0.5mm]
 \frac{\By^\beta_{a_1\cdots a_6\By,b_1\cdots b_5\By}\,\epsilon_{\beta\alpha}}{\sqrt{6!\,5!}} \\[-0.5mm]
 \frac{\By_{a_1\cdots a_6\By,b_1\cdots b_4}}{\sqrt{6!\,4!}} \\ \hline
 \frac{\By_{a_1\cdots a_6\By,b_1\cdots b_6\By,a}}{\sqrt{6!\,6!}} \\[-0.5mm]
 \frac{\By^\beta_{a_1\cdots a_6\By,b_1\cdots b_6}\,\epsilon_{\beta\alpha}}{\sqrt{6!\,6!}}
\end{pmatrix}}_{S^I{}_\sfI\,\sfx^\sfI}
 \quad \Leftrightarrow \quad 
 \underbrace{\begin{pmatrix}
 \sfx^a \\[-0.5mm] \sfx^{\By} \\ \hline
 \By^\alpha_a \\[-0.5mm] \By^\alpha_{\By} \\ \hline
 \frac{\By_{a_1a_2a_3}}{\sqrt{3!}} \\[-0.5mm]
 \frac{\By_{a_1a_2 \By}}{\sqrt{2!}} \\ \hline
 \frac{\By^\alpha_{a_1\cdots a_5}}{\sqrt{5!}} \\[-0.5mm]
 \frac{\By^\alpha_{a_1\cdots a_4\By}}{\sqrt{4!}} \\ \hline
 \frac{\By_{a_1\cdots a_6,a}}{\sqrt{6!}} \\[-0.5mm]
 \frac{\By_{a_1\cdots a_6,\By}}{\sqrt{6!}} \\[-0.5mm]
 \frac{\By_{a_1\cdots a_5\By,a}}{\sqrt{5!}} \\[-0.5mm]
 \frac{\By_{a_1\cdots a_5\By,\By}}{\sqrt{5!}} \\ \hline
 \frac{\By^{\alpha\beta}_{a_1\cdots a_6\By}}{\sqrt{6!}}\\ \hline
 \frac{\By^\alpha_{a_1\cdots a_6\By,b_1b_2}}{\sqrt{6!\,2!}}\\[-0.5mm]
 \frac{\By^\alpha_{a_1\cdots a_6\By,a \By}}{\sqrt{6!}}\\ \hline
 \frac{\By_{a_1\cdots a_6\By,b_1\cdots b_4}}{\sqrt{6!\,4!}}\\[-0.5mm]
 \frac{\By_{a_1\cdots a_6\By,b_1b_2b_3\By}}{\sqrt{6!\,3!}}\\ \hline
 \frac{\By^\alpha_{a_1\cdots a_6\By,b_1\cdots b_6}}{\sqrt{6!\,6!}}\\[-0.5mm]
 \frac{\By^\alpha_{a_1\cdots a_6\By,b_1\cdots b_5\By}}{\sqrt{6!\,5!}}\\ \hline
 \frac{\By_{a_1\cdots a_6\By,b_1\cdots b_6\By,a}}{\sqrt{6!\,6!}}\\[-0.5mm]
 \frac{\By_{a_1\cdots a_6\By,b_1\cdots b_6\By,\By}}{\sqrt{6!\,6!}}
 \end{pmatrix}}_{\sfx^\sfI}
 = \underbrace{\begin{pmatrix}
 x^a \\[-0.5mm]
 y_{\Ay\Az} \\ \hline
 \epsilon^{\alpha\beta}\,y_{a\beta} \\[-0.5mm]
 x^\alpha \\ \hline
 \frac{y_{a_1a_2a_3\Ay\Az}}{\sqrt{3!}} \\[-0.5mm]
 \frac{y_{a_1a_2}}{\sqrt{2!}} \\ \hline
 \frac{\epsilon^{\alpha\beta}\,y_{a_1\cdots a_5 \Ay\Az,\beta}}{\sqrt{5!}} \\[-0.5mm]
 \frac{\epsilon^{\alpha\beta}\,y_{a_1\cdots a_4\beta}}{\sqrt{4!}} \\ \hline
 \frac{y_{a_1\cdots a_6\Ay\Az,a\Ay\Az}}{\sqrt{6!}} \\[-0.5mm]
 Y_{a_1\cdots a_6} \\[-0.5mm]
 Y_{a_1\cdots a_5,a} \\[-0.5mm]
 \frac{y_{a_1\cdots a_5}}{\sqrt{5!}} \\ \hline
 \frac{\epsilon^{\alpha\gamma}\epsilon^{\beta\delta}\,y_{a_1\cdots a_6(\gamma,\delta)}}{\sqrt{6!}}\\ \hline
 \frac{\epsilon^{\alpha\beta}\,y_{a_1\cdots a_6\Ay\Az,b_1b_2\beta}}{\sqrt{6!\,2!}} \\[-0.5mm]
 \frac{\epsilon^{\alpha\beta}\,y_{a_1\cdots a_6\beta,a}}{\sqrt{6!}} \\ \hline
 \frac{y_{a_1\cdots a_6\Ay\Az,b_1\cdots b_4\Ay\Az}}{\sqrt{6!\,4!}} \\[-0.5mm]
 \frac{y_{a_1\cdots a_6\Ay\Az,b_1b_2b_3}}{\sqrt{6!\,3!}} \\ \hline
 \frac{\epsilon^{\alpha\beta}\,y_{a_1\cdots a_6\Ay\Az,b_1\cdots b_6\Ay\Az,\beta}}{\sqrt{6!\,6!}} \\[-0.5mm]
 \frac{\epsilon^{\alpha\beta}\,y_{a_1\cdots a_6\Ay\Az,b_1\cdots b_5\beta}}{\sqrt{6!\,5!}} \\ \hline
 \frac{y_{a_1\cdots a_6\Ay\Az,b_1\cdots b_6\Ay\Az,a}}{\sqrt{6!\,6!}} \\[-0.5mm]
 \frac{y_{a_1\cdots a_6\Ay\Az,b_1\cdots b_6}}{\sqrt{6!\,6!}}
 \end{pmatrix}}_{(S^{-1})^\sfI{}_I\,x^I},
}
\end{align*}
where
\begin{align}
 \begin{pmatrix}
 \sfY_{a_1\cdots a_6} \\ \sfY_{a_1\cdots a_5,a}
 \end{pmatrix}
 &\equiv 
 \begin{pmatrix}
 \frac{9\sqrt{2}+1}{14}\,\frac{\bdelta^{b_1\cdots b_6}_{a_1\cdots a_6}}{\sqrt{6!\,6!}} & \frac{3\sqrt{2}-2}{28}\,\frac{\bdelta^{b_1\cdots b_5b}_{a_1\cdots a_6}}{\sqrt{6!\,5!}} \\
 \frac{3\sqrt{2}-2}{28}\,\frac{\bdelta_{a_1\cdots a_5a}^{b_1\cdots b_6}}{\sqrt{5!\,6!}} & \frac{\bdelta_{a_1\cdots a_5}^{b_1\cdots b_5}\,\delta_a^b -\frac{3\sqrt{2}+5}{28}\, \bdelta_{a_1\cdots a_5a}^{b_1\cdots b_5b}}{\sqrt{5!\,5!}}
 \end{pmatrix}
 \begin{pmatrix}
  \frac{\By_{b_1\cdots b_6,\By}}{\sqrt{6!}} \\
  \frac{\By_{b_1\cdots b_5\By,b}}{\sqrt{5!}}
 \end{pmatrix} ,
\\
 \begin{pmatrix}
 Y_{a_1\cdots a_6} \\ Y_{a_1\cdots a_5,a}
 \end{pmatrix}
 &\equiv 
 \begin{pmatrix}
 \frac{9\sqrt{2}+1}{14}\,\frac{\bdelta^{b_1\cdots b_6}_{a_1\cdots a_6}}{\sqrt{6!\,6!}} & \frac{3\sqrt{2} -2}{28}\,\frac{\bdelta^{b_1\cdots b_5b}_{a_1\cdots a_6}}{\sqrt{6!\,5!}} \\
 \frac{3\sqrt{2}-2}{28}\,\frac{\bdelta_{a_1\cdots a_5a}^{b_1\cdots b_6}}{\sqrt{5!\,6!}} & \frac{\bdelta_{a_1\cdots a_5}^{b_1\cdots b_5}\,\delta_a^b -\frac{3\sqrt{2}+5}{28}\, \bdelta_{a_1\cdots a_5a}^{b_1\cdots b_5b}}{\sqrt{5!\,5!}}
 \end{pmatrix}
 \begin{pmatrix}
  \frac{\epsilon^{\alpha\beta}\,y_{b_1\cdots b_6[\alpha,\beta]}}{\sqrt{2!\,6!}} \\
  \frac{y_{b_1\cdots b_5\Ay\Az,b}}{\sqrt{5!}}
 \end{pmatrix}
\end{align}
with $\bdelta_{i_1\cdots i_n}^{j_1\cdots j_n}\equiv n!\,\delta_{i_1\cdots i_n}^{j_1\cdots j_n}$\,.
The constant matrix $S^I{}_\sfJ$ can be read off from the above map between the coordinates. 
We can check that the matrix $S^I{}_\sfI$ satisfies the property
\begin{align}
 S^I{}_\sfK \, (S^\GT)^\sfK{}_J = \delta^I_J\,,\qquad 
 (S^\GT)^\sfI{}_K\,S^K{}_\sfJ = \delta^\sfI_\sfJ\,,
\end{align}
under the generalized transpose, which is defined for a matrix $A=(A^I{}_J)$ as
\begin{align}
 (A^\GT)^I{}_J \equiv \delta^{IK}\,(A^\rmT)_K{}^L \,\delta_{LJ}
 \equiv \delta^{IK}\,A^L{}_K\,\delta_{LJ} \,,
\end{align}
namely the standard matrix transpose ${}^\rmT$ followed by a flip in the position of the indices. 
This property shows that the flat metric is preserved under the linear map:
\begin{align}
 \delta_{\sfI\sfJ} = S^K{}_\sfI \, S^L{}_\sfJ\, \delta_{KL} \,. 
\end{align}

Now, the constant matrix $S^I{}_\sfJ$ has been completely determined and the relation \eqref{eq:linear-map-metric} connects the two parameterizations. 
By comparing both sides, we can express the M-theory fields in terms of the type IIB fields, and vice versa. 
In the case $n\leq 7$\,, the generalized metric does not contain the dual graviton, but it appears in $n\geq 8$ and here we consider the generalized metric in $E_8$ EFT. 

\subsection{Connection between the two parameterizations}

By comparing the two parameterizations \eqref{eq:linear-map-metric} of the $E_8$ generalized metric, we find the following relation between the M-theory fields and type IIB fields:
\begin{align}
 \bigl(\MG_{ij}\bigr)
  &\ \, = \ \begin{pmatrix}
 \MG_{ab} & \MG_{a\beta} \\ \MG_{\alpha b} & \MG_{\alpha\beta}
 \end{pmatrix}
\nn\\
 &\overset{\text{M--B}}{=} \Exp{-\frac{2}{3}\,\BPhi}\,\BG_{\By\By}^{1/3}
 \begin{pmatrix}
 \delta_a^c & -\bm{\BA}_{a\By}^\gamma \\ 0 & \delta_\alpha^\gamma
 \end{pmatrix}
\begin{pmatrix}
 \frac{2\,\BG_{c\By, d\By}}{\BG_{\By\By}} & 0 \\ 0 & \frac{\Exp{\BPhi}}{\BG_{\By\By}}\, \Bm_{\gamma\delta}
 \end{pmatrix}
 \begin{pmatrix}
 \delta^d_b & 0 \\ -\bm{\BA}_{b\By}^\delta & \delta^\delta_\beta
 \end{pmatrix} \,,
\\
 \bm{\MA}_{a \Ay\Az} &\overset{\text{M--B}}{=} \frac{\BG_{a \By}}{\BG_{\By\By}} \,,
\\
 \bm{\MA}_{ab \alpha} &\overset{\text{M--B}}{=} \Bigl(\bm{\BA}^\beta_{ab} - \frac{2\,\bm{\BA}^\beta_{[a|\By|}\,\BG_{b]\By}}{\BG_{\By\By}}\Bigr)\,\epsilon_{\beta\alpha}\,, 
\\
 \bm{\MA}_{abc} &\overset{\text{M--B}}{=} \bm{\BA}_{abc \By} -\frac{3}{2}\,\epsilon_{\gamma\delta}\,\bm{\BA}^\gamma_{[ab}\,\bm{\BA}^\delta_{c]\By}
 - \frac{3\,\epsilon_{\gamma\delta}\,\bm{\BA}^\gamma_{[a|\By|}\,\bm{\BA}^\delta_{b|\By|}\,\BG_{c]\By}}{\BG_{\By\By}} \,,
\\
 \bm{\MA}_{a_1\cdots a_4 \Ay\Az}
 &\overset{\text{M--B}}{=} \bm{\BA}_{a_1\cdots a_4}- \Bigl(\frac{2\,\bm{\BA}_{[a_1a_2a_3|\By|}\, \BG_{a_4]\By}}{\BG_{\By\By}} + \frac{3\,\epsilon_{\gamma\delta}\,\bm{\BA}^\gamma_{[a_1a_2}\,\bm{\BA}^\delta_{a_3|\By|}\,\BG_{a_4]\By}}{\BG_{\By\By}} \Bigr)\,,
\\
 \bm{\MA}_{a_1\cdots a_5 \alpha}
 &\overset{\text{M--B}}{=}
 \Bigl(\bm{\BA}^\beta_{a_1\cdots a_5\By}+ 5\,\bm{\BA}_{[a_1a_2a_3|\By|}\,\bm{\BA}^\beta_{a_4a_5]}
 +\frac{5}{2}\, \epsilon_{\gamma\delta}\,\bm{\BA}^\gamma_{[a_1a_2}\,\bm{\BA}^\delta_{a_3|\By|}\,\bm{\BA}^\beta_{a_4a_5]}
\nn\\
 &\quad\ -\frac{10\,\bm{\BA}_{[a_1a_2a_3|\By|}\,\bm{\BA}^\beta_{a_4|\By|}\BG_{a_5]\By}}{\BG_{\By\By}}
 -\frac{15\,\epsilon_{\gamma\delta}\,\bm{\BA}^\gamma_{[a_1a_2}\,\bm{\BA}^\delta_{a_3|\By|}\,\bm{\BA}^\beta_{a_4|\By|}\,\BG_{a_5]\By}}{\BG_{\By\By}} \Bigr) \,\epsilon_{\beta\alpha}
\\
 \bm{\MA}_{a_1\cdots a_6}
 &\overset{\text{M--B}}{=}
  \bm{\BA}_{a_1\cdots a_6\By, \By}
 -15\,\epsilon_{\gamma\delta}\,\bm{\BA}_{[a_1a_2a_3|\By|}\,\Bigl(\bm{\BA}^\gamma_{a_4a_5}\,\bm{\BA}^\delta_{a_6]\By}
 +\frac{2\,\bm{\BA}^\gamma_{a_4|\By|}\,\bm{\BA}^\delta_{a_5|\By|}\,\BG_{a_6]\By}}{\BG_{\By\By}}\Bigr) \,,
\\
 \bm{\MA}_{a_1\cdots a_6 \Ay\Az, \alpha}
 &\overset{\text{M--B}}{=}
 \biggl[\bm{\BA}^\beta_{a_1\cdots a_6}
 +\frac{\bm{\BA}^\beta_{[a_1a_2}\,\bigl(20\,\bm{\BA}_{a_3a_4a_5|\By|}+30\,\epsilon_{\gamma\delta}\,\bm{\BA}^\gamma_{a_3a_4}\,\bm{\BA}^\delta_{a_5|\By|}\bigr)\,\BG_{a_6]\By}}{\BG_{\By\By}} \biggr]\,\epsilon_{\beta\alpha} \,,
\\
 \bm{\MA}_{a_1\cdots a_6 \Ay\Az, b}
 &\overset{\text{M--B}}{=}
 \bm{\BA}_{a_1\cdots a_6\By, b}
 -\frac{15}{2}\,\bm{\BA}_{[a_1\cdots a_4}\, \bm{\BA}_{a_5a_6] b\By}
 -10\, \epsilon_{\alpha\beta}\, \bm{\BA}_{[a_1a_2a_3|\By|}\, \bm{\BA}^\alpha_{a_4a_5}\,\bm{\BA}^\beta_{a_6]b}
\nn\\
 &\quad +\frac{15}{2}\,\epsilon_{\alpha\beta}\,\epsilon_{\gamma\delta}\,\bm{\BA}^\alpha_{[a_1a_2}\,\bm{\BA}^\beta_{a_3|b|}\,\bm{\BA}^\gamma_{a_4a_5}\,\bm{\BA}^\delta_{a_6]\By}
\nn\\
 &\quad +\frac{20 \,\bm{\BA}_{[a_1a_2a_3|\By|}\, \bm{\BA}_{a_4a_5 |b\By|}\, \BG_{a_6]\By}}{\BG_{\By\By}}
 - \frac{20\,\epsilon_{\alpha\beta}\,\bm{\BA}_{[a_1a_2a_3|\By|}\,\bm{\BA}^\alpha_{a_4 |b|}\,\bm{\BA}^\beta_{a_5|\By|}\, \BG_{a_6]\By}}{\BG_{\By\By}}
\nn\\
 &\quad - \frac{10\,\epsilon_{\alpha\beta}\,\bm{\BA}_{[a_1a_2a_3|\By|}\,\bm{\BA}^\alpha_{a_4a_5}\,\bm{\BA}^\beta_{|b\By|}\, \BG_{a_6]\By}}{\BG_{\By\By}}
 + \frac{30\,\epsilon_{\alpha\beta}\,\bm{\BA}_{[a_1a_2|b\By|}\, \bm{\BA}^\alpha_{a_3a_4}\,\bm{\BA}^\beta_{a_5|\By|}\,\BG_{a_6]\By}}{\BG_{\By\By}}
\nn\\
 &\quad +\frac{15\,\epsilon_{\alpha\beta}\,\epsilon_{\gamma\delta}\,\bigl(\bm{\BA}^\alpha_{[a_1a_2}\,\bm{\BA}^\beta_{a_3|b|}\,\bm{\BA}^\gamma_{a_4|\By|}\,\bm{\BA}^\delta_{a_5|\By|}
 - \bm{\BA}^\alpha_{[a_1a_2}\,\bm{\BA}^\beta_{a_3|\By|}\,\bm{\BA}^\gamma_{a_4a_5}\,\bm{\BA}^\delta_{|b\By|}\bigr)\,\BG_{a_6]\By}}{\BG_{\By\By}} \,. 
\end{align}

By making identifications
\begin{align}
\begin{split}
 \bm{\MA}_{\hat{M}_1\hat{M}_2\hat{M}_3} &= \MA_{\hat{M}_1\hat{M}_2\hat{M}_3}\,,\qquad 
 \bm{\MA}_{\hat{M}_1\cdots \hat{M}_6} = \MA_{\hat{M}_1\cdots \hat{M}_6}\,,
\\
 \bm{\MA}_{\hat{M}_1\cdots \hat{M}_8, \hat{N}}
 &\simeq \MA_{\hat{M}_1\cdots \hat{M}_8, \hat{N}} - 28\,\MA_{[\hat{M}_1\cdots \hat{M}_6}\,\MA_{\hat{M}_7\hat{M}_8]\hat{N}} 
\end{split}
\label{eq:MA-bmMA}
\end{align}
for M-theory fields and
\begin{align}
 \bm{\BA}^\alpha_{M_1M_2} &= \BA^\alpha_{M_1M_2}\,,\qquad 
 \bm{\BA}_{M_1\cdots M_4} = \BA_{M_1\cdots M_4} \,,\qquad 
 \bm{\BA}^\alpha_{M_1\cdots M_6} = \BA^\alpha_{M_1\cdots M_6}\,,
\\
 \bm{\BA}_{M_1\cdots M_7, N} &\simeq \BA_{M_1\cdots M_7, N} + 7\,\bigl(\BB_{[M_1\cdots M_6}\,\BB_{M_7]N} - \BB_{[M_1\cdots M_6}\,\BB_{M_7N]} \bigr)
\nn\\
 &\quad +\frac{105}{4}\, \bigl[ \BC_{[M_1\cdots M_4}\, \bigl(\BB_{M_5M_6}\,\BC_{M_7]N}-3\,\BC_{M_5M_6}\,\BB_{M_7] N} \bigr)
\nn\\
 &\quad\qquad\quad - \BC_{[M_1\cdots M_4}\, \bigl(\BB_{M_5M_6}\,\BC_{M_7N]} -3\,\BC_{M_5M_6}\,\BB_{M_7N]} \bigr) \bigr]
\nn\\
 &\quad +\frac{315}{8}\,\bigl(\BC_{[M_1M_2}\,\BC_{M_3M_4}\, \BB_{M_5M_6}\,\BB_{M_7]N} - \BC_{[M_1M_2}\,\BC_{M_3M_4}\, \BB_{M_5M_6}\,\BB_{M_7N]}\bigr) 
\label{eq:BA-bmBA}
\end{align}
for type IIB fields, and by using the 11D--10D map, these relations are precisely the $T$-duality rules obtained in Section \ref{sec:T-dual}.
The $S$-duality rule for the new dual graviton is simply
\begin{align}
 \bm{\BA}'_{M_1\cdots M_7, N} = \bm{\BA}_{M_1\cdots M_7, N} \,,
\end{align}
which is consistent with \eqref{eq:dual-graviton-S-dual} under the identification \eqref{eq:BA-bmBA}. 

Note that, in order to obtain the duality rules for the higher mixed-symmetry potentials, we need to consider the $E_n$ generalized metric with $n\geq 9$\,. 

\subsection{1-form $\cA_\mu^{I}$ as the generalized graviphoton}

In Section \ref{sec:1-form}, we found that the 1-form gauge field $\cA_\mu^{I}$ has a simple structure in terms of the tensors $\MN$ and $\BN$:
\begin{align}
 \cA_\mu^{I} = \MN_\mu^{I} + \hat{A}_\mu^k\,\MN_k^{I}\quad (\text{M-theory})\,,\qquad 
 \bm{\cA}_\mu^{\sfI} = \BN_\mu^{\sfI} + \BA_\mu^\sfp\,\BN_\sfp^{\sfI}\quad (\text{type IIB})\,. 
\label{eq:Amu-structure}
\end{align}
In fact, this combination has a clear origin. 
The basic idea is as follows. 

\paragraph*{Generalized graviphoton in DFT:}

In type IIB theory, the graviphoton is given by
\begin{align}
 \BA_\mu^\sfm = \bm{\Bg}_{\mu\nu}\,\Bg^{\nu \sfm} \,. 
\end{align}
We can consider a generalization of this graviphoton in double field theory (DFT). 
In DFT, the inverse of the generalized metric has the form,\footnote{Note that the $B$-field in this paper has the opposite sign to the one usually used in DFT.}
\begin{align}
 \cH^{\hat{I}\hat{J}} = \begin{pmatrix} \Mg^{MN} & \Mg^{MK}\,\hat{B}_{KN} \\ -\hat{B}_{MK}\,\Mg^{KN} & (\Mg-\hat{B}\,\Mg^{-1}\,\hat{B})_{MN}
\end{pmatrix},\qquad 
 (x^{\hat{I}})\equiv (x^M,\,\tilde{x}_M)\,.
\end{align}
We decompose the physical coordinates as $(x^M)=(x^\mu,\,x^m)$ and define the generalized coordinates for the compact directions as $(x^I)\equiv (x^m,\,\tilde{x}_m)$ ($m=1,\dotsc,n-1$). 
Then, we find
\begin{align}
\begin{split}
 \cH^{\mu I} &= \bigl(\Mg^{\mu m},\, \Mg^{\mu K}\,\hat{B}_{Km}\bigr) = \bigl(\Mg^{\mu m},\, \Mg^{\mu\nu}\,\hat{B}_{\nu m} + \Mg^{\mu p}\,\hat{B}_{pm}\bigr)
\\
 &= \Mg^{\mu\nu}\, \bigl(A_\nu^m,\, \hat{B}_{\nu m} + A_\nu^p \,\hat{B}_{pm}\bigr)\,.\end{split}
\end{align}
This leads us to define the generalized graviphoton as
\begin{align}
 \cA_\mu^I \equiv \bm{g}_{\mu\nu}\,\cH^{\nu I} = \begin{pmatrix} A_\mu^m \\ \hat{B}_{\mu m} + A_\mu^p\,\hat{B}_{p m} \end{pmatrix} \qquad 
 \bigl[\bm{g}\equiv (\Mg^{\mu\nu})^{-1} \bigr]\,,
\end{align}
which transforms covariantly under $\OO(n-1,n-1)$ transformations, and is sometimes used in the double sigma model (see, e.g., \cite{hep-th:0406102,1802.00442}). 
In the context of DFT, $A_\mu^M$ has been studied in \cite{1307.0039}, where it is called the Kaluza--Klein vector. 
By using
\begin{align}
 N_N{}^I \equiv \begin{pmatrix} \delta_N^m \\ \hat{B}_{Nm} \end{pmatrix} ,
\end{align}
we observe that the generalized graviphoton can be expressed as
\begin{align}
 \cA_\mu^{I} = N_\mu{}^{I} + A_\mu^p\,N_p{}^{I}\,,
\end{align}
which has the same structure as \eqref{eq:Amu-structure}. 

\paragraph*{Generalized graviphoton in EFT:}

We now consider the case of EFT starting with the generalized metric $\cM_{\hat{I}\hat{J}}$ in $E_{11}$ EFT. 
Denoting the inverse matrix of $\cM^{\mu\nu}$ by $\bm{m}_{\mu\nu}$\,, we define the generalized graviphoton as
\begin{align}
 \cA_\mu^I \equiv \bm{m}_{\mu\nu}\,\cM^{\nu I} \,. 
\end{align}
In the following, we show that this $\cA_\mu^I$ is precisely the 1-form considered in Section \ref{sec:1-form}. 
To this end, let us recall that the generalized metric has the structure
\begin{align}
 \cM_{IJ} = (L^\rmT\,\hat{\cM}\,L)_{IJ}\,,\qquad 
 \cM^{IJ} = (L^{-1}\,\hat{\cM}^{-1}\,L^{-\rmT})^{IJ} \,.
\end{align}
By using the fact that the matrix $L$ has a lower-triangular form, we find
\begin{align}
 \cM^{\mu\nu} &= \hat{\cM}^{\mu\nu} = \abs{\Mg}^{\frac{1}{2}}\, \bm{g}^{\mu\nu} \equiv \bm{m}^{\mu\nu}\,,
\\
 \cM^{\mu I} &= (\hat{\cM}^{-1}\,L^{-\rmT})^{\mu J} = \hat{\cM}^{\mu N}\,(L^{-\rmT})_{N}{}^{J} = \abs{\Mg}^{\frac{1}{2}} \, \bm{g}^{\mu\nu} \, \bigl[(L^{-\rmT})_{\nu}{}^I + A_\nu^k\,(L^{-\rmT})_{k}{}^I \bigr] \,,
\end{align}
where we have used
\begin{align}
 (\hat{\cM}^{\hat{M}\hat{N}}) = \abs{\Mg}^{\frac{1}{2}} \begin{pmatrix} \delta^\mu_\rho & 0 \\ \MA^i_\rho & \delta^i_k \end{pmatrix}
 \begin{pmatrix} \bm{g}^{\rho\sigma} & 0 \\ 0 & \MG^{kl} \end{pmatrix}
 \begin{pmatrix} \delta_\sigma^\mu & \MA_\sigma^j \\ 0 & \delta_l^j \end{pmatrix}.
\end{align}
Then, we obtain
\begin{align}
 \cA_\mu^I = \bm{m}_{\mu\nu}\, \cM^{\nu I} = (L^{-\rmT})_{\mu}{}^I + \MA_\mu^k\,(L^{-\rmT})_{k}{}^I\,. 
\end{align}

In order to show that this is the same as the 1-form considered in Section \ref{sec:1-form}, let us compute the explicit form of $(L^{-\rmT})_{\mu}{}^I$ in M-theory/type IIB parameterizations. 
In the M-theory parameterization, $L$ is defined as \eqref{eq:L-M} and by using the matrix representations of the $E_{11}$ generators given in Appendix \ref{app:generators}, we obtain
\begin{align}
 (L^{-\rmT})_{\hat{N}}{}^I \simeq 
 \begin{pmatrix}
 \delta^i_{\hat{N}} \\
 \frac{\bm{\MA}_{\hat{N} i_1i_2}}{\sqrt{2!}}
\\
 \frac{\bm{\MA}_{\hat{N} i_1\cdots i_5}-5\,\bm{\MA}_{\hat{N}[i_1i_2}\,\bm{\MA}_{i_3i_4i_5]}}{\sqrt{5!}} \\
 \frac{\bm{\MA}_{\hat{N}i_1\cdots i_7,i} - 7\,\bm{\MA}_{\hat{N}i[i_1\cdots i_4}\,\bm{\MA}_{i_5i_6i_7]} +35\,\bm{\MA}_{\hat{N}[i_1i_2}\,\bm{\MA}_{i_3i_4i_5}\,\bm{\MA}_{i_6i_7]i}}{\sqrt{7!}}
\\
 \vdots
 \end{pmatrix} ,
\label{eq:L-T-M}
\end{align}
where $i\in \{i_1,\dotsc ,i_7\}$ has been assumed for the fourth row. 
By using the identification \eqref{eq:MA-bmMA}, $(L^{-\rmT})_{\hat{N}}{}^I$ is precisely the same as $\MN_{\hat{N}}{}^I$ given in \eqref{eq:N-M} and the generalized graviphoton is the same as the 1-form \eqref{eq:cA-M}. 

On the other hand, in the type IIB parameterization, $\sfL$ is defined as \eqref{eq:L-B} and we obtain
\begin{align}
 (\sfL^{-\rmT})_N{}^{\sfI} \simeq 
 \begin{pmatrix}
 \delta_\mu^\sfm 
\\
 \bm{\BA}_{N\sfm}^\alpha 
\\
 \frac{\bm{\BA}_{N\sfm_1\sfm_2\sfm_3} - \frac{3}{2}\,\epsilon_{\gamma\delta}\,\bm{\BA}^\gamma_{N[\sfm_1}\,\bm{\BA}^\delta_{\sfm_2\sfm_3]}}{\sqrt{3!}}
\\
 \frac{\bm{\BA}^\alpha_{N\sfm_1 \cdots \sfm_5} + 5\,\bm{\BA}^\alpha_{N[\sfm_1}\, \bm{\BA}_{\sfm_2\cdots \sfm_5]}
 + 5\,\epsilon_{\gamma\delta}\,\bm{\BA}^\gamma_{N[\sfm_1}\,\bm{\BA}^\delta_{\sfm_2\sfm_3}\,\bm{\BA}^\alpha_{\sfm_4\sfm_5]}}{\sqrt{5!}}
\\
 \frac{\Bigl[\genfrac{}{}{0pt}{1}{\bm{\BA}_{N \sfm_1\cdots \sfm_6, \sfm} + \epsilon_{\gamma\delta}\, \bm{\BA}^\gamma_{N\sfm}\,\bm{\BA}^\delta_{\sfm_1\cdots \sfm_6}+10\,\bm{\BA}_{N[\sfm_1\sfm_2\sfm_3}\,\bm{\BA}_{\sfm_4\sfm_5\sfm_6]\sfm}~~~~~~~~~~~~~~}{-30\,\epsilon_{\gamma\delta}\,\bm{\BA}^\gamma_{N[\sfm_1}\,\bm{\BA}^\delta_{\sfm_2\sfm_3}\,\bm{\BA}_{\sfm_4\sfm_5\sfm_6]\sfm}+\frac{15}{2}\,\epsilon_{\alpha\beta}\,\epsilon_{\gamma\delta}\,\bm{\BA}^\alpha_{N[\sfm_1}\,\bm{\BA}^\beta_{\sfm_2\sfm_3}\,\bm{\BA}^\gamma_{\sfm_4\sfm_5}\,\bm{\BA}^\delta_{\sfm_6]\sfm}}\Bigr]}{\sqrt{6!}}
\\
 \vdots
\end{pmatrix}, 
\label{eq:L-T-B}
\end{align}
where $\sfm\in \{\sfm_1,\dotsc, \sfm_6\}$ has been assumed for the fifth row. 
Again by using the identification \eqref{eq:BA-bmBA} and $\sfm\in \{\sfm_1,\dotsc, \sfm_6\}$\,, $(\sfL^{-\rmT})_{N}{}^{\sfI}$ matches $\BN_{N}{}^\sfI$ given in \eqref{eq:N-B} and the generalized graviphoton in the type IIB parameterization is precisely the 1-form \eqref{eq:cA-B}. 

Here, let us comment on the relation to the series of papers \cite{1403.7198,1409.6314,1412.2768}. 
The standard wave solution in 11D supergravity has the non-vanishing flux associated with the graviphoton $\cA_\mu^i$\,. 
In \cite{1403.7198,1409.6314,1412.2768}, the wave solution was embedded into EFT, which has non-vanishing $\cA_\mu^I$\,. 
Then, by rotating the duality frames, various brane solutions were obtained in EFT. 
Particularly in \cite{1412.2768}, the 1-form $\cA_\mu^I$ was regarded as the graviphoton in the $(4+56)$-dimensional exceptional space. 
Since all of their brane solutions in EFT couple to the generalized graviphoton $\cA_\mu^I$\,, branes were interpreted as a kind of generalized wave in the exceptional space. 
Although the explicit parameterization of $\cA_\mu^I$ was not determined there, conceptually, their idea is closely related to the result obtained here. 

\section{Parameterization of $\cA_p^{I_p}$}
\label{sec:p-forms}

In this section, we study the parameterization of the higher $p$-form fields $\cA_p^{I_p}$\,. 

\subsection*{2-form $\cA_2^{I_2}$}

The 2-form gauge field $\cA_2^{I_2}$ transforms in the string multiplet, characterized by the Dynkin label $[0,\dotsc,0,1,0]$\,. 
It is decomposed as
\begin{align}
{\footnotesize
 \cA_{\mu\nu}^{I_2}
 = \begin{pmatrix}
 \MA_{[\mu\nu]; i} \\
 \frac{1}{\sqrt{4!}}\,\MA_{[\mu\nu]; i_1\cdots i_4} \\
 \frac{1}{\sqrt{6!}}\,\MA_{[\mu\nu]; i_1\cdots i_6,i} \\
 \vdots 
 \end{pmatrix},\qquad 
 \bm{\cA}_{\mu\nu}^{I_2}
 = \begin{pmatrix}
 \BA^\alpha_{\mu\nu} \\
 \frac{1}{\sqrt{2!}}\,\BA_{[\mu\nu]; \sfm_1\sfm_2} \\
 \frac{1}{\sqrt{4!}}\,\BA^\alpha_{[\mu\nu]; \sfm_1\cdots \sfm_4} \\
 \frac{1}{\sqrt{6!}}\,\BA^\alpha_{[\mu\nu]; \sfm_1\cdots \sfm_5,\sfm} \\
 \vdots 
 \end{pmatrix}, }
\end{align}
and, e.g., the first component in each parameterization can be expanded as
\begin{align}
 \MA_{[\mu\nu]; i} &= \MA_{\mu\nu i} + m_1\, \MA_{[\mu}^k\, \MA_{\nu] ki} + m_2\, \MA_{[\mu}^k\,\MA_{\nu]}^l\, \MA_{kl i} \,,
\\
 \BA^\alpha_{\mu\nu} &= \BA^\alpha_{\mu\nu} + b_1\, \BA_{[\mu}^\sfp\, \BA^\alpha_{\nu] \sfp} + b_2\, \BA_{[\mu}^\sfp\,\BA_{\nu]}^\sfq\, \BA^\alpha_{\sfp\sfq} \,,
\end{align}
by introducing parameters $m_1,\,m_2,\,b_1$, and $b_2$\,. 
We already have the $T$-duality rules, and by following the same procedure as the 1-form, we can determine these parameters. 

Repeating the procedure, we find the parameterization
{\footnotesize
\begin{align}
 \cA_{\mu\nu}^{I_2}
 &= \begin{pmatrix}
 \MN_{[\mu;\nu] i} + \MA_{[\mu|}^k\,\MN_{k;|\nu]i}^{\phantom{k}} \\
 \frac{1}{\sqrt{4!}}\bigl(\MN_{[\mu;\nu] i_1\cdots i_4} + \MA_{[\mu|}^k\,\MN_{k;|\nu]i_1\cdots i_4}^{\phantom{k}}\bigr) \\
 \frac{1}{\sqrt{6!}}\, \bigl(\MN_{[\mu;\nu] i_1\cdots i_6, i} + \MA_{[\mu|}^k\,\MN_{k;|\nu] i_1\cdots i_6,i}\bigr)\\
 \vdots 
 \end{pmatrix},
\\
 \bm{\cA}_{\mu\nu}^{I_2}
 &= \begin{pmatrix}
 \BN^\alpha_{[\mu;\nu]} + \BA_{[\mu|}^\sfp\,\BN^\alpha_{\sfp;|\nu]} \\
 \frac{1}{\sqrt{2!}}\bigl(\BN_{[\mu;\nu] m_1m_2} + \BA_{[\mu|}^\sfp\,\BN_{\sfp;|\nu]\sfm_1\sfm_2}^{\phantom{p}}\bigr) \\
 \frac{1}{\sqrt{4!}}\bigl(\BN^\alpha_{[\mu;\nu] \sfm_1\cdots \sfm_4} + \BA_{[\mu|}^\sfp\,\BN^\alpha_{\sfp;|\nu]\sfm_1\cdots \sfm_4}\bigr) \\
 \frac{1}{\sqrt{5!}}\bigl(\BN_{[\mu;\nu] \sfm_1\cdots \sfm_5,\sfm} + \BA_{[\mu|}^\sfp\,\BN_{\sfp;|\nu]\sfm_1\cdots \sfm_5,\sfm}\bigr) \\
 \vdots 
 \end{pmatrix}.
\end{align}}\noindent
Interestingly, the tensors $\MN$ and $\BN$ are precisely the same as those defined in Section \ref{sec:1-form-results}. 
The origin of this simple structure can be understood as follows. 

For example, let us consider the map \eqref{eq:3-form-2-form}
\begin{align}
 \MA_{\mu a\alpha} + \MA_\mu^k\,\MA_{ka\alpha} \overset{\text{M--B}}{=} \bigl(\BA_{\mu a}^\beta + \BA_\mu^\sfp\,\BA^\beta_{\sfp a}\bigr)\,\epsilon_{\beta\alpha} \,,
\end{align}
in which both sides are connected through $T$-duality. 
However, the $T$-duality rule is 9D covariant, and even if we replace the index $a$ by the 9D index $A=(\mu,\,a)$\,, the above relation is still satisfied. 
Then, choosing $A=\nu$ and antisymmetrizing $\mu$ and $\nu$\,, we get
\begin{align}
 \MA_{\mu\nu \alpha} + \MA_{[\mu}^k\,\MA_{|k|\nu] \alpha} \overset{\text{M--B}}{=} \bigl(\BA_{\mu\nu}^\beta + \BA_{[\mu}^\sfp\,\BA^\beta_{|\sfp|\nu]}\bigr)\,\epsilon_{\beta\alpha} \,,
\end{align}
which connects the first row of $\cA_{\mu\nu}^{I_2}$ and the first row of $\bm{\cA}_{\mu\nu}^{I_2}$\,. 
In this manner, simply by replacing an internal index $a$ with an external index $\nu$ and acting the antisymmetrization, we obtain the parameterization of the 2-form from the result of the 1-form.

In the literature, several components of the 1-form and 2-form have been studied in, e.g., \cite{1312.0614,1506.01065}. 
By following the notation of \cite{1802.00442}, their M-theory parameterizations are
\begin{align}
 \cA_\mu{}^m = A_\mu{}^m\,,\qquad
 \cA_{\mu mn} = \frac{\hat{C}_{\mu mn}{}^\beta - A_\mu{}^k\, \hat{C}_{k mn}{}^\beta}{\sqrt{2}} \,, \qquad
 \cB_{\mu\nu m} = \frac{\hat{C}_{\mu\nu m} - A_{[\mu}{}^k \,\hat{C}_{\nu] m k}}{\sqrt{10}}\,,
\end{align}
while the type IIB parameterizations are
\begin{align}
 \cA_\mu{}^i = A_\mu{}^i\,,\qquad
 \cA_{\mu i\alpha} = \epsilon_{\alpha\beta}\, \bigl(\hat{C}_{\mu i}{}^\beta - A_\mu{}^k\, \hat{C}_{k i}{}^\beta \bigr) \,, \qquad
 \cB_{\mu\nu}{}^\alpha = \frac{C_{\mu\nu}{}^\alpha - A_{[\mu}{}^j\,\hat{C}_{|j|\nu]}{}^\alpha}{\sqrt{10}}\,.
\end{align}
By comparing, e.g., $\cA_{\mu mn}$ with $\cB_{\mu\nu m}$\,, we find that their results also follow the antisymmetrization rule and seem to be consistent with our results up to conventions. 

\subsection*{$3$-form and higher $p$-form}

Similar to the case of the 2-form, a parameterization of a general $p$-form can be obtained by acting the antisymmetrization on that of the 1-form. 
In the case of the 3-form, we obtain
{\footnotesize
\begin{align}
 \cA_{\mu\nu\rho}^{I_2}
 &= \begin{pmatrix}
 \MN_{[\mu;\nu\rho]} + \MA_{[\mu|}^k\,\MN_{k;|\nu\rho]}^{\phantom{k}} \\
 \frac{1}{\sqrt{3!}}\bigl(\MN_{[\mu;\nu\rho] i_1i_2i_3} + \MA_{[\mu|}^k\,\MN_{k;|\nu\rho]i_1i_2i_3}^{\phantom{k}}\bigr) \\
 \frac{1}{\sqrt{5!}}\, \bigl(\MN_{[\mu;\nu\rho] i_1\cdots i_5, i} + \MA_{[\mu|}^k\,\MN_{k;|\nu\rho] i_1\cdots i_5,i}\bigr)\\
 \vdots 
 \end{pmatrix},
\\
 \bm{\cA}_{\mu\nu\rho}^{I_2}
 &= \begin{pmatrix}
 \BN_{[\mu;\nu\rho] \sfm} + \BA_{[\mu|}^\sfp\,\BN_{\sfp;|\nu\rho]\sfm}^{\phantom{p}} \\
 \frac{1}{\sqrt{3!}}\bigl(\BN^\alpha_{[\mu;\nu\rho] \sfm_1\sfm_2\sfm_3} + \BA_{[\mu|}^\sfp\,\BN^\alpha_{\sfp;|\nu\rho] \sfm_1\sfm_2\sfm_3}\bigr) \\
 \frac{1}{\sqrt{4!}}\bigl(\BN_{[\mu;\nu\rho] \sfm_1\cdots \sfm_4,\sfm} + \BA_{[\mu|}^\sfp\,\BN_{\sfp;|\nu\rho]\sfm_1\cdots \sfm_4,\sfm}\bigr) \\
 \vdots 
 \end{pmatrix}.
\end{align}}\noindent
Compared to the 2-form, the first component in the type IIB side $\BN^\alpha_{\mu;\nu}$ has disappeared because the number of indices is not enough to account for a 3-form. 
The 4-form is
{\footnotesize
\begin{align}
 \cA_{\mu_1\cdots\mu_4}^{I_2}
 &= \begin{pmatrix}
 \frac{1}{\sqrt{2!}}\bigl(\MN_{[\mu_1;\mu_2\cdots\mu_4] i_1i_2} + \MA_{[\mu_1|}^k\,\MN_{k;|\mu_2\mu_3\mu_4]i_1i_2}^{\phantom{k}}\bigr) \\
 \frac{1}{\sqrt{4!}}\, \bigl(\MN_{[\mu_1;\mu_2\cdots\mu_4] i_1\cdots i_4, i} + \MA_{[\mu_1|}^k\,\MN_{k;|\mu_2\mu_3\mu_4] i_1\cdots i_4,i}\bigr)\\
 \vdots 
 \end{pmatrix},
\\
 \bm{\cA}_{\mu_1\cdots\mu_4}^{I_2}
 &= \begin{pmatrix}
 \BN_{[\mu_1;\mu_2\cdots\mu_4]} + \BA_{[\mu_1|}^\sfp\,\BN_{\sfp;|\mu_2\mu_3\mu_4]}^{\phantom{p}} \\
 \frac{1}{\sqrt{2!}}\bigl(\BN^\alpha_{[\mu_1;\mu_2\cdots\mu_4] \sfm_1\sfm_2} + \BA_{[\mu_1|}^\sfp\,\BN^\alpha_{\sfp;|\mu_2\mu_3\mu_4] \sfm_1\sfm_2}\bigr) \\
 \frac{1}{\sqrt{3!}}\bigl(\BN_{[\mu_1;\mu_2\cdots\mu_4]\sfm_1\sfm_2\sfm_3,\sfm} + \BA_{[\mu_1|}^\sfp\,\BN_{\sfp;|\mu_2\mu_3\mu_4]\sfm_1\sfm_2\sfm_3,\sfm}\bigr) \\
 \vdots 
 \end{pmatrix},
\end{align}}\noindent
and higher $p$-forms are also obtained in a similar manner. 

We note that if there exists a certain invariant tensor $f_{I_p I_q}{}^{I_r}$ with symmetry $f_{I_p I_q}{}^{I_r}=(-1)^{pq}\,f_{I_q I_p}{}^{I_r}$\,, we can redefine an $r$-form $\cA^{I_r}_r$ as
\begin{align}
 \cA^{I_r}_r \ \to \ \cA'^{I_r}_r = \cA^{I_r}_r + f_{I_p I_q}{}^{I_r}\,\cA^{I_p}_p\wedge \cA_q^{I_q} \,.
\label{eq:re-define-A}
\end{align}
In such case, the $r$-form field is not unique and we cannot fix the parameterization unambiguously. 

\section{Summary and discussion}
\label{sec:conclusions}

In this paper, we have proposed a systematic way to determine the parameterization of the $p$-form field $\cA^{I_p}_p$\,. 
As a demonstration, we have determined how the dual graviton enters the $p$-form field. 
We have also determined the duality rules for the dual graviton, which have been partially studied in the literature. 
Our procedure is based on the (factorized) $T$-duality and $S$-duality transformations, which form a subgroup of the full $U$-duality group. 
Accordingly, our procedure cannot determine the contribution of the mixed-symmetry potentials that do not couple to any supersymmetric branes. 
However, we have provided another approach to determine the parameterization of $\cA^{I_p}_p$\,. 
We have found that the 1-form field is precisely the generalized graviphoton $\cA_\mu^I = \bm{m}_{\mu\nu}\, \cM^{\nu I}$ defined by the $E_{11}$ generalized metric. 
By following the procedure of \cite{hep-th:0104081,hep-th:0307098,1009.2624}, we can in principle determine the parameterization of the $E_{11}$ generalized metric level by level. 
We can then determine the full parameterization of the 1-form field. 
As we have shown, once the parameterization of the 1-form field is determined, we can easily obtain the parameterization of the $p$-form field by antisymmetrizing the indices. 

As future directions, it would be interesting to revisit the worldvolume actions of exotic branes. 
In the case of exotic branes, the Wess--Zumino term contains the mixed-symmetry potentials, but at present, the explicit forms of the brane actions are known for a few examples \cite{hep-th:9908094,1309.2653,1404.5442,1601.05589}. 
A manifestly $U$-duality-covariant Wess--Zumino term, which employs the $p$-form fields $\cA^{I_p}_p$, has been proposed in \cite{1009.4657} and it is important to clarify the connection to the results of \cite{hep-th:9908094,1309.2653,1404.5442,1601.05589} by using the concrete parameterization of $\cA^{I_p}_p$\,. 
It would be also interesting to develop another $U$-duality-manifest approach to brane actions \cite{1607.04265,1712.10316} (see also \cite{1712.07115,1802.00442} for a similar approach). 

It would also be useful to study the duality transformation rules for more mixed-symmetry potentials beyond the dual graviton. 
By following the procedure proposed in this paper, it is a straightforward task to determine such duality rules. 
Recently, a $T$-duality manifest formulation for mixed-symmetry potentials has been studied in detail in \cite{1903.05601}, which aims to be more useful for determining the $T$-duality rules. 
Nevertheless, in order to consider the $S$-duality rule or the M-theory uplifts, our $U$-duality-based procedure would potentially prove more useful. 

\subsection*{Acknowledgments}

The work of JJFM is supported by Plan Propio de Investigaci\'on of the University of Murcia R-957/2017 and Fundaci\'on S\'eneca (21257/PI/19 and 20949/PI/18).
The work of YS is supported by JSPS KAKENHI Grant Numbers 18K13540 and 18H01214.

\appendix

\section{Notation}
\label{app:notation}

In this appendix, we summarize the notation that has been used in this work to denote various fields corresponding to each theory and each dimension, as well as the different types of indices.

M-theory and type IIA/IIB theory are defined in $D$ dimensions, where $D=11,10$ respectively. 
Upon a dimensional reduction on a torus, we have a $d$-dimensional supergravity theory, with a global symmetry group $E_n$, where $n=D-d$. 
According to this, all the splittings of the M-theory and type IIB coordinates and the higher/lower-dimensional indices that have been used are shown in Figure \ref{fig:indices}. 
The $D$-dimensional coordinates in M-theory and type IIB theory are denoted by $x^{\hat{M}}$ and $\sfx^M$\,, respectively.

In addition, indices for the $p$-form multiplet are denoted as $I_p$ in M-theory and as $\sfI_p$ in type IIB theory. 
In particular, for the 1-form, we denote $I\equiv I_1$ and $\sfI_1\equiv \sfI$\,. 
In type IIB theory, the index of the vector representation of the $\SL(2)$ $S$-duality group is represented by $\alpha=1,2$.

\begin{figure}[bp]
\centering
\begin{align*}
\hat M \left[
\phantom{\begin{array}{c}
\mbox{}\\\mbox{}\\[8pt]\mbox{}\\\mbox{}\\\mbox{}\\\mbox{}\\\mbox{}\\\mbox{}
\end{array}
}
\right. 
\hspace{-12pt}
\begin{array}{l}
M \left[
\phantom{\begin{array}{c}
\mbox{}\\[8pt]\mbox{}\\\mbox{}\\\mbox{}\\\mbox{}\\\mbox{}\\\mbox{}
\end{array}
}
\right.
\\
\phantom{\mbox{}}
\end{array}
\hspace{-12pt}
\begin{array}{l}
A \left[
\phantom{\begin{array}{c}
\mbox{}\\\mbox{}\\\mbox{}\\\mbox{}\\\mbox{}\\\mbox{}
\end{array}
}
\right.
\\
\phantom{\mbox{}}
\\
\phantom{\mbox{}}
\end{array}
\hspace{-12pt}
\begin{array}{l}
\phantom{\mbox{}}
\\
\phantom{\mbox{}}
\\
\phantom{\mbox{}}
\\ 
i \left[
\phantom{\begin{array}{c}
\mbox{}\\[8pt]\mbox{}\\\mbox{}\\\mbox{}\\\mbox{}
\end{array}
}
\right.
\end{array}
\hspace{-12pt}
\begin{array}{l}
\phantom{\mbox{}}
\\
\phantom{\mbox{}}
\\
\phantom{\mbox{}}
\\ 
m \left[
\phantom{\begin{array}{c}
\mbox{}\\[8pt]\mbox{}\\\mbox{}\\\mbox{}
\end{array}
}
\right.
\\
\phantom{\mbox{}}
\end{array}
\hspace{-12pt}
\begin{array}{r}
\mu \left[
\phantom{\begin{array}{c}
\mbox{}\\\mbox{}\\[5pt]\mbox{}
\end{array}
}
\right.
\\[8pt]
a \left[
\phantom{\begin{array}{c}
\mbox{}\\\mbox{}\\\mbox{}
\end{array}
}
\right.
\\[8pt]
\alpha \left[
\phantom{\begin{array}{c}
\mbox{}\\\mbox{}
\end{array}
}
\right.
\end{array}
\hspace{-12pt}
\left(
\begin{array}{c}
x^0\\\vdots\\ x^{d-1}
\\
x^d\\\vdots\\ x^{8}
\\
x^y\\x^z
\end{array}
\right)
\hspace{-12pt}
\begin{array}{c}
\mbox{}\\\mbox{}\mbox{}\\\mbox{}\\\mbox{}\\\mbox{}\\[5pt] \mbox{}\\ \quad \leftarrow \ T \text{-duality} \ \rightarrow \quad \\S^1 
\end{array}
\hspace{-12pt}
\begin{array}{r}
\left(
\begin{array}{c}
\sfx^0\\\vdots\\ \sfx^{d-1}
\\
\sfx^d\\\vdots\\ \sfx^{8}
\\
\sfx^\sfy
\end{array}
\right)
\\
\phantom{\mbox{}}
\end{array}
\hspace{-12pt}
\begin{array}{l}
\left.
\phantom{\begin{array}{c}
\mbox{}\\\mbox{}\\\mbox{}
\end{array}
}
\right] \mu 
\\ 
\left.
\phantom{\begin{array}{c}
\mbox{}\\\mbox{}\\\mbox{}
\end{array}
}
\right] a
\\
\phantom{\begin{array}{c}
\mbox{}\\\mbox{}
\end{array}
}
\end{array}
\hspace{-12pt}
\begin{array}{l}
\phantom{\mbox{}}
\\[8pt]
\phantom{\mbox{}}
\\
\phantom{\mbox{}}
\\ 
\left.
\phantom{\begin{array}{c}
\mbox{}\\\mbox{}\\\mbox{}\\\mbox{}
\end{array}
}
\right] \sfm
\\
\phantom{\mbox{}}
\end{array}
\hspace{-12pt}
\begin{array}{l}
\left.
\phantom{\begin{array}{c}
\mbox{}\\\mbox{}\\\mbox{}\\\mbox{}\\\mbox{}\\\mbox{}
\end{array}
}
\right] A
\\
\phantom{\mbox{}}
\\
\phantom{\mbox{}}
\end{array}
\hspace{-12pt}
\begin{array}{l}
\left.
\phantom{\begin{array}{c}
\mbox{}\\[8pt]\mbox{}\\\mbox{}\\\mbox{}\\\mbox{}\\\mbox{}\\\mbox{}
\end{array}
}
\right] M
\\
\phantom{\mbox{}}
\end{array}
\end{align*}
\caption{\textit{Left: splittings of the M-theory coordinates $(x^{\hat M})$ and their index notation. A compactification on a circle $S^1$ along the direction $x^z$ and a $T$-duality transformation along the $x^y$ coordinate are considered. Right: splittings of the type IIB coordinates $(\sfx^\sfM)$ and their notations are shown. A $T$-duality transformation is taken along the coordinate $\sfx^\sfy$.}}
\label{fig:indices}
\end{figure}

\begin{table}[t]
\centering
\begin{tabular}{cccc}
Field & M-theory & Type IIB & Type IIA
\\
\hline
$U$-duality-covariant $p$-form & $\cA_p^{I_p}$ & $\bm{\cA}_p^{\sfI_p}$ & --
\\
generalized metric & $\cM$ & $\sfM$ & --
\\
generalized vielbein & $\cE$ & $\sfE$ & --
\\
twist matrix & $L$ & $\sfL$ & --
\\
$D$-dim.\ metric & $\Mg$ & $\Bg$ & $g$
\\
$D$-dim.\ fields & $\MA$ & $\BA$, $\BB$, $\BC$ & $\AA$, $\AB$, $\AC$
\\
$D$-dim.\ fields (Section \ref{sec:generalized-metric})& $\bm{\MA}$ & $\bm{\BA}$ & --
\\
spacetime metric & $\bm{\Mg}$ & $\bm{\Bg}$ & $\bm{g}$
\\
internal metric & $\MG$ & $\BG$ & $G$
\\
\end{tabular}
\caption{\textit{
Summary of the fields that have been used in this work. While the first four lines correspond to $U$-duality multiplets, the rest correspond to standard supergravity fields. In the last two lines we show the $d$-dimensional fields that appear after compactification.
}}
\label{tab:fields}
\end{table}

In Table \ref{tab:fields}, we summarize the notation that we have used to represent the fields of various theories. 
Fields transforming as $U$-duality multiplets are considered. 
Similarly, standard supergravity fields of M-theory and type II theories and the lower-dimensional fields that arise after compactification are considered.

\section{$E_n$ generators}
\label{app:generators}

In this appendix, we show the explicit matrix representation of the $E_n$ generators in the vector representation. 
In the M-theory parameterization, our matrices are consistent with \cite{1303.2035}. 
Through the linear map from M-theory parameterization to type IIB parameterization, we also find the matrix representations in the type IIB parameterization, which is new. 

Here, we show the results for $E_8$\,, but the $E_n$ generators with $n\leq 7$ can be easily obtained through a truncation. 
For example, an $E_8$ generator $R^{i_{1\cdots8},i}$ disappears in $E_7$ because the index $i$ ranges over seven directions and $i_{1\cdots8}$ automatically vanishes. 
Conversely, our $E_8$ generators can be understood as a truncation of the $E_{11}$ generators. 
In $E_{11}$\,, the matrix representation becomes infinite dimensional, but the first several blocks are the same as the $E_8$ generators. 
Accordingly, although we have computed the matrix $(L^{-\rmT})$ in \eqref{eq:L-T-M} by using the $E_8$ generators, the first four rows do not change even if we use the matrix representation of the $E_{11}$ generators.\footnote{To be more precise, in our matrix representations in M-theory, in the fourth row and below that, we have used Schouten-like identities; i.e., terms with antisymmetrized nine indices $(\cdots)_{[i_1\cdots i_7 i j]}$ have been dropped because they disappear automatically in $n\leq 8$\,. However, this does not affect the computation of $(L^{-\rmT})$ in \eqref{eq:L-T-M} because the restriction rule $i\in \{i_1,\dotsc ,i_7\}$ has been assumed there and terms with the structure $(\cdots)_{[i_1\cdots i_7 i j]}$ disappear even for $n=11$\,. In this sense, \eqref{eq:L-T-M} can be understood as being obtained from the $E_{11}$ generalized metric.}
In that sense, the results given in this appendix can be understood as a truncation of the $E_{11}$ generators. 

\subsection{M-theory parameterization}

In the M-theory parameterization, the $E_n$ generators are decomposed as
\begin{align}
 \{T_{\hat{\alpha}}\} = \bigl\{ K^i{}_j,\, R_{i_{123}} ,\, R_{i_{1\cdots6}} ,\, R_{i_{1\cdots8},i} ,\, R^{i_{123}} ,\, R^{i_{1\cdots6}} ,\, R^{i_{1\cdots8},i} ,\,\dotsc\bigr\}\,,
\end{align}
where the ellipses disappear for $n\leq 8$\,. 
In this appendix, we may use a short-hand notation for the multiple indices:
\begin{align}
 (\cdots)_{i_{1\cdots p}} \equiv (\cdots)_{i_1\cdots i_p}\,.
\end{align}
We are also using the notation
\begin{align}
 \bdelta_{i_1\cdots i_n}^{j_1\cdots j_n}\equiv n!\,\delta_{i_1\cdots i_n}^{j_1\cdots j_n}\,. 
\end{align}
If we restrict ourselves to the case $n\leq 8$\,, they satisfy the following commutation relations:\footnote{If we consider the $E_{11}$ algebra, for example $\bigl[R^{i_1i_2i_3},\, R_{j_1 \cdots j_8,j}\bigr]$ needs to be modified as $\bigl[R^{i_1i_2i_3},\, R_{j_1 \cdots j_8,j}\bigr] = \frac{1}{5!}\, \bigl(\bdelta^{i_1i_2i_3 r_1\cdots r_5}_{j_1 \cdots j_8}\, R_{r_1\cdots r_5 j} - \bdelta^{i_1i_2i_3 r_1\cdots r_5}_{[j_1 \cdots j_8|}\, R_{r_1\cdots r_5 |j]}\bigr)$\,. However, the second term on the right-hand side identically vanishes for $n\leq 8$\,. In this sense, the commutation relations shown here are valid only in the case $n\leq 8$\,.}
\begin{align}
 \bigl[K^i{}_j,\, K^k{}_l \bigr]
 &= \delta_j^k\,K^i{}_l - \delta_l^i\, K^k{}_j \,,
\\
 \bigl[K^i{}_j,\, R^{k_1k_2k_3} \bigr]
 &= \frac{1}{2!}\,\bdelta_{jr_1r_2}^{k_1k_2k_3} \,R^{ir_1r_2}\,,
\\
 \bigl[K^i{}_j,\, R^{k_1\cdots k_6} \bigr]
 &= \frac{1}{5!}\,\bdelta_{jr_1\cdots r_5}^{k_1\cdots k_6} \, R^{ir_1\cdots r_5}\,,
\\
 \bigl[K^i{}_j,\,R^{k_1\cdots k_8,k}\bigr]
 &= \frac{1}{7!}\,\bdelta^{k_1\cdots k_8}_{jr_1\cdots r_7}\, R^{ir_1 \cdots r_7,k} + \delta_j^k\,R^{k_1\cdots k_8,i}\,, 
\\
 \bigl[K^i{}_j,\, R_{k_1k_2k_3} \bigr]
 &= -\frac{1}{2!}\,\bdelta_{k_1k_2k_3}^{ir_1r_2} \, R_{jr_1r_2}\,,
\\
 \bigl[K^i{}_j,\, R_{k_1\cdots k_6} \bigr]
 &= -\frac{1}{5!}\,\bdelta^{ir_1\cdots r_5}_{k_1\cdots k_6}\, R_{jr_1\cdots r_5}\,,
\\
 \bigl[K^i{}_j,\,R_{k_1\cdots k_8,k}\bigr]
 &= -\frac{1}{7!}\,\bdelta_{k_1\cdots k_8}^{ir_1\cdots r_7}\, R_{jr_1 \cdots r_7,k} - \delta_k^i\,R_{k_1\cdots k_8,j}\,,
\\
 \bigl[R^{i_1i_2i_3},\, R^{j_1j_2j_3}\bigr]
 &= - R^{i_1i_2i_3 j_1j_2j_3}\,,
\\
 \bigl[R^{i_1i_2i_3},\, R^{j_1 \cdots j_6}\bigr]
 &= - \frac{1}{5!}\, \bdelta^{j_1 \cdots j_6}_{r_1\cdots r_5 s}\, R^{i_1i_2i_3 r_1\cdots r_5,s}\,,
\\
 \bigl[ R^{i_1i_2i_3},\,R_{j_1j_2j_3} \bigr]
 &= \frac{1}{2!}\,\bdelta_{r_1r_2 s}^{i_1i_2i_3}\,\bdelta_{j_1j_2j_3}^{r_1r_2 t}\, K^s{}_t - \frac{1}{3}\,\bdelta_{j_1j_2j_3}^{i_1i_2i_3}\,\delta^t_s\, K^s{}_t \,, 
\\
 \bigl[ R^{i_1i_2i_3},\, R_{j_1\cdots j_6} \bigr]
 &= \frac{1}{3!}\, \bdelta_{j_1\cdots j_6}^{i_1i_2i_3 r_1r_2r_3}\,R_{r_1r_2r_3} \,, 
\\
 \bigl[R^{i_1i_2i_3},\, R_{j_1 \cdots j_8,j}\bigr]
 &= \frac{1}{5!}\, \bdelta^{i_1i_2i_3 r_1\cdots r_5}_{j_1 \cdots j_8}\, R_{r_1\cdots r_5 j} \,,
\\
 \bigl[R^{i_1\cdots i_6},\, R_{j_1j_2j_3}\bigr]
 &= \frac{1}{3!}\,\bdelta^{i_1\cdots i_6}_{j_1j_2j_3 r_1r_2r_3}\,R^{r_1r_2r_3} \,, 
\\
 \bigl[R^{i_1\cdots i_6},\,R_{j_1\cdots j_6} \bigr]
 &= \frac{1}{5!}\, \bdelta^{i_1\cdots i_6}_{r_1\cdots r_5 s}\, \bdelta_{j_1\cdots j_6}^{r_1\cdots r_5 t}\, K^s{}_t - \frac{2}{3}\,\bdelta_{j_1\cdots j_6}^{i_1\cdots i_6}\,\delta^t_s \,K^s{}_t \,,
\\
 \bigl[R^{i_1 \cdots i_6},\, R_{j_1 \cdots j_8,j}\bigr]
 &= \frac{1}{2!}\,\bdelta^{i_1 \cdots i_6 r_1r_2}_{j_1 \cdots j_8}\, R_{r_1r_2j} \,,
\\
 \bigl[R^{i_1\cdots i_8,i},\, R_{j_1j_2j_3}\bigr]
 &= \frac{1}{5!}\, \bdelta^{i_1 \cdots i_8}_{j_1j_2j_3 r_1\cdots r_5}\, R^{r_1\cdots r_5 i} \,,
\\
 \bigl[R^{i_1 \cdots i_8,i}, R_{j_1 \cdots j_6}\bigr]
 &= \frac{1}{2!}\, \bdelta^{i_1 \cdots i_8}_{j_1\cdots j_6 r_1r_2}\, R^{r_1r_2 i} \,,
\\
 \bigl[R^{i_1 \cdots i_8,i}, R_{j_1 \cdots j_8,j}\bigr]
 &= \bdelta^{i_1\cdots i_8}_{j_1 \cdots j_8}\, K^i{}_j \,,
\\
 \bigl[R_{i_1i_2i_3},\,R_{j_1j_2j_3}\bigr]
 &= R_{i_1i_2i_3j_1j_2j_3}\,,
\\
 \bigl[R_{i_1i_2i_3},\, R_{j_1 \cdots j_6}\bigr]
 &= \frac{1}{5!}\, \bdelta_{j_1\cdots j_6}^{r_1\cdots r_5 s}\, R_{i_1i_2i_3r_1\cdots r_5,s}\,. 
\end{align}
We note that our convention will be related to that of \cite{1303.2035} as follows:
\begin{align*}
 \begin{array}{|c||c|c|c|c|c|c|c|} \hline
 \text{Here} & K^i{}_j & R_{i_{123}} & R^{i_{123}} & R_{i_{1\cdots 6}} & R^{i_{1\cdots 6}} & R_{i_{1\cdots 8},i} & R^{i_{1\cdots 8},i}
 \\ \hline
 \text{\cite{1303.2035}} & K^i{}_j & R_{i_{123}} & R^{i_{123}} & 2\,R_{i_{1\cdots 6}} & -2\,R^{i_{1\cdots 6}} & 2 \,R_{i_{1\cdots 8},i} & 2\, R^{i_{1\cdots 8},i} \\ \hline
 \end{array} \,.
\end{align*}

Now, we show the matrix representations of these generators in the vector representation. 
In the M-theory parameterization, the vector representation (for $n\leq 8$) is decomposed as
\begin{align}
 \{x^I\} &= \Bigl\{x^i,\,\tfrac{y_{i_{12}}}{\sqrt{2!}},\,\tfrac{y_{i_{1\cdots 5}}}{\sqrt{5!}},\,\tfrac{y_{i_{1\cdots 7},i}}{\sqrt{7!}},\,\tfrac{y_{i_{1\cdots 8}}}{\sqrt{8!}},\,\tfrac{y_{i_{1\cdots 8},k_{123}}}{\sqrt{8!\,3!}},\,\tfrac{y_{i_{1\cdots 8},k_{1\cdots 6}}}{\sqrt{8!\,6!}},\,\tfrac{y_{i_{1\cdots 8},k_{1\cdots 8},i}}{\sqrt{8!\,8!}} \Bigr\},
\end{align}
where $y_{[i_{1\cdots 7},i]}=0$\,. 
In this paper, in order to reduce the matrix size, we have combined $y_{i_{1\cdots 7},i}$ and $y_{i_{1\cdots 8}}$\,, and our $y_{i_{1\cdots 7},i}$ do not satisfy $y_{[i_{1\cdots 7},i]}=0$\,. 
We then find that the following matrices $(T_{\hat{\alpha}})^I{}_J$ satisfy the above $E_8$ algebra:
\begin{align}
 K^{p}{}_{q} &\equiv {\footnotesize {\arraycolsep=0.2mm \underset{7\times7}{\mathrm{diag}} 
 \begin{pmatrix}
 -\delta_{q}^i \delta_j^{p} \\
 \frac{\bdelta_{i_{12}}^{pr} \bdelta_{qr}^{j_{12}}}{\sqrt{2!\,2!}} \\
 \frac{\bdelta_{i_{1\cdots5}}^{pr_{1\cdots4}} \bdelta_{qr_{1\cdots4}}^{j_{1\cdots5}}}{4!\sqrt{5!\,5!}} \\
 \frac{\frac{1}{6!}\bdelta_{i_{1\cdots7}}^{pr_{1\cdots6}} \bdelta_{qr_{1\cdots6}}^{j_{1\cdots 7}}\delta_i^j +\bdelta_{i_{1\cdots7}}^{j_{1\cdots 7}} \delta_{i}^{p}\delta_{q}^j}{\sqrt{7!\,7!}} \\
 \frac{\frac{1}{7!}\bdelta_{i_{1\cdots8}}^{pr_{1\cdots7}} \bdelta_{qr_{1\cdots7}}^{j_{1\cdots8}}\bdelta_{k_{123}}^{l_{123}} +\frac{1}{2!}\bdelta_{i_{1\cdots8}}^{j_{1\cdots8}} \bdelta_{k_{123}}^{pr_{12}}\bdelta_{qr_{12}}^{l_{123}}}{\sqrt{8!\,3!\,8!\,3!}} \\
 \frac{\frac{1}{7!}\bdelta_{i_{1\cdots8}}^{pr_{1\cdots7}} \bdelta_{qr_{1\cdots7}}^{j_{1\cdots8}}\bdelta_{k_{1\cdots6}}^{l_{1\cdots6}} +\frac{1}{5!}\bdelta_{i_{1\cdots8}}^{j_{1\cdots8}} \bdelta_{k_{1\cdots6}}^{pr_{1\cdots5}}\bdelta_{qr_{1\cdots5}}^{l_{1\cdots6}}}{\sqrt{8!\,6!\,8!\,6!}} \\
 \frac{\frac{1}{7!}\bdelta_{i_{1\cdots8}}^{pr_{1\cdots7}} \bdelta_{qr_{1\cdots7}}^{j_{1\cdots8}}\bdelta_{k_{1\cdots8}}^{l_{1\cdots8}}\delta_i^j + \frac{1}{7!}\bdelta_{i_{1\cdots8}}^{j_{1\cdots8}} \bdelta_{k_{1\cdots8}}^{pr_{1\cdots7}}\bdelta_{qr_{1\cdots7}}^{l_{1\cdots8}}\delta_i^j + \bdelta_{i_{1\cdots8}}^{j_{1\cdots8}} \bdelta_{k_{1\cdots8}}^{l_{1\cdots8}} \delta^{p}_i \delta_{q}^j}{\sqrt{8!\,8!\,8!\,8!}}
 \end{pmatrix} - \frac{\delta_{q}^{p} \,\delta^I_J}{9-n}}} \,, 
\\
 R^{p_{123}} &\equiv {\footnotesize 
 {\arraycolsep=0mm \begin{pmatrix}
 0 & 0 & 0 & 0 & 0 & 0 & 0 \\
 \frac{-\bdelta_{j i_{12}}^{p_{123}}}{\sqrt{2!}}\!\! & 0 & 0 & 0 & 0 & 0 & 0 \\
 0 & \frac{\bdelta^{j_{12} p_{123}}_{i_{1\cdots5}}}{\sqrt{5!\,2!}} & 0 & 0 & 0 & 0 & 0 \\
 0 & 0 & \!\!\!\!\!\!\frac{\frac{1}{2!}\bdelta^{j_{1\cdots5} r_{12}}_{i_{1\cdots 7}} \bdelta^{p_{123}}_{r_{12}i} -\frac{1}{4} \bdelta^{j_{1\cdots5} p_{123}}_{i_{1\cdots 7}i}}{\sqrt{7!\,5!}}\!\!\!\!\!\!\!\!\!\!\!\! & 0 & 0 & 0 & 0 \\
 0 & 0 & 0 & \!\!\!\!\!\!\!\!\!\!\!\!\!\!\!\!\!\!\!\frac{\frac{1}{2!} \bdelta^{j_{1\cdots7} r}_{i_{1\cdots8}} \bdelta^{p_{123}}_{rs_{12}} \bdelta^{s_{12}j}_{k_{123}}-\frac{1}{4}\bdelta^{j_{1\cdots7}j}_{i_{1\cdots8}} \bdelta^{p_{123}}_{k_{123}}}{\sqrt{8!\,3!\,7!}}\!\!\!\!\!\!\!\!\! & 0 & 0 & 0 \\
 0 & 0 & 0 & 0 & \!\!\!\!\!\!\!\!\frac{\bdelta^{j_{1\cdots8}}_{i_{1\cdots8}} \bdelta^{l_{123}p_{123}}_{k_{1\cdots6}}}{\sqrt{8!\,6!\,8!\,3!}}\!\!\!\! & 0 & 0 \\
 0 & 0 & 0 & 0 & 0 & \!\!\!\!\!\!\!\!\frac{\bdelta^{j_{1\cdots8}}_{i_{1\cdots8}} \bdelta^{l_{1\cdots6}r_{12}}_{k_{1\cdots8}} \bdelta^{p_{123}}_{i r_{12}}}{2!\sqrt{8!\,8!\,8!\,6!}} & ~0~ 
 \end{pmatrix} } } , 
\\
 R_{p_{123}} &\equiv {\footnotesize 
 {\arraycolsep=-1.0mm \begin{pmatrix}
 ~0~ & ~~\frac{-\bdelta^{i j_{12}}_{p_{123}}}{\sqrt{2!}}~ & 0 & 0 & 0 & 0 & 0 \\
 0 & 0 & \!\!\frac{\bdelta_{i_{12} p_{123}}^{j_{1\cdots5}}}{\sqrt{2!\,5!}}\! & 0 & 0 & 0 & 0 \\
 0 & 0 & 0 & \frac{\frac{1}{2!}\bdelta_{i_{1\cdots5} r_{12}}^{j_{1\cdots 7}} \bdelta_{p_{123}}^{r_{12}j} -\frac{1}{4} \bdelta_{i_{1\cdots5} p_{123}}^{j_{1\cdots 7}j}}{\sqrt{5!\,7!}}\!\!\!\!\!\!\!\!\!\!\!\!\!\!\! 
 & 0 & 0 & 0 \\
 0 & 0 & 0 & 0& \!\!\!\!\!\!\!\!\!\!\!\!\!\!\frac{\frac{1}{2!} \bdelta_{i_{1\cdots7} r}^{j_{1\cdots8}} \bdelta_{p_{123}}^{rs_{12}} \bdelta_{s_{12}i}^{l_{123}}-\frac{1}{4}\bdelta_{i_{1\cdots7}i}^{j_{1\cdots8}} \bdelta_{p_{123}}^{l_{123}}}{\sqrt{7!\,8!\,3!}}\!\!\!\!\!\!\!\!\! & 0 & 0 \\
 0 & 0 & 0 & 0& 0 & \!\!\frac{\bdelta_{i_{1\cdots8}}^{j_{1\cdots8}} \bdelta_{k_{123}p_{123}}^{l_{1\cdots6}}}{\sqrt{8!\,3!\,8!\,6!}}\!\! & 0 \\
 0 & 0 & 0 & 0& 0 & 0 & \!\!\!\!\!\frac{\bdelta_{i_{1\cdots8}}^{j_{1\cdots8}} \bdelta_{k_{1\cdots6}r_{12}}^{l_{1\cdots8}} \bdelta_{p_{123}}^{j r_{12}}}{2!\sqrt{8!\,6!\,8!\,8!}} \\
 0 & 0 & 0 & 0& 0 & 0 & 0 
 \end{pmatrix} } }
\nn\\
 &= (R^{p_{123}})^\GT \,, 
\\
 R^{p_{1\cdots6}} &\equiv {\footnotesize {\arraycolsep=0mm \begin{pmatrix}
 0 & 0 & 0 & 0 & 0 & 0 & 0 \\
 0 & 0 & 0 & 0 & 0 & 0 & 0 \\
 \frac{\bdelta_{i_{1\cdots5} j}^{p_{1\cdots6}}}{\sqrt{5!}} & 0 & 0 & 0 & 0 & 0 & 0 \\
 0 & \!\! \frac{\bdelta^{j_{12}}_{ir} \bdelta^{p_{1\cdots6} r}_{i_{1\cdots 7}} +\frac{1}{2} \bdelta^{p_{1\cdots6} j_{12}}_{i_{1\cdots 7}i}}{\sqrt{7!\,2!}}\!\!\!\!\!\! & 0 & 0 & 0 & 0 & 0 \\
 0 & 0 & \!\!\!\!\!\!\!\!\frac{\bdelta^{j_{1\cdots5} r_{123}}_{i_{1\cdots8}} \bdelta^{p_{1\cdots6}}_{k_{123} r_{123}}}{3!\sqrt{8!\,3!\,5!}}\!\!\!\!\!\!\!\! & 0 & 0 & 0 & 0 \\
 0 & 0 & 0 & \!\!\!\!\!\!\!\!\frac{\bdelta^{r j_{1\cdots7}}_{i_{1\cdots8}} \bdelta^{p_{1\cdots6}j}_{k_{1\cdots6} r} + \frac{1}{2} \bdelta^{j_{1\cdots7}j}_{i_{1\cdots8}} \bdelta^{p_{1\cdots6}}_{k_{1\cdots6}}}{\sqrt{8!\,6!\,7!}}\!\!\!\!\!\!\!\! & 0 & 0 & 0 \\
 0 & 0 & 0 & 0 & \!\!\!\!\frac{-\bdelta^{j_{1\cdots8}}_{i_{1\cdots8}} \bdelta^{p_{1\cdots6} r_{12}}_{k_{1\cdots8}} \bdelta^{l_{123}}_{i r_{12}}}{2!\sqrt{8!\,8!\,8!\,3!}} & ~0~ & ~0~ \end{pmatrix} } } ,
\\
 R_{p_{1\cdots6}} &\equiv {\footnotesize {\arraycolsep=0mm \begin{pmatrix}
 0 & ~0~ & ~\frac{\bdelta^{j_{1\cdots5} i}_{p_{1\cdots6}}}{\sqrt{5!}}~ & 0 & 0 & 0 & 0 \\
 0 & 0 & 0 & \!\!\!\!\!\! \frac{\bdelta_{i_{12}}^{jr} \bdelta_{p_{1\cdots6} r}^{j_{1\cdots 7}} +\frac{1}{2} \bdelta_{p_{1\cdots6} i_{12}}^{j_{1\cdots 7}j}}{\sqrt{2!\,7!}} \!\!\!\!\!\!\!\! & 0 & 0 & 0 \\
 0 & 0 & 0 & 0 & \frac{\bdelta_{i_{1\cdots5} r_{123}}^{j_{1\cdots8}} \bdelta_{p_{1\cdots6}}^{l_{123} r_{123}}}{3!\sqrt{5!\,8!\,3!}} & 0 & 0 \\
 0 & 0 & 0 & 0& 0 & \!\!\!\!\!\!\!\! \frac{\bdelta_{r i_{1\cdots7}}^{j_{1\cdots8}} \bdelta_{p_{1\cdots6}i}^{l_{1\cdots6} r} + \frac{1}{2} \bdelta_{i_{1\cdots7}i}^{j_{1\cdots8}} \bdelta_{p_{1\cdots6}}^{l_{1\cdots6}}}{\sqrt{7!\,8!\,6!}} \!\!\!\!\!\!\!\!\!\! & 0 \\
 0 & 0 & 0 & 0& 0 & 0 & \!\!\!\!\!\!\!\! \frac{-\bdelta_{i_{1\cdots8}}^{j_{1\cdots8}} \bdelta_{p_{1\cdots6} r_{12}}^{l_{1\cdots8}} \bdelta_{k_{123}}^{j r_{12}}}{2!\sqrt{8!\,3!\,8!\,8!}} \\
 0 & 0 & 0 & 0& 0 & 0 & 0 \\
 0 & 0 & 0 & 0& 0 & 0 & 0 \end{pmatrix}}}
\nn\\
 &= (R^{p_{1\cdots6}})^\GT \,, 
\\
 R^{p_{1\cdots8},p} &\equiv {\footnotesize {\arraycolsep=0.5mm \begin{pmatrix}
 0 & 0 & 0 & 0 & 0 & 0 & 0 \\
 0 & 0 & 0 & 0 & 0 & 0 & 0 \\
 0 & 0 & 0 & 0 & 0 & 0 & 0 \\
 \frac{\bdelta_{i_{1\cdots 7}j}^{p_{1\cdots8}}\delta_i^p+\frac{1}{4} \bdelta^{p_{1\cdots8}}_{i_{1\cdots 7}i}\delta^p_j}{\sqrt{7!}} & 0 & 0 & 0 & 0 & 0 & 0 \\
 0 & -\frac{\bdelta_{k_{123}}^{j_{12}p}\bdelta^{p_{1\cdots8}}_{i_{1\cdots8}}}{\sqrt{8!\,3!\,2!}} & 0 & 0 & 0 & 0 & 0 \\
 0 & 0 & \frac{\bdelta^{j_{1\cdots5}p}_{k_{1\cdots6}}\bdelta^{p_{1\cdots8}}_{i_{1\cdots8}}}{\sqrt{8!\,6!\,5!}} & 0 & 0 & 0 & 0 \\
 0 & 0 & 0 & \!\!\frac{\bdelta^{j_{1\cdots7} p}_{i_{1\cdots8}}\bdelta^{p_{1\cdots8}}_{k_{1\cdots8}}\delta^j_i + \frac{1}{4}\bdelta^{j_{1\cdots7}j}_{i_{1\cdots8}}\bdelta^{p_{1\cdots8}}_{k_{1\cdots8}}\delta^p_i}{\sqrt{8!\,8!\,7!}} & 0 & 0 & 0 \end{pmatrix}}} ,
\\
 R_{p_{1\cdots8},p} &\equiv {\footnotesize {\arraycolsep=0.5mm \begin{pmatrix}
 0 & 0 & 0 & \frac{\bdelta^{j_{1\cdots 7}i}_{p_{1\cdots8}}\delta^j_p+\frac{1}{4} \bdelta_{p_{1\cdots8}}^{j_{1\cdots 7}j}\delta_p^i}{\sqrt{7!}} & 0 & 0 & 0 \\
 0 & 0 & 0 & 0 & -\frac{\bdelta^{l_{123}}_{i_{12}p}\bdelta_{p_{1\cdots8}}^{j_{1\cdots8}}}{\sqrt{2!\,8!\,3!}} & 0 & 0 \\
 0 & 0 & 0 & 0 & 0 & \frac{\bdelta_{i_{1\cdots5}p}^{l_{1\cdots6}}\bdelta_{p_{1\cdots8}}^{j_{1\cdots8}}}{\sqrt{5!\,8!\,6!}} & 0 \\
 0 & 0 & 0 & 0& 0 & 0 & \frac{\bdelta_{i_{1\cdots7} p}^{j_{1\cdots8}}\bdelta_{p_{1\cdots8}}^{l_{1\cdots8}}\delta_i^j + \frac{1}{4}\bdelta_{i_{1\cdots7}i}^{j_{1\cdots8}}\bdelta_{p_{1\cdots8}}^{l_{1\cdots8}}\delta_p^j}{\sqrt{7!\,8!\,8!}} \\
 0 & 0 & 0 & 0& 0 & 0 & 0 \\
 0 & 0 & 0 & 0& 0 & 0 & 0 \\
 0 & 0 & 0 & 0& 0 & 0 & 0 \end{pmatrix}}}
\nn\\
 &= (R^{p_{1\cdots8},p})^\GT \,.
\end{align}
We can identify the Cartan generators as
\begin{align}
 \{H_k\} = \{K^d{}_d - K^{d+1}{}_{d+1},\, \dotsc,\, K^9{}_9 - K^z{}_z,\, K^8{}_8 + K^9{}_9 + K^z{}_z + \tfrac{1}{3}\,D \} \,,
\end{align}
and the positive/negative simple-root generators are
\begin{align}
 \{E_k\} = \{K^d{}_{d+1} ,\, \dotsc,\, K^9{}_z,\, R^{89z} \} \,,\qquad
 \{F_k\} = \{K^{d+1}{}_d ,\, \dotsc,\, K^z{}_9,\, R_{89z} \} \,. 
\end{align}
They satisfy the relations
\begin{align}
\begin{split}
 &[H_k,\,E_l]=A_{kl}\,E_l\,,\qquad
 [H_k,\,F_l]=-A_{kl}\,F_l\,,\qquad
 [E_k,\,F_l]= \delta_{kl}\,H_l\,, 
\\
 &\frac{1}{\alpha_n}\,\tr(H_k\,H_l) = A_{kl}\,,\qquad 
 \frac{1}{\alpha_n}\,\tr(E_k\,F_l) = \delta_{kl}\,,
\end{split}
\end{align}
where
\begin{align}
 (A_{kl})= {\footnotesize\begin{pmatrix}
  2 & -1 & & & & \\
  -1 & 2 & \ddots & & & \\
  & \ddots & \ddots & -1 & & -1 \\
  & & -1 & 2 & -1 & 0 \\
  & & & -1 & 2 & 0 \\
  & & -1 & 0 & 0 & 2
\end{pmatrix}},\qquad\quad
 \begin{array}{|c||c|c|c|c|c|} \hline
 n&4&5&6&7&8 \\ \hline
 \alpha_n&3&4&6&12&60 \\ \hline
\end{array} \,.
\end{align}
The set of positive/negative root generators can be obtained by taking commutators of the simple-root generators $E_k$/$F_k$\,, and they can be summarized as
\begin{align}
\begin{split}
 \{E_{\bm{\alpha}}\} &= \{K^i{}_j\ (i<j),\, R_{i_{123}},\,R_{i_{1\cdots 6}},\,R_{i_{1\cdots 8},i}\}\,,
\\
 \{F_{\bm{\alpha}}\} &= \{K^i{}_j\ (i>j),\, R^{i_{123}},\,R^{i_{1\cdots 6}},\,R^{i_{1\cdots 8},i}\}\,.
\end{split}
\end{align}

\subsection{Type IIB parameterization}

We can transform the $E_n$ generators of the M-theory parameterization into the type IIB parameterization by using the linear map \eqref{eq:E8-linear-map}. 
Namely, we act the following operation to the matrix representations of the generators:
\begin{align}
 \cT(T_{\hat{\alpha}})^\sfI{}_\sfJ \equiv (S^\GT)^\sfI{}_K\, (T_{\hat{\alpha}})^K{}_L\, S^L{}_\sfJ\,.
\end{align}
Then, $\cT(T_{\hat{\alpha}})^\sfI{}_\sfJ$ is the matrix representations in the type IIB parameterization. 
The explicit form of the constant matrix $S^I{}_\sfJ$ has been determined such that the algebra of the type IIB generators is closed. 
We also change the name of the generators such that the $\SL(7)\times \SL(2)$ symmetry is manifest. 
Concretely, we convert the non-positive-level generators of the M-theory parameterization into those of the type IIB parameterization as follows:
\begin{align}
\begin{gathered}
 \xymatrix@R-20pt{
 & \cT(K^a{}_b) \ar@{=}[r] & \sfK^a{}_b \ar@/^/[rd] & \\
 \cT(K^i{}_j) \ar@/^/[ru] \ar[r] \ar@/_/[rd] \ar@/_/[rdd] & \cT\bigl(\tfrac{K^a{}_a-2\, K^\alpha{}_\alpha}{3}\bigr) \ar@{=}[r] & \sfK^\By{}_\By \ar[r] & \sfK^\sfm{}_\sfn \\
 & \cT\bigl(K^\alpha{}_\beta -\tfrac{1}{2}\,\delta^\alpha_\beta\,K^\gamma{}_\gamma \bigr) \ar@{=}[r] & \epsilon^{\alpha\gamma}\,\sfR_{\gamma\beta} \ar[r] & \sfR_{\alpha\beta} \\
 & \cT(K^\alpha{}_a) \ar@{=}[r] & \sfR^\alpha_{a\By} \ar@/^/[rd] & \\
 & \cT(R_{a_1\Ay\Az}) \ar@{=}[r] & \sfK^\By{}_{a_1} \ar@/_/[ruuu] & \sfR^\alpha_{\sfm_1\sfm_2} \\
  \cT(R_{i_1i_2i_3}) \ar@/^/[ur] \ar[r] \ar@/_/[rd] & \cT(R_{a_1a_2\alpha}) \ar@{=}[r] & \epsilon^\rmT_{\alpha\beta}\,\sfR^\beta_{a_1a_2} \ar@/_/[ru] & \\
 & \cT(R_{a_1a_2a_3}) \ar@{=}[r] & \sfR_{a_1a_2a_3\By} \ar[r] & \sfR_{\sfm_1\cdots\sfm_4} \\
 & \cT(R_{a_1\cdots a_4\Ay\Az}) \ar@{=}[r] & \sfR_{a_1\cdots a_4} \ar@/_/[ru] & \\
  \cT(R_{i_1\cdots i_6}) \ar@/^/[ur] \ar[r] \ar@/_/[rd] & \cT(R_{a_1\cdots a_5\alpha}) \ar@{=}[r] & \epsilon^\rmT_{\alpha\beta}\,\sfR^\beta_{a_1\cdots a_5\By} \ar[r] & \sfR^\alpha_{\sfm_1\cdots \sfm_6} \\
 & \cT(R_{a_1\cdots a_6}) \ar@{=}[r] & \sfR_{a_1\cdots a_6\By,\By} \ar@/^/[rd] & \\
  \cT(R_{i_1\cdots i_8,i}) \ar[r] \ar@/_/[rd] & \cT(R_{a_1\cdots a_6\Ay\Az,\alpha}) \ar@{=}[r] & \epsilon^\rmT_{\alpha\beta}\,\sfR^\beta_{a_1\cdots a_6} \ar@/_/[ruu] & \sfR_{\sfm_1\cdots\sfm_7,\sfm} \\
 & \cT(R_{a_1\cdots a_6\Ay\Az,a}) \ar@{=}[r] & \sfR_{a_1\cdots a_6\By,a} \ar@/_/[ru]& 
 }
\end{gathered}
\end{align}
A similar map for the positive-level generators can be found by taking the generalized transpose; e.g.,
\begin{align}
 \cT(R^{a_1a_2a_3}) = \bigl[\cT(R_{a_1a_2a_3})\bigr]^\GT = \bigl[\sfR_{a_1a_2a_3\By}\bigr]^\GT = \sfR^{a_1a_2a_3\By}\,. 
\end{align}
We then obtain the $E_n$ generators ($n\leq 8$) in the type IIB parameterization:
\begin{align}
 \{\sfT_{\bm{\alpha}}\} = \{\sfK^\sfm{}_\sfn ,\, \sfR_{\alpha\beta},\,\sfR_\alpha^{\sfm_{12}},\,\sfR^{\sfm_{1\cdots 4}},\,\sfR^{\sfm_{1\cdots6}}_\alpha,\,\sfR^{\sfm_{1\cdots 7},\sfm}\,\sfR^\alpha_{\sfm_{12}},\,\sfR_{\sfm_{1\cdots 4}},\,\sfR_{\sfm_{1\cdots6}}^\alpha,\,\sfR_{\sfm_{1\cdots 7},\sfm}\}\,. 
\end{align}
By using the notations
\begin{align}
 \bdelta_{\sfm_1\cdots \sfm_n}^{\sfn_1\cdots \sfn_n}\equiv n!\,\delta_{\sfm_1\cdots \sfm_n}^{\sfn_1\cdots \sfn_n}\,,\qquad 
 (\cdots)_{\sfm_{1\cdots p}} \equiv (\cdots)_{\sfm_1\cdots \sfm_p}\,,\qquad 
 \delta^{\alpha\beta}_{\gamma\delta}\equiv \delta^{(\alpha}_{(\gamma}\delta^{\beta)}_{\delta)}\,,
\end{align}
their matrix representations are found as follows:
\begin{align}
 \sfK^{\sfr}{}_{\sfs} &\equiv {\arraycolsep=0.2mm \underset{10\times10}{\mathrm{diag}} 
 \begin{pmatrix}
 -\delta_{\sfs}^\sfm \delta_\sfn^{\sfr} \\[-1mm]
 \delta^\alpha_\beta \delta_{\sfm}^{\sfr} \delta_{\sfs}^{\sfn} \\
 \frac{\frac{1}{2!}\bdelta_{\sfm_{123}}^{\sfr\sft_{12}} \bdelta_{\sfs\sft_{12}}^{\sfn_{123}}}{\sqrt{3!\,3!}}\\[1mm]
 \frac{\frac{1}{4!}\delta^\alpha_\beta \bdelta_{\sfm_{1\cdots 5}}^{\sfr\sft_{1\cdots 4}} \bdelta_{\sfs\sft_{1\cdots 4}}^{\sfn_{1\cdots 5}}}{\sqrt{5!\,5!}} \\[1mm]
 \frac{\frac{1}{5!}\bdelta_{\sfm_{1\cdots 6}}^{\sfr\sft_{1\cdots 5}} \bdelta_{\sfs\sft_{1\cdots 5}}^{\sfn_{1\cdots 6}}\delta_\sfm^\sfn +\bdelta_{\sfm_{1\cdots 6}}^{\sfn_{1\cdots 6}} \delta_{\sfm}^{\sfr}\delta_{\sfs}^\sfn}{\sqrt{6!\,6!}} \\[1mm]
 \frac{\frac{1}{6!}\delta^{\alpha_{12}}_{\beta_{12}}\bdelta_{\sfm_{1\cdots 7}}^{\sfr\sft_{1\cdots 6}} \bdelta_{\sfs\sft_{1\cdots 6}}^{\sfn_{1\cdots 7}}}{\sqrt{7!\,7!}} \\[1mm]
 \frac{\delta^\alpha_\beta(\frac{1}{6!}\bdelta_{\sfm_{1\cdots 7}}^{\sfr\sft_{1\cdots 6}} \bdelta_{\sfs\sft_{1\cdots 6}}^{\sfn_{1\cdots 7}}\bdelta_{\sfp_{12}}^{\sfq_{12}} + \bdelta_{\sfm_{1\cdots 7}}^{\sfn_{1\cdots 7}} \bdelta_{\sfp_{12}}^{\sfr\sft}\bdelta_{\sfs\sft}^{\sfq_{12}})}{\sqrt{7!\,2!\,7!\,2!}} \\[1mm]
 \frac{\frac{1}{6!}\bdelta_{\sfm_{1\cdots 7}}^{\sfr\sft_{1\cdots 6}} \bdelta_{\sfs\sft_{1\cdots 6}}^{\sfn_{1\cdots 7}}\bdelta_{\sfq_{1\cdots 4}}^{\sfp_{1\cdots 4}} +\frac{1}{3!}\bdelta_{\sfm_{1\cdots 7}}^{\sfn_{1\cdots 7}} \bdelta_{\sfp_{1\cdots 4}}^{\sfr\sft_{123}}\bdelta_{\sfs\sft_{123}}^{\sfq_{1\cdots 4}}}{\sqrt{7!\,4!\,7!\,4!}} \\[1mm]
 \frac{\delta^\alpha_\beta (\frac{1}{6!}\bdelta_{\sfm_{1\cdots 7}}^{\sfr\sft_{1\cdots 6}} \bdelta_{\sfs\sft_{1\cdots 6}}^{\sfn_{1\cdots 7}}\bdelta_{\sfp_{1\cdots 6}}^{\sfq_{1\cdots 6}} + \frac{1}{5!}\bdelta_{\sfm_{1\cdots 7}}^{\sfn_{1\cdots 7}} \bdelta_{\sfp_{1\cdots 6}}^{\sfr\sft_{1\cdots 5}}\bdelta_{\sfs\sft_{1\cdots 5}}^{\sfq_{1\cdots 6}})}{\sqrt{7!\,6!\,7!\,6!}} \\[1mm]
 \frac{\frac{1}{6!}\bdelta_{\sfm_{1\cdots 7}}^{\sfr\sft_{1\cdots 6}} \bdelta_{\sfs\sft_{1\cdots 6}}^{\sfn_{1\cdots 7}}\bdelta_{\sfp_{1\cdots 7}}^{\sfq_{1\cdots 7}}\delta_\sfm^\sfn + \frac{1}{6!}\bdelta_{\sfm_{1\cdots 7}}^{\sfn_{1\cdots 7}} \bdelta_{\sfp_{1\cdots 7}}^{\sfr\sft_{1\cdots 6}}\bdelta_{\sfs\sft_{1\cdots 6}}^{\sfq_{1\cdots 7}}\delta_\sfm^\sfn + \bdelta_{\sfm_{1\cdots 7}}^{\sfn_{1\cdots 7}} \bdelta_{\sfp_{1\cdots 7}}^{\sfq_{1\cdots 7}} \delta^{\sfr}_\sfm \delta_{\sfs}^\sfn}{\sqrt{7!\,7!\,7!\,7!}}
\end{pmatrix} - \frac{\delta_{\sfs}^{\sfr}\,\delta^\sfI_\sfJ}{9-n}} \,, 
\\
 \sfR_{\gamma\delta} &\equiv {\footnotesize {\arraycolsep=0.5mm 
 \begin{pmatrix}
 0 & 0 & 0 & 0 & 0 & 0 & 0 & 0 & 0 & 0 \\
 0 & \delta^\alpha_{(\gamma} \epsilon_{\delta)\beta}\delta_\sfm^\sfn & 0 & 0 & 0 & 0 & 0 & 0 & 0 & 0 \\
 0 & 0 & 0 & 0 & 0 & 0 & 0 & 0 & 0 & 0 \\
 0 & 0 & 0 & \frac{\delta^\alpha_{(\gamma} \epsilon_{\delta)\beta}\bdelta^{\sfn_{1\cdots 5}}_{\sfm_{1\cdots 5}}}{\sqrt{5!\,5!}} & 0 & 0 & 0 & 0 & 0 & 0 \\
 0 & 0 & 0 & 0 & 0 & 0 & 0 & 0 & 0 & 0 \\
 0 & 0 & 0 & 0 & 0 & \frac{(\sfR_{\gamma\delta})^{\alpha_{12}}_{\beta_{12}}\,\bdelta^{\sfn_{1\cdots 7}}_{\sfm_{1\cdots 7}}}{\sqrt{7!\,7!}} & 0 & 0 & 0 & 0 \\
 0 & 0 & 0 & 0 & 0 & 0 & \frac{\delta^\alpha_{(\gamma} \epsilon_{\delta)\beta}\bdelta^{\sfn_{1\cdots 7}}_{\sfm_{1\cdots 7}}\,\bdelta^{\sfq_{12}}_{\sfp_{12}}}{\sqrt{7!\,2!\,7!\,2!}} & 0 & 0 & 0 \\
 0 & 0 & 0 & 0 & 0 & 0 & 0 & 0 & 0 & 0 \\
 0 & 0 & 0 & 0 & 0 & 0 & 0 & 0 & \frac{\delta^\alpha_{(\gamma} \epsilon_{\delta)\beta}\bdelta^{\sfn_{1\cdots 7}}_{\sfm_{1\cdots 7}}\,\bdelta^{\sfq_{1\cdots 6}}_{\sfp_{1\cdots 6}}}{\sqrt{7!\,6!\,7!\,6!}} & 0 \\
 0 & 0 & 0 & 0 & 0 & 0 & 0 & 0 & 0 & 0 
\end{pmatrix}}}
\nn\\
 &\quad \Bigl[(\sfR_{\gamma\delta})^{\alpha_{12}}_{\beta_{12}} \equiv 
 \delta^{\alpha_1\alpha_2}_{\beta_1(\gamma}\,\epsilon_{\delta)\beta_2}
 + \delta^{\alpha_1\alpha_2}_{\beta_2(\gamma}\,\epsilon_{\delta)\beta_1}\Bigr] \,,
\\
 \sfR_\gamma^{\sfr_{12}}
 &\equiv {\footnotesize{\arraycolsep=0.2mm 
 \begin{pmatrix}
 0 & 0 & 0 & 0 & 0 & 0 & 0 & 0 & 0 & 0 \\
 \delta_\gamma^\alpha\bdelta_{\sfm\sfn}^{\sfr_{12}}\!\!\!\! & 0 & 0 & 0 & 0 & 0 & 0 & 0 & 0 & 0 \\
 0 & \!\!\!\!\frac{\epsilon_{\beta\gamma}\bdelta_{\sfm_{123}}^{\sfn\sfr_{12}}}{\sqrt{3!}}\!\!\!\! & 0 & 0 & 0 & 0 & 0 & 0 & 0 & 0 \\
 0 & 0 & \!\!\!\!\frac{\delta_\gamma^\alpha\bdelta_{\sfm_{1\cdots 5}}^{\sfn_{123}\sfr_{12}}}{\sqrt{5!\,3!}}\!\!\!\! & 0 & 0 & 0 & 0 & 0 & 0 & 0 \\
 0 & 0 & 0 & \!\!\!\!\!\!\frac{\epsilon_{\beta\gamma} \Bigl[\genfrac{}{}{0pt}{1}{c_2\,\bdelta^{\sfn_{1\cdots 5}\sfr_{12}}_{\sfm_{1\cdots 6}\sfm}}{+\bdelta_{\sfm_{1\cdots 6}}^{\sfn_{1\cdots 5}\sft} \bdelta_{\sfm\sft}^{\sfr_{12}}}\Bigr]}{\sqrt{6!\,5!}}\!\!\!\!\!\!\!\!\!\! & 0 & 0 & 0 & 0 & 0 & 0 \\
 0 & 0 & 0 & \!\!\!\!\frac{\delta^{\alpha_{12}}_{\beta\gamma}\bdelta^{\sfn_{1\cdots 5}\sfr_{12}}_{\sfm_{1\cdots 7}}}{\sqrt{7!\,5!}}\!\!\!\! & 0 & 0 & 0 & 0 & 0 & 0 \\
 0 & 0 & 0 & 0 & \!\!\!\!\!\!\!\!\!\!\!\!\frac{\delta_\gamma^\alpha \Bigl[\genfrac{}{}{0pt}{1}{c_2\,\bdelta_{\sfm_{1\cdots 7}}^{\sfn_{1\cdots 6}\sfn} \bdelta_{\sfp_{12}}^{\sfr_{12}}}{+\bdelta_{\sfm_{1\cdots 7}}^{\sfn_{1\cdots 6}\sfs} \bdelta_{\sfs\sft}^{\sfr_{12}} \bdelta_{\sfp_{12}}^{\sft\sfn}}\Bigr]}{\sqrt{7!\,2!\,6!}} & \frac{\delta_{(\beta_1}^\alpha \epsilon_{\beta_2)\gamma} \bdelta_{\sfm_{1\cdots 7}}^{\sfn_{1\cdots 7}}\bdelta_{\sfp_{12}}^{\sfr_{12}}}{\sqrt{7!\,2!\,7!}}\!\!\!\! & 0 & 0 & 0 & 0 \\
 0 & 0 & 0 & 0 & 0 & 0 & \!\!\!\!\!\!\!\!\!\!\!\!\!\!\!\! \frac{\epsilon_{\beta\gamma}\bdelta^{\sfn_{1\cdots 7}}_{\sfm_{1\cdots 7}}\bdelta^{\sfq_{12}\sfr_{12}}_{\sfp_{1\cdots 4}}}{\sqrt{7!\,4!\,7!\,2!}}\!\!\!\!\!\!\!\! & 0 & 0 & 0 \\
 0 & 0 & 0 & 0 & 0 & 0 & 0 & \!\!\!\!\frac{\delta_\gamma^\alpha\bdelta^{\sfn_{1\cdots 7}}_{\sfm_{1\cdots 7}}\bdelta^{\sfp_1\cdots\sfq_4\sfr_{12}}_{\sfp_{1\cdots 6}}}{\sqrt{7!\,6!\,7!\,4!}}\!\!\!\!\!\!\!\!\!\!\!\! & 0 & 0 \\
 0 & 0 & 0 & 0 & 0 & 0 & 0 & 0 & \!\!\!\!\!\!\!\!\!\frac{\epsilon_{\beta\gamma}\bdelta^{\sfn_{1\cdots 7}}_{\sfm_{1\cdots 7}}\bdelta^{\sfq_{1\cdots 6}\sft}_{\sfp_{1\cdots 7}}\bdelta^{\sfr_{12}}_{\sfm\sft}}{\sqrt{7!\,7!\,7!\,6!}} & 0 
\end{pmatrix}}}
\nn\\
 &\quad \biggl[c_2 \equiv \frac{4+\sqrt{2}}{14}\biggr]\,,
\\
 \sfR^\gamma_{\sfr_{12}}
 &\equiv {\footnotesize{\arraycolsep=0.2mm 
 \begin{pmatrix}
 0 & \,\delta^\gamma_\beta\bdelta^{\sfn\sfm}_{\sfr_{12}}\!\! & 0 & 0 & 0 & 0 & 0 & 0 & 0 & 0 \\
 0 & 0 & \!\!\!\!\frac{\epsilon^{\alpha\gamma}\bdelta^{\sfn_{123}}_{\sfm\sfr_{12}}}{\sqrt{3!}}\!\!\!\! & 0 & 0 & 0 & 0 & 0 & 0 & 0 \\
 0 & 0 & 0 & \!\!\frac{\delta^\gamma_\beta\bdelta^{\sfn_{1\cdots 5}}_{\sfm_{123}\sfr_{12}}}{\sqrt{3!\,5!}}\!\!\!\!\!\!& 0 & 0 & 0 & 0 & 0 & 0 \\
 0 & 0 & 0 & 0 & \!\!\!\!\frac{\epsilon^{\alpha\gamma} \Bigl[\genfrac{}{}{0pt}{1}{c_2\,\bdelta_{\sfm_{1\cdots 5}\sfr_{12}}^{\sfn_{1\cdots 6}\sfn}}{+\bdelta^{\sfn_{1\cdots 6}}_{\sfm_{1\cdots 5}\sft} \bdelta^{\sfn\sft}_{\sfr_{12}}}\Bigr]}{\sqrt{5!\,6!}} & \frac{\delta_{\beta_{12}}^{\alpha\gamma}\bdelta_{\sfm_{1\cdots 5}\sfr_{12}}^{\sfn_{1\cdots 7}}}{\sqrt{5!\,7!}}\!\!\!\!\!\!\!\!\! & 0 & 0 & 0 & 0 \\
 0 & 0 & 0 & 0 & 0 & 0 & \!\!\!\!\!\!\!\! \frac{\delta^\gamma_\beta \Bigl[\genfrac{}{}{0pt}{1}{c_2\,\bdelta^{\sfn_{1\cdots 7}}_{\sfm_{1\cdots 6}\sfm} \bdelta^{\sfq_{12}}_{\sfr_{12}}}{+\bdelta^{\sfn_{1\cdots 7}}_{\sfm_{1\cdots 6}\sfs} \bdelta^{\sfs\sft}_{\sfr_{12}} \bdelta^{\sfq_{12}}_{\sft\sfm}}\Bigr]}{\sqrt{6!\,7!\,2!}}\!\!\!\!\!\!\!\! & 0 & 0 & 0 \\
 0 & 0 & 0 & 0 & 0 & 0 & \!\!\!\frac{\delta^{(\alpha_1}_\beta \epsilon^{\alpha_2)\gamma} \bdelta^{\sfn_{1\cdots 7}}_{\sfm_{1\cdots 7}}\bdelta^{\sfq_{12}}_{\sfr_{12}}}{\sqrt{7!\,7!\,2!}}\!\!\!\!\!\! & 0 & 0 & 0 \\
 0 & 0 & 0 & 0 & 0 & 0 & 0 & \!\!\!\!\!\!\!\!\!\frac{\epsilon^{\alpha\gamma}\bdelta_{\sfm_{1\cdots 7}}^{\sfn_{1\cdots 7}}\bdelta_{\sfp_{12}\sfr_{12}}^{\sfq_{1\cdots 4}}}{\sqrt{7!\,2!\,7!\,4!}}\!\!\!\! & 0 & 0 \\
 0 & 0 & 0 & 0 & 0 & 0 & 0 & 0 & \!\!\!\!\!\!\!\!\!\frac{\delta^\gamma_\beta\bdelta_{\sfm_{1\cdots 7}}^{\sfn_{1\cdots 7}}\bdelta_{\sfp_1\cdots\sfp_4\sfr_{12}}^{\sfq_{1\cdots 6}}}{\sqrt{7!\,4!\,7!\,6!}}\!\!\!\!\!\!\!\!\! & 0 \\
 0 & 0 & 0 & 0 & 0 & 0 & 0 & 0 & 0 & \!\!\!\!\!\!\!\!\!\!\!\frac{\epsilon^{\alpha\gamma}\bdelta_{\sfm_{1\cdots 7}}^{\sfn_{1\cdots 7}}\bdelta_{\sfp_{1\cdots 6}\sft}^{\sfq_{1\cdots 7}}\bdelta_{\sfr_{12}}^{\sfn\sft}}{\sqrt{7!\,6!\,7!\,7!}}\!\! \\
 0 & 0 & 0 & 0 & 0 & 0 & 0 & 0 & 0 & 0 
\end{pmatrix}}}
\nn\\
 &= (\sfR_\gamma^{\sfr_{12}})^\GT\,,
\\
 \sfR^{\sfr_{1\cdots 4}} &\equiv {\footnotesize {\arraycolsep=0.2mm 
 \begin{pmatrix}
 0 & 0 & 0 & 0 & 0 & 0 & 0 & 0 & 0 & 0 \\
 0 & 0 & 0 & 0 & 0 & 0 & 0 & 0 & 0 & 0 \\
 \frac{\bdelta^{\sfr_{1\cdots 4}}_{\sfm_{123}\sfn}}{\sqrt{3!}} & 0 & 0 & 0 & 0 & 0 & 0 & 0 & 0 & 0 \\
 0 & \!\!\!\!\frac{-\delta_\beta^\alpha \bdelta^{\sfn\sfr_{1\cdots 4}}_{\sfm_{1\cdots 5}}}{\sqrt{5!}}\!\!\!\! & 0 & 0 & 0 & 0 & 0 & 0 & 0 & 0 \\
 0 & 0 & \!\!\!\!\frac{\Bigl[\genfrac{}{}{0pt}{1}{\frac{1}{3!}\bdelta^{\sfn_{123}\sft_{123}}_{\sfm_{1\cdots6}}\bdelta^{\sfr_{1\cdots4}}_{\sfm\sft_{123}}}{+c_4\,\bdelta^{\sfn_{123}\sfr_{1\cdots 4}}_{\sfm_{1\cdots6}\sfm}}\Bigr]}{\sqrt{6!\,3!}}\!\!\!\! & 0 & 0 & 0 & 0 & 0 & 0 & 0 \\
 0 & 0 & 0 & 0 & 0 & 0 & 0 & 0 & 0 & 0 \\
 0 & 0 & 0 & \!\!\!\!\!\!\!\!\!\!\!\!\frac{\delta_\beta^\alpha\bdelta^{\sfn_{1\cdots 5}\sft_{12}}_{\sfm_{1\cdots 7}}\bdelta^{\sfr_{1\cdots 4}}_{\sft_{12}\sfp_{12}}}{2!\sqrt{7!\,2!\,5!}}\!\!\!\!\!\!\!\!\!\!\!\! & 0 & 0 & 0 & 0 & 0 & 0 \\
 0 & 0 & 0 & 0 & \!\frac{\Bigl[\genfrac{}{}{0pt}{1}{\frac{1}{3!}\bdelta^{\sfn_{1\cdots 6}\sfs}_{\sfm_{1\cdots 7}}\bdelta^{\sft_{123}\sfn}_{\sfp_{1\cdots 4}}\bdelta^{\sfr_{1\cdots 4}}_{\sfs\sft_{123}}}{+c_4\,\bdelta^{\sfn_{1\cdots 6}\sfn}_{\sfm_{1\cdots 7}}\bdelta^{\sfr_{1\cdots 4}}_{\sfp_{1\cdots 4}}}\Bigr]}{\sqrt{7!\,4!\,6!}} & 0 & 0 & 0 & 0 & 0 \\
 0 & 0 & 0 & 0 & 0 & 0 & \frac{-\delta_\beta^\alpha\bdelta_{\sfm_{1\cdots 7}}^{\sfn_{1\cdots 7}}\bdelta_{\sfp_{1\cdots 6}}^{\sfq_{12}\sfr_{1\cdots 4}}}{\sqrt{7!\,6!\,7!\,2!}}\!\!\!\!\!\!\!\! & 0 & 0 & 0 \\
 0 & 0 & 0 & 0 & 0 & 0 & 0 & \!\!\!\!\!\!\!\!\frac{\bdelta_{\sfm_{1\cdots 7}}^{\sfn_{1\cdots 7}}\bdelta_{\sfp\sft_{123}}^{\sfq_{1\cdots 4}}\bdelta_{\sfp_{1\cdots 7}}^{\sft_{123}\sfr_{1\cdots 4}}}{3!\sqrt{7!\,7!\,7!\,4!}} & 0 & 0 
\end{pmatrix}}}
\nn\\
 &\quad \biggl[c_4\equiv \frac{4+\sqrt{2}}{7}\biggr]\,,
\\
 \sfR_{\sfr_{1\cdots 4}} &\equiv {\footnotesize {\arraycolsep=0.2mm 
 \begin{pmatrix}
 0 & 0 & \frac{\bdelta_{\sfr_{1\cdots 4}}^{\sfn_{123}\sfm}}{\sqrt{3!}} & 0 & 0 & 0 & 0 & 0 & 0 & 0 \\
 0 & 0 & 0 & \!\!\!\frac{-\delta^\alpha_\beta \bdelta_{\sfm\sfr_{1\cdots 4}}^{\sfn_{1\cdots 5}}}{\sqrt{5!}}\!\!\!\!\!\!\!\!\!\! & 0 & 0 & 0 & 0 & 0 & 0 \\
 0 & 0 & 0 & 0 & \frac{\Bigl[\genfrac{}{}{0pt}{1}{\frac{1}{3!}\bdelta_{\sfm_{123}\sft_{123}}^{\sfn_{1\cdots6}}\bdelta_{\sfr_{1\cdots4}}^{\sfn\sft_{123}}}{+c_4\,\bdelta_{\sfm_{123}\sfr_{1\cdots 4}}^{\sfn_{1\cdots6}\sfn}}\Bigr]}{\sqrt{3!\,6!}} & 0 & 0 & 0 & 0 & 0 \\
 0 & 0 & 0 & 0 & 0 & 0 & \frac{\delta^\alpha_\beta\bdelta_{\sfm_{1\cdots 5}\sft_{12}}^{\sfn_{1\cdots 7}}\bdelta_{\sfr_{1\cdots 4}}^{\sft_{12}\sfq_{12}}}{2!\sqrt{5!\,7!\,2!}}\!\!\!\!\!\!\!\! & 0 & 0 & 0 \\
 0 & 0 & 0 & 0 & 0 & 0 & 0 & \!\!\!\!\!\!\!\!\!\frac{\Bigl[\genfrac{}{}{0pt}{1}{\frac{1}{3!}\bdelta_{\sfm_{1\cdots 6}\sfs}^{\sfn_{1\cdots 7}}\bdelta_{\sft_{123}\sfm}^{\sfq_{1\cdots 4}}\bdelta_{\sfr_{1\cdots 4}}^{\sfs\sft_{123}}}{+c_4\,\bdelta_{\sfm_{1\cdots 6}\sfm}^{\sfn_{1\cdots 7}}\bdelta_{\sfr_{1\cdots 4}}^{\sfq_{1\cdots 4}}}\Bigr]}{\sqrt{6!\,7!\,4!}}\!\!\!\!\!\!\!\!\! & 0 & 0 \\
 0 & 0 & 0 & 0 & 0 & 0 & 0 & 0 & 0 & 0 \\
 0 & 0 & 0 & 0 & 0 & 0 & 0 & 0 & \!\!\!\!\!\!\!\!\!\frac{-\delta^\alpha_\beta\bdelta^{\sfn_{1\cdots 7}}_{\sfm_{1\cdots 7}}\bdelta^{\sfq_{1\cdots 6}}_{\sfp_{12}\sfr_{1\cdots 4}}}{\sqrt{7!\,2!\,7!\,6!}}\!\!\!\!\!\!\! & 0 \\
 0 & 0 & 0 & 0 & 0 & 0 & 0 & 0 & 0 & \!\!\!\!\!\frac{\bdelta^{\sfn_{1\cdots 7}}_{\sfm_{1\cdots 7}}\bdelta^{\sfq\sft_{123}}_{\sfp_{1\cdots 4}}\bdelta^{\sfq_{1\cdots 7}}_{\sft_{123}\sfr_{1\cdots 4}}}{3!\sqrt{7!\,4!\,7!\,7!}} \\
 0 & 0 & 0 & 0 & 0 & 0 & 0 & 0 & 0 & 0 \\
 0 & 0 & 0 & 0 & 0 & 0 & 0 & 0 & 0 & 0 
\end{pmatrix}}}
\nn\\
 &= (\sfR^{\sfr_{1\cdots 4}})^\GT\,,
\\
 \sfR_\gamma^{\sfr_{1\cdots 6}} &\equiv {\footnotesize {\arraycolsep=0.2mm 
 \begin{pmatrix}
 0 & 0 & 0 & 0 & 0 & 0 & 0 & 0 & 0 & 0 \\
 0 & 0 & 0 & 0 & 0 & 0 & 0 & 0 & 0 & 0 \\
 0 & 0 & 0 & 0 & 0 & 0 & 0 & 0 & 0 & 0 \\
 \frac{\delta_\gamma^\alpha\bdelta^{\sfr_{1\cdots6}}_{\sfm_{1\cdots 5}\sfn}}{\sqrt{5!}}\!\!\!\! & 0 & 0 & 0 & 0 & 0 & 0 & 0 & 0 & 0 \\
 0 & \!\!\!\!\!\!\!\!\frac{\epsilon_{\beta\gamma}\Bigl[\genfrac{}{}{0pt}{1}{c_6\,\bdelta^{\sfn\sfr_{1\cdots 6}}_{\sfm_{1\cdots 6}\sfm}}{-\bdelta^{\sfr_{1\cdots 6}}_{\sfm_{1\cdots 6}}\delta_\sfm^\sfn}\Bigr]}{\sqrt{6!}}\!\!\!\!\!\!\!\! & 0 & 0 & 0 & 0 & 0 & 0 & 0 & 0 \\
 0 & \!\!\!\!\frac{-\delta^{\alpha_{12}}_{\beta\gamma}\bdelta^{\sfn\sfr_{1\cdots 6}}_{\sfm_{1\cdots 7}}}{\sqrt{7!}}\!\!\!\! & 0 & 0 & 0 & 0 & 0 & 0 & 0 & 0 \\
 0 & 0 & \!\!\!\!\!\!\!\!\frac{-\delta_\gamma^\alpha\bdelta^{\sfr_{1\cdots 6}\sft}_{\sfm_{1\cdots 7}}\bdelta^{\sfn_{123}}_{\sfp_{12}\sft}}{\sqrt{7!\,2!\,3!}}\!\!\!\!\!\!\! & 0 & 0 & 0 & 0 & 0 & 0 & 0 \\
 0 & 0 & 0 & \!\!\!\!\!\!\!\!\frac{\epsilon_{\beta\gamma}\bdelta^{\sfr_{1\cdots 6}\sft}_{\sfm_{1\cdots 7}}\bdelta^{\sfn_{1\cdots 5}}_{\sfp_{1\cdots 4}\sft}}{\sqrt{7!\,4!\,5!}}\!\!\!\!\!\!\!\!\!\!\!\! & 0 & 0 & 0 & 0 & 0 & 0 \\
 0 & 0 & 0 & 0 & \!\!\!\!\!\!\!\!\frac{-\delta_\gamma^\alpha\Bigl[\genfrac{}{}{0pt}{1}{c_6\,\bdelta^{\sfn_{1\cdots 6}\sfn}_{\sfm_{1\cdots 7}}\bdelta^{\sfr_{1\cdots 6}}_{\sfp_{1\cdots 6}}}{-\bdelta^{\sfn_{1\cdots 6}}_{\sfp_{1\cdots 6}}\bdelta^{\sfn\sfr_{1\cdots 6}}_{\sfm_{1\cdots 7}}}\Bigr]}{\sqrt{7!\,6!\,6!}} & \frac{\delta^\alpha_{(\beta_1} \epsilon_{\beta_2)\gamma} \bdelta^{\sfn_{1\cdots 7}}_{\sfm_{1\cdots 7}}\bdelta^{\sfr_{1\cdots 6}}_{\sfp_{1\cdots 6}}}{\sqrt{7!\,7!\,6!}}\!\!\!\!\!\!\!\!\!\!\!\! & 0 & 0 & 0 & 0 \\
 0 & 0 & 0 & 0 & 0 & 0 & \!\!\!\!\!\!\!\!\!\!\!\!\!\!\frac{\epsilon_{\beta\gamma}\bdelta^{\sfn_{1\cdots 7}}_{\sfm_{1\cdots 7}}\bdelta_{\sfp_{1\cdots 7}}^{\sfr_{1\cdots 6}\sft}\bdelta^{\sfq_{12}}_{\sfm\sft}}{\sqrt{7!\,7!\,7!\,2!}} & 0 & 0 & 0 
\end{pmatrix}}}
\nn\\
 &\quad \Bigl[c_6 \equiv \frac{2-3\sqrt{2}}{14}\Bigr] \,,
\\
 \sfR^\gamma_{\sfr_{1\cdots 6}} &\equiv {\footnotesize {\arraycolsep=0.2mm 
 \begin{pmatrix}
 0 & 0 & 0 & \frac{\delta^\gamma_\beta\bdelta_{\sfr_{1\cdots6}}^{\sfn_{1\cdots 5}\sfm}}{\sqrt{5!}}\!\! & 0 & 0 & 0 & 0 & 0 & 0 \\
 0 & 0 & 0 & 0 & \!\!\!\!\!\!\frac{\epsilon^{\alpha\gamma}\Bigl[\genfrac{}{}{0pt}{1}{c_6\,\bdelta_{\sfm\sfr_{1\cdots 6}}^{\sfn_{1\cdots 6}\sfn}}{-\bdelta_{\sfr_{1\cdots 6}}^{\sfn_{1\cdots 6}}\delta^\sfn_\sfm}\Bigr]}{\sqrt{6!}} & \frac{-\delta_{\beta_{12}}^{\alpha\gamma}\bdelta_{\sfm\sfr_{1\cdots 6}}^{\sfn_{1\cdots 7}}}{\sqrt{7!}} & 0 & 0 & 0 & 0 \\
 0 & 0 & 0 & 0 & 0 & 0 & \!\!\!\!\!\!\!\!\!\!\!\!\frac{-\delta^\gamma_\beta\bdelta_{\sfr_{1\cdots 6}\sft}^{\sfn_{1\cdots 7}}\bdelta_{\sfm_{123}}^{\sfq_{12}\sft}}{\sqrt{3!\,7!\,2!}}\!\!\!\!\!\!\!\!\!\!\!\! & 0 & 0 & 0 \\
 0 & 0 & 0 & 0 & 0 & 0 & 0 & \frac{\epsilon^{\alpha\gamma}\bdelta_{\sfr_{1\cdots 6}\sft}^{\sfn_{1\cdots 7}}\bdelta_{\sfm_{1\cdots 5}}^{\sfq_{1\cdots 4}\sft}}{\sqrt{5!\,7!\,4!}}\!\!\!\!\!\!\!\! & 0 & 0 \\
 0 & 0 & 0 & 0 & 0 & 0 & 0 & 0 & \!\!\!\!\!\!\!\!\!\!\!\!\!\!\!\!\frac{-\delta^\gamma_\beta\Bigl[\genfrac{}{}{0pt}{1}{c_6\,\bdelta_{\sfm_{1\cdots 6}\sfm}^{\sfn_{1\cdots 7}}\bdelta_{\sfr_{1\cdots 6}}^{\sfq_{1\cdots 6}}}{-\bdelta_{\sfm_{1\cdots 6}}^{\sfq_{1\cdots 6}}\bdelta_{\sfm\sfr_{1\cdots 6}}^{\sfn_{1\cdots 7}}}\Bigr]}{\sqrt{6!\,7!\,6!}}\!\!\!\!\!\!\!\!\!\!\!\!\!\!\!\! & 0 \\
 0 & 0 & 0 & 0 & 0 & 0 & 0 & 0 & \!\!\!\!\!\!\!\!\frac{\delta_\beta^{(\alpha_1} \epsilon^{\alpha_2)\gamma} \bdelta_{\sfm_{1\cdots 7}}^{\sfn_{1\cdots 7}}\bdelta_{\sfr_{1\cdots 6}}^{\sfq_{1\cdots 6}}}{\sqrt{7!\,7!\,6!}}\!\!\!\!\!\!\!\! & 0 \\
 0 & 0 & 0 & 0 & 0 & 0 & 0 & 0 & 0 & \!\!\!\!\!\!\!\!\!\!\!\!\frac{\epsilon^{\alpha\gamma}\bdelta_{\sfm_{1\cdots 7}}^{\sfn_{1\cdots 7}}\bdelta^{\sfq_{1\cdots 7}}_{\sfr_{1\cdots 6}\sft}\bdelta_{\sfp_{12}}^{\sfn\sft}}{\sqrt{7!\,2!\,7!\,7!}} \\
 0 & 0 & 0 & 0 & 0 & 0 & 0 & 0 & 0 & 0 \\
 0 & 0 & 0 & 0 & 0 & 0 & 0 & 0 & 0 & 0 \\
 0 & 0 & 0 & 0 & 0 & 0 & 0 & 0 & 0 & 0 
\end{pmatrix}}}
\nn\\
 &= (\sfR_\gamma^{\sfr_{1\cdots 6}})^\GT\,,
\\
 \sfR^{\sfr_{1\cdots 7},\sfr} &\equiv {\footnotesize {\arraycolsep=0.2mm 
 \begin{pmatrix}
 0 & 0 & 0 & 0 & 0 & 0 & 0 & 0 & 0 & 0 \\
 0 & 0 & 0 & 0 & 0 & 0 & 0 & 0 & 0 & 0 \\
 0 & 0 & 0 & 0 & 0 & 0 & 0 & 0 & 0 & 0 \\
 0 & 0 & 0 & 0 & 0 & 0 & 0 & 0 & 0 & 0 \\
 \frac{\Bigl[\genfrac{}{}{0pt}{1}{c_{7,1}\,\delta_\sfn^\sfr\bdelta^{\sfr_{1\cdots 7}}_{\sfm_{1\cdots 6}\sfm}}{-\bdelta^{\sfr_{1\cdots 7}}_{\sfn\sfm_{1\cdots 6}}\delta^\sfr_\sfm}\Bigr]}{\sqrt{6!}} & 0 & 0 & 0 & 0 & 0 & 0 & 0 & 0 & 0 \\
 0 & 0 & 0 & 0 & 0 & 0 & 0 & 0 & 0 & 0 \\
 0 & \frac{\delta_\beta^\alpha \bdelta^{\sfn\sfr}_{\sfp_{12}} \bdelta^{\sfr_{1\cdots 7}}_{\sfm_{1\cdots 7}}}{\sqrt{7!\,2!}} & 0 & 0 & 0 & 0 & 0 & 0 & 0 & 0 \\
 0 & 0 & \frac{\bdelta^{\sfn_{123}\sfr}_{\sfp_{1\cdots 4}} \bdelta^{\sfr_{1\cdots 7}}_{\sfm_{1\cdots 7}}}{\sqrt{7!\,4!\,3!}} & 0 & 0 & 0 & 0 & 0 & 0 & 0 \\
 0 & 0 & 0 & \frac{\delta_\beta^\alpha \bdelta^{\sfn_{1\cdots 5}\sfr}_{\sfp_{1\cdots 6}} \bdelta^{\sfr_{1\cdots 7}}_{\sfm_{1\cdots 7}}}{\sqrt{7!\,6!\,5!}} & 0 & 0 & 0 & 0 & 0 & 0 \\
 0 & 0 & 0 & 0 & \!\frac{\Bigl[\genfrac{}{}{0pt}{1}{c_{7,1}\,\bdelta^{\sfn_{1\cdots 6}\sfn}_{\sfm_{1\cdots 7}}\bdelta^{\sfr_{1\cdots 7}}_{\sfp_{1\cdots 7}}\delta_\sfm^\sfr}{-\bdelta^{\sfn_{1\cdots 6}\sfr}_{\sfm_{1\cdots 7}}\bdelta^{\sfr_{1\cdots 7}}_{\sfp_{1\cdots 7}}\delta^\sfn_\sfm}\Bigr]}{\sqrt{7!\,7!\,6!}} & 0 & 0 & 0 & 0 & 0 
\end{pmatrix}}}
\nn\\
 &\quad \biggl[c_{7,1}\equiv \frac{4+\sqrt{2}}{7\sqrt{2}}\biggr] \,,
\\
 \sfR_{\sfr_{1\cdots 7},\sfr} &\equiv {\footnotesize {\arraycolsep=0.2mm 
 \begin{pmatrix}
 0 & 0 & 0 & 0 & \frac{\Bigl[\genfrac{}{}{0pt}{1}{c_{7,1}\,\delta^\sfm_\sfr\bdelta_{\sfr_{1\cdots 7}}^{\sfn_{1\cdots 6}\sfn}}{-\bdelta_{\sfr_{1\cdots 7}}^{\sfm\sfn_{1\cdots 6}}\delta_\sfr^\sfn}\Bigr]}{\sqrt{6!}} & 0 & 0 & 0 & 0 & 0 \\
 0 & 0 & 0 & 0 & 0 & 0 & \frac{\delta^\alpha_\beta \bdelta_{\sfm\sfr}^{\sfq_{12}} \bdelta_{\sfr_{1\cdots 7}}^{\sfn_{1\cdots 7}}}{\sqrt{7!\,2!}} & 0 & 0 & 0 \\
 0 & 0 & 0 & 0 & 0 & 0 & 0 & \frac{\bdelta_{\sfm_{123}\sfr}^{\sfq_{1\cdots 4}} \bdelta_{\sfr_{1\cdots 7}}^{\sfn_{1\cdots 7}}}{\sqrt{3!\,7!\,4!}} & 0 & 0 \\
 0 & 0 & 0 & 0 & 0 & 0 & 0 & 0 & \frac{\delta^\alpha_\beta \bdelta_{\sfm_{1\cdots 5}\sfr}^{\sfq_{1\cdots 6}} \bdelta_{\sfr_{1\cdots 7}}^{\sfn_{1\cdots 7}}}{\sqrt{5!\,7!\,6!}} & 0 \\
 0 & 0 & 0 & 0 & 0 & 0 & 0 & 0 & 0 & \frac{\Bigl[\genfrac{}{}{0pt}{1}{c_{7,1}\,\bdelta_{\sfm_{1\cdots 6}\sfm}^{\sfn_{1\cdots 7}}\bdelta_{\sfr_{1\cdots 7}}^{\sfq_{1\cdots 7}}\delta^\sfn_\sfr}{-\bdelta_{\sfm_{1\cdots 6}\sfr}^{\sfn_{1\cdots 7}}\bdelta_{\sfr_{1\cdots 7}}^{\sfq_{1\cdots 7}}\delta_\sfm^\sfn}\Bigr]}{\sqrt{6!\,7!\,7!}} \\
 0 & 0 & 0 & 0 & 0 & 0 & 0 & 0 & 0 & 0 \\
 0 & 0 & 0 & 0 & 0 & 0 & 0 & 0 & 0 & 0 \\
 0 & 0 & 0 & 0 & 0 & 0 & 0 & 0 & 0 & 0 \\
 0 & 0 & 0 & 0 & 0 & 0 & 0 & 0 & 0 & 0 \\
 0 & 0 & 0 & 0 & 0 & 0 & 0 & 0 & 0 & 0 
\end{pmatrix}}}
\nn\\
 &=(\sfR^{\sfr_{1\cdots 7},\sfr})^\GT\,.
\end{align}
Again, we can identify the Cartan generators as
\begin{align}
 \{\sfH_\sfk\} = \{\sfK^d{}_d - \sfK^{d+1}{}_{d+1},\, \dotsc,\, \sfK^7{}_7 - \sfK^8{}_8,\, \sfK^8{}_8 + \sfK^9{}_9 -\tfrac{1}{4}\,\sfD -\sfR_{12},\, 2\,\sfR_{12},\, \sfK^8{}_8 - \sfK^9{}_9 \} \,,
\end{align}
and the positive/negative simple-root generators are
\begin{align}
 \{\sfE_\sfk\} = \{\sfK^d{}_{d+1} ,\, \dotsc,\, \sfK^7{}_8,\, \sfR_1^{89},\,\sfR_{22},\,\sfK^8{}_9 \} \,,\qquad
 \{\sfF_\sfk\} = \{\sfK^{d+1}{}_d ,\, \dotsc,\, \sfK^8{}_7,\, \sfR^1_{67},\,-\sfR_{11},\,\sfK^9{}_8 \} \,. 
\end{align}
The set of positive/negative root generators can be summarized as
\begin{align}
\begin{split}
 \{\sfE_{\bm{\alpha}}\} &= \{\sfK^\sfm{}_\sfn\ (\sfm<\sfn),\, \sfR_{22},\,\sfR_\alpha^{\sfm_{12}},\,\sfR^{\sfm_{1\cdots 4}},\,\sfR^{\sfm_{1\cdots6}}_\alpha,\,\sfR^{\sfm_{1\cdots 7},\sfm}\}\,,
\\
 \{\sfF_{\bm{\alpha}}\} &= \{\sfK^\sfm{}_\sfn\ (\sfm>\sfn),\, -\sfR_{11},\,\sfR^\alpha_{\sfm_{12}},\,\sfR_{\sfm_{1\cdots 4}},\,\sfR_{\sfm_{1\cdots6}}^\alpha,\,\sfR_{\sfm_{1\cdots 7},\sfm}\}\,. 
\end{split}
\end{align}

We have checked that the obtained type IIB generators satisfy the following $E_8$ algebra:
\begin{align}
 \bigl[\sfK^\sfm{}_\sfn,\, \sfK^\sfp{}_\sfq \bigr]
 &= \delta_\sfn^\sfp\,\sfK^\sfm{}_\sfq - \delta_\sfq^\sfm\, \sfK^\sfp{}_\sfn \,,
\\
 \bigl[\sfK^\sfm{}_\sfn,\, \sfR^{\sfp_{12}}_\alpha \bigr]
 &= \bdelta_{\sfn\sfr}^{\sfp_{12}} \,\sfR^{\sfm\sfr}_\alpha\,,
\\
 \bigl[\sfK^\sfm{}_\sfn,\, \sfR^{\sfp_{1\cdots 4}} \bigr]
 &= \frac{1}{3!}\,\bdelta_{\sfn\sfr_{123}}^{\sfp_{1\cdots 4}} \, \sfR^{\sfm\sfr_{123}}\,,
\\
 \bigl[\sfK^\sfm{}_\sfn,\, \sfR^{\sfp_{1\cdots 6}}_\alpha \bigr]
 &= \frac{1}{5!}\,\bdelta_{\sfn\sfr_{1\cdots 5}}^{\sfp_{1\cdots 6}} \, \sfR^{\sfm\sfr_{1\cdots 5}}_\alpha\,,
\\
 \bigl[\sfK^\sfm{}_\sfn,\, \sfR^{\sfp_{1\cdots 7},\sfp} \bigr]
 &= \frac{1}{6!}\,\bdelta_{\sfn\sfr_{1\cdots 6}}^{\sfp_{1\cdots 7}} \, \sfR^{\sfm\sfr_{1\cdots 6},\sfp} + \delta_\sfn^\sfp\,\sfR^{\sfp_{1\cdots 7},\sfm}\,,
\\
 \bigl[\sfK^\sfm{}_\sfn,\, \sfR_{\sfp_{12}}^\alpha \bigr]
 &= - \bdelta^{\sfm\sfr}_{\sfp_{12}} \,\sfR_{\sfn\sfr}^\alpha\,,
\\
 \bigl[\sfK^\sfm{}_\sfn,\, \sfR_{\sfp_{1\cdots 4}} \bigr]
 &= - \frac{1}{3!}\,\bdelta^{\sfm\sfr_{123}}_{\sfp_{1\cdots 4}} \, \sfR_{\sfn\sfr_{123}}\,,
\\
 \bigl[\sfK^\sfm{}_\sfn,\, \sfR_{\sfp_{1\cdots 6}}^\alpha \bigr]
 &= - \frac{1}{5!}\,\bdelta^{\sfm\sfr_{1\cdots 5}}_{\sfp_{1\cdots 6}} \, \sfR_{\sfn\sfr_{1\cdots 5}}^\alpha\,,
\\
 \bigl[\sfK^\sfm{}_\sfn,\, \sfR_{\sfp_{1\cdots 7},\sfp} \bigr]
 &= - \frac{1}{6!}\,\bdelta^{\sfm\sfr_{1\cdots 6}}_{\sfp_{1\cdots 7}} \, \sfR_{\sfn\sfr_{1\cdots 6},\sfp} - \delta^\sfm_\sfp\,\sfR_{\sfp_{1\cdots 7},\sfn}\,,
\\
%%%%%
 \bigl[\sfR_{\alpha\beta},\, \sfR_{\gamma\delta} \bigr]
 &= \delta^\sigma_{(\alpha}\epsilon_{\beta)\gamma}\,\sfR_{\sigma\delta} + \delta^\sigma_{(\alpha}\epsilon_{\beta)\delta}\,\sfR_{\gamma\sigma}\,, 
\\
 \bigl[\sfR_{\alpha\beta},\, \sfR^{\sfm_{12}}_\gamma \bigr]
 &= \delta^\sigma_{(\alpha}\epsilon_{\beta)\gamma}\,\sfR^{\sfm_{12}}_\sigma\,,
\\
 \bigl[\sfR_{\alpha\beta},\, \sfR^{\sfm_{1\cdots 6}}_\gamma \bigr]
 &= \delta^\sigma_{(\alpha}\epsilon_{\beta)\gamma}\,\sfR^{\sfm_{1\cdots 6}}_\sigma\,,
\\
 \bigl[\sfR_{\alpha\beta},\, \sfR_{\sfm_{12}}^\gamma \bigr]
 &= -\delta^\gamma_{(\alpha}\epsilon_{\beta)\sigma}\,\sfR_{\sfm_{12}}^\sigma\,,
\\
 \bigl[\sfR_{\alpha\beta},\, \sfR_{\sfm_{1\cdots 6}}^\gamma \bigr]
 &= -\delta^\gamma_{(\alpha}\epsilon_{\beta)\sigma}\,\sfR_{\sfm_{1\cdots 6}}^\sigma\,,
\\
%%%%%
 \bigl[\sfR^{\sfm_{12}}_\alpha,\, \sfR^{\sfn_{12}}_\beta \bigr]
 &= -\epsilon_{\alpha\beta}\,\sfR^{\sfm_{12}\sfn_{12}} \,,
\\
 \bigl[\sfR^{\sfm_{12}}_\alpha,\, \sfR^{\sfn_{1\cdots 4}} \bigr]
 &= \sfR_\alpha^{\sfm_{12}\sfn_{1\cdots 4}} \,,
\\
 \bigl[\sfR^{\sfm_{12}}_\alpha,\, \sfR^{\sfn_{1\cdots 6}}_\beta \bigr]
 &= -\frac{1}{5!}\,\epsilon_{\alpha\beta}\, \bdelta^{\sfn_{1\cdots 6}}_{\sfr_{1\cdots 5}\sfs}\,\sfR^{\sfm_{12}\sfr_{1\cdots 4},\sfs} \,,
\\
 \bigl[\sfR^{\sfm_{12}}_\alpha,\, \sfR_{\sfn_{12}}^\beta \bigr]
 &= \delta_\alpha^\beta\,\bdelta^{\sfm_{12}}_{\sfp\sfr}\,\bdelta_{\sfn_{12}}^{\sfq\sfr}\, \sfK^\sfp{}_\sfq
 - \frac{1}{4}\,\delta_\alpha^\beta\,\bdelta^{\sfm_{12}}_{\sfn_{12}}\,\delta_\sfp^\sfq\, \sfK^\sfp{}_\sfq
 - \epsilon^{\beta\gamma}\,\bdelta^{\sfm_{12}}_{\sfn_{12}}\,\sfR_{\alpha\gamma}\,,
\\
 \bigl[\sfR^{\sfm_{12}}_\alpha,\, \sfR_{\sfn_{1\cdots 4}} \bigr]
 &= \frac{1}{2!}\,\epsilon_{\alpha\beta}\,\bdelta^{\sfm_{12}\sfr_{12}}_{\sfn_{1\cdots 4}}\,\sfR^\beta_{\sfr_{12}} \,,
\\
 \bigl[\sfR^{\sfm_{12}}_\alpha,\, \sfR_{\sfn_{1\cdots 6}}^\beta \bigr]
 &= -\frac{1}{4!}\,\delta_\alpha^\beta\, \bdelta_{\sfn_{1\cdots 6}}^{\sfm_{12}\sfr_{1\cdots 4}}\,\sfR_{\sfr_{1\cdots 4}} \,,
\\
 \bigl[\sfR^{\sfm_{12}}_\alpha,\, \sfR_{\sfn_{1\cdots 7},\sfn} \bigr]
 &= \frac{1}{5!}\,\epsilon_{\alpha\beta}\, \bdelta_{\sfn_{1\cdots 7}}^{\sfm_{12}\sfr_{1\cdots 5}}\,\sfR^\beta_{\sfr_{1\cdots 5}\sfn} \,,
\\
%%%%%
 \bigl[\sfR^{\sfm_{1\cdots 4}},\, \sfR^{\sfn_{1\cdots 4}} \bigr]
 &= \frac{1}{3!}\,\bdelta^{\sfn_{1\cdots 4}}_{\sfr_{123}\sfs}\,\sfR^{\sfm_{1\cdots 4}\sfr_{123},\sfs} \,,
\\
 \bigl[\sfR^{\sfm_{1\cdots 4}},\, \sfR_{\sfn_{12}}^\alpha\bigr]
 &= \frac{1}{2!}\,\epsilon^{\alpha\beta}\,\bdelta_{\sfn_{12}\sfr_{12}}^{\sfm_{1\cdots 4}}\,\sfR_\beta^{\sfr_{12}} \,,
\\
 \bigl[\sfR^{\sfm_{1\cdots 4}},\, \sfR_{\sfn_{1\cdots 4}} \bigr]
 &= \frac{1}{3!}\,\bdelta^{\sfm_{1\cdots 4}}_{\sfp\sfr_{123}}\,\bdelta_{\sfn_{1\cdots 4}}^{\sfq\sfr_{123}}\, \sfK^\sfp{}_\sfq
 - \frac{1}{2}\, \bdelta^{\sfm_{1\cdots 4}}_{\sfn_{1\cdots 4}}\,\delta_\sfp^\sfq\, \sfK^\sfp{}_\sfq
\\
 \bigl[\sfR^{\sfm_{1\cdots 4}},\, \sfR_{\sfn_{1\cdots 6}}^\alpha \bigr]
 &= \frac{1}{2!}\,\bdelta^{\sfm_{1\cdots 4}\sfr_{12}}_{\sfn_{1\cdots 6}}\, \sfR^\alpha_{\sfr_{12}}\,,
\\
 \bigl[\sfR^{\sfm_{1\cdots 4}},\, \sfR_{\sfn_{1\cdots 7},\sfn} \bigr]
 &= -\frac{1}{3!}\,\bdelta^{\sfm_{1\cdots 4}\sfr_{123}}_{\sfn_{1\cdots 7}}\, \sfR_{\sfr_{123}\sfn}
\\
%%%%%
 \bigl[\sfR^{\sfm_{1\cdots 6}}_\alpha,\, \sfR_{\sfn_{12}}^\beta \bigr]
 &= -\frac{1}{4!}\,\delta^\beta_\alpha\, \bdelta^{\sfm_{1\cdots 6}}_{\sfn_{12}\sfr_{1\cdots 4}}\,\sfR^{\sfr_{1\cdots 4}} \,,
\\
 \bigl[\sfR^{\sfm_{1\cdots 6}}_\alpha,\, \sfR_{\sfn_{1\cdots 4}} \bigr]
 &= \frac{1}{2!}\,\bdelta_{\sfn_{1\cdots 4}\sfr_{12}}^{\sfm_{1\cdots 6}}\, \sfR_\alpha^{\sfr_{12}}\,,
\\
 \bigl[\sfR^{\sfm_{1\cdots 6}}_\alpha,\, \sfR_{\sfn_{1\cdots 6}}^\beta \bigr]
 &= \frac{1}{5!}\,\delta_\alpha^\beta\,\bdelta^{\sfm_{1\cdots 6}}_{\sfp\sfr_{1\cdots 5}}\,\bdelta_{\sfn_{1\cdots 6}}^{\sfq\sfr_{1\cdots 5}}\, \sfK^\sfp{}_\sfq
 - \frac{3}{4}\,\delta_\alpha^\beta\,\bdelta^{\sfm_{1\cdots 6}}_{\sfn_{1\cdots 6}}\,\delta_\sfp^\sfq\, \sfK^\sfp{}_\sfq
 - \epsilon^{\beta\gamma}\,\bdelta^{\sfm_{1\cdots 6}}_{\sfn_{1\cdots 6}}\,\sfR_{\alpha\gamma}\,,
\\
 \bigl[\sfR^{\sfm_{1\cdots 6}}_\alpha,\, \sfR_{\sfn_{1\cdots 7},\sfn} \bigr]
 &= -\epsilon_{\alpha\beta}\,\bdelta^{\sfm_{1\cdots 6}\sfr}_{\sfn_{1\cdots 7}}\, \sfR^\beta_{\sfr\sfn}\,,
\\
%%%%%
 \bigl[\sfR^{\sfm_{1\cdots 7},\sfm},\, \sfR_{\sfn_{12}}^\alpha \bigr]
 &= \frac{1}{5!}\,\epsilon^{\alpha\beta}\, \bdelta^{\sfm_{1\cdots 7}}_{\sfn_{12}\sfr_{1\cdots 5}}\,\sfR_\beta^{\sfr_{1\cdots 5}\sfn} \,,
\\
 \bigl[\sfR^{\sfm_{1\cdots 7},\sfm},\, \sfR_{\sfn_{1\cdots 4}} \bigr]
 &= -\frac{1}{3!}\,\bdelta_{\sfn_{1\cdots 4}\sfr_{123}}^{\sfm_{1\cdots 7}}\, \sfR^{\sfr_{123}\sfm}
\\
 \bigl[\sfR^{\sfm_{1\cdots 7},\sfm} ,\, \sfR_{\sfn_{1\cdots 6}}^\alpha\bigr]
 &= -\epsilon_{\alpha\beta}\,\bdelta_{\sfn_{1\cdots 6}\sfr}^{\sfm_{1\cdots 7}}\, \sfR_\beta^{\sfr\sfm}\,,
\\
 \bigl[\sfR^{\sfm_{1\cdots 7},\sfm} ,\, \sfR_{\sfn_{1\cdots 7},\sfn}\bigr]
 &= \bdelta_{\sfn_{1\cdots 7}}^{\sfm_{1\cdots 7}}\, \sfK^\sfm{}_\sfn\,,
\\
%%%%%
 \bigl[\sfR_{\sfm_{12}}^\alpha,\, \sfR_{\sfn_{12}}^\beta \bigr]
 &= \epsilon^{\alpha\beta}\,\sfR_{\sfm_{12}\sfn_{12}} \,,
\\
 \bigl[\sfR_{\sfm_{12}}^\alpha,\, \sfR_{\sfn_{1\cdots 4}} \bigr]
 &= - \sfR_{\sfm_{12}\sfn_{1\cdots 4}}^\alpha \,,
\\
 \bigl[\sfR_{\sfm_{12}}^\alpha,\, \sfR_{\sfn_{1\cdots 6}}^\beta \bigr]
 &= \frac{1}{5!}\,\epsilon^{\alpha\beta}\, \bdelta_{\sfn_{1\cdots 6}}^{\sfr_{1\cdots 5}\sfs}\,\sfR_{\sfm_{12}\sfr_{1\cdots 5},\sfs} \,,
\\
 \bigl[\sfR_{\sfm_{1\cdots 4}},\, \sfR_{\sfn_{1\cdots 4}} \bigr]
 &= -\frac{1}{3!}\,\bdelta_{\sfn_{1\cdots 4}}^{\sfr_{123}\sfs}\,\sfR_{\sfm_{1\cdots 4}\sfr_{123},\sfs} \,.
\end{align}

\end{document}